\documentclass[fleqn,usenatbib]{mnras}

\usepackage{newtxtext,newtxmath}

\usepackage[T1]{fontenc}
\usepackage{ae,aecompl}
\usepackage{pdflscape}

\usepackage{graphicx}	
\usepackage{amsmath}	
\usepackage{flexisym}	
\usepackage{gensymb}    
\usepackage[flushleft]{threeparttable}

\newcommand{\psra}{PSR\,J0023$+$0923}
\newcommand{\psrb}{PSR\,J0251$+$2606}
\newcommand{\psrc}{PSR\,J0636$+$5129}
\newcommand{\psrd}{PSR\,J0952$-$0607}
\newcommand{\psre}{PSR\,J1544$+$4937}
\newcommand{\psrf}{PSR\,J1641$+$8049}

\title[HiPERCAM observations of black widows]{A black widow population dissection through HiPERCAM multi-band light curve modelling}

\author[D. Mata S\'anchez et al.]{D. Mata S\'anchez$^{1,2,3}$\thanks{E-mail: matasanchez.astronomy@gmail.com}, M. R. Kennedy$^{1,4}$,  C. J. Clark$^{1,5,6}$, R. P. Breton$^{1}$, \newauthor  V. S. Dhillon$^{2,7}$, G. Voisin$^{8,1}$, F. Camilo$^{9}$, S. Littlefair$^{7}$, T. R. Marsh$^{10}$, J. Stringer$^{1}$ \\
$^{1}$Jodrell Bank Centre for Astrophysics, Department of Physics and Astronomy, University of Manchester, Manchester M13 9PL, UK\\
$^{2}$Instituto de Astrof\'{i}sica de Canarias, E-38205 La Laguna, Tenerife, Spain \\
$^{3}$Departamento de astrof\'{i}sica, Universidad de La Laguna, E-38206 La Laguna, Tenerife, Spain \\
$^{4}$ Department of Physics, University College Cork, Cork T12 ND89, Ireland \\
$^{5}$ Max Planck Institute for Gravitational Physics (Albert Einstein Institute), Hannover, Callinstra{\ss}e 38, D-30167 Hannover, Germany\\
$^{6}$ Leibniz Universit\"{a}t Hannover, 30167 Hannover, Germany\\
$^{7}$Department of Physics and Astronomy, University of Sheffield, Sheffield S3 7RH, UK \\
$^{8}$LUTH, Observatoire de Paris, PSL Research University, CNRS, 5 place Jules Janssen, 92195 Meudon, France\\
$^{9}$South African Radio Astronomy Observatory, 2 Fir Street, Observatory 7925, South Africa \\
$^{10}$Department of Physics, University of Warwick, Gibbet Hill Road, Coventry
CV4 7AL, UK
}

\date{Accepted 2023 January 17. Received 2022 September 12; in original form 2022 August 01}

\pubyear{2022}

\begin{document}
\label{firstpage}
\pagerange{\pageref{firstpage}--\pageref{lastpage}}
\maketitle

\begin{abstract}

Black widows are extreme millisecond pulsar binaries where the pulsar wind ablates their low-mass companion stars. In the optical range, their light curves vary periodically due to the high irradiation and tidal distortion of the companion, which allows us to infer the binary parameters. We present simultaneous multi-band observations obtained with the HIPERCAM instrument at the 10.4-m GTC telescope for six of these systems. The combination of this five-band ($u_s g_s r_s i_s z_s$) fast photometer with the world's largest optical telescope enables us to inspect the light curve range near minima. We present the first light curve for \psrf{}, as well as attain a significant increase in signal-to-noise and cadence compared with previous publications for the remaining 5 targets: \psra{}, \psrb{}, \psrc{}, \psrd{} and \psre{}. We report on the results of the light curve modelling with the \textsc{Icarus} code for all six systems, which reveals some of the hottest and densest companion stars known. We compare the parameters derived with the limited but steadily growing black widow population for which optical modelling is available. We find some expected correlations, such as that between the companion star mean density and the orbital period of the system, which can be attributed to the high number of Roche-lobe filling companions. On the other hand, the positive correlation between the orbital inclination and the irradiation temperature of the companion is puzzling. We propose such a correlation would arise if pulsars with magnetic axis orthogonal to their spin axis are capable of irradiating their companions to a higher degree.

\end{abstract}




\section{Introduction}

The study of rapidly rotating neutron stars (millisecond pulsars, MSPs) in binary star systems allows us to address some of the most important questions related to these compact objects, such as how neutron stars are spun up to millisecond periods, or what the maximum mass of a neutron star can be. Depending on the nature of the companion star, the evolutionary history of the binary system can be dramatically different (see, e.g., \citealt{2006csxs.book..623T}). This is particularly true of the ``black widow'' (BW) and ``redback'' (RB) systems, where the neutron star primary is ablating either an ultra-low mass degenerate companion ($M_{\rm c}\sim 0.01~\mathrm{M}_{\odot}$; BWs; \citealt{Fruchter1988+J1957}) or a low mass semi-degenerate companion ($0.1\lesssim M_{\rm c}\lesssim 0.5~ \mathrm{M}_{\odot}$; RBs; \citealt{Roberts2011}). While the radio emission of these sources is driven by the pulsar, the optical light curves are dominated by the secondary star, often showing periodic modulations: at twice the orbital frequency due to the orbital motion of the tidally distorted companion (ellipsoidal modulation), and/or at the orbital frequency due to the changing viewing angle of the severely heated inner facing hemisphere of the secondary star (the ``day-time'' side, irradiated by the pulsar wind). As such, their optical light curves are ideal for high precision modelling. They solve the degeneracies left by pulsar timing in the orbital solution concerning inclination and component masses, as well as allows us to study the companion's characteristics and its interaction with the pulsar wind (see, e.g., \citealt{Breton2013+Modelling}).

The currently known population of Galactic spiders consists of $ \sim 20$ RBs and $\sim 30$ BWs (see, e.g., \citealt{2019ApJ...872...42S} and \citealt{Draghis2019} for reviews on RBs and BWs, respectively). Among them, the fastest spinning MSPs (\citealt{2017ApJ...846L..20B, Hessels2006}) and some of the most massive neutron stars (e.g. \citealt{vanKerkwijk2011,2018ApJ...859...54L}) have been found. These are just a few of the many remarkable properties exhibited by this population, which include long-lasting radio eclipses (e.g. \citealt{2020MNRAS.494.2948P}), gamma-ray pulsations (e.g. \citealt{Nieder2019}), and the discovery of systems transitioning between a pulsar state and an accretion state among the RB kind (transitional MSPs, \citealt{2009Sci...324.1411A}). In this work, we will focus on the BW subclass, which due to their intrinsically fainter companions require state-of-the-art instruments mounted on the largest telescopes to be characterised.

\section{Systems included in the study}

We present high-time resolution, multi-band photometry of 6 BWs, obtained using the HiPERCAM instrument \citep{Dhillon2016+HCAM,Dhillon2021} mounted on the 10.4m Gran Telescopio Canarias (GTC, La Palma, Spain). The main limitation of the study on all of these sources is the quality of their optical light curves, particularly close to minimum light, where they all fade below 26 mag in SDSS-$g'$. Such a faint magnitude at the orbital minimum makes it impossible to obtain phase-resolved spectroscopy, and hampers light curve modelling, as often the determination of the temperature of the cool side of the star requires a detection at this phase (see, e.g., Figure 3 of \citealt{Kaplan2018+J0636}). Table \ref{tab:target_details} summaries the known binary parameters from radio and $\gamma$-ray observations of each target. Below is a description of these sources, reviewing some of their unique properties.

\begin{table*}
	\centering
	\caption{Summary of the properties of the BW systems in this study derived only from radio observations. This includes the pulsar's projected semi-major axis ($x$, reported in light-seconds), spin period (P$_{\rm spin}$), proper motion ($\mu$), spin-down luminosity ($\dot{|E|}$, before the Shklovskii correction; see Sec. \ref{sec:spindown}) and orbital period (P$_{\rm orb}$). The minimum companion mass (M$_{c,{\rm min}}$) comes from assuming a binary inclination of 90\degree\ and a neutron star mass of 1.4 M$_{\odot}$. We also inform about previous detections of radio eclipses.}	
	\begin{tabular}{l c c c c c c c l}
		\hline
		Target		            & $x$         & $P_{\rm spin}$ & $\mu$ & $\dot{|E|}$   & $P_{\rm orb}$ & $M_{c,{\rm min}}$  & Eclipses  & References\\
		& (lts)         & (ms) & $\rm (m.a.s\, yr^{-1})$  & $(10^{34}\rm{erg} \, s^{-1})$              & (hr)          & (M$_{\odot}$) & & \\
		\hline\hline
		\psra{}          & 0.035         & 3.05              & $13.88\pm 0.10$ & $1.6$ & 3.33         & 0.017      & N   & \cite{Hessels2011, BakNielsen2020}\\
		\psrb{}         & 0.066         & 2.54              & $17\pm 3$ & 1.8 & 4.86         & 0.025        & Y & \cite{Cromartie2016, Deneva2021}\\
		\psrc{}          & 0.009         & 2.87              & $3.63\pm 0.07$ & 0.6 & 1.60     & 0.010        & N & \cite{Stovall2014+GBNCC,Alam2021}\\
		\psrd{}          & 0.063         & 1.41              & $-$ & 6.7 & 6.42         & 0.019        & N & \cite{2017ApJ...846L..20B,Nieder2019}\\
		\psre{}          & 0.033         & 2.16              & $-$ & 1.2 & 2.9         & 0.017        & Y  & \cite{Bhattacharyya2013}\\
		\psrf{}          & 0.064         & 2.02              &$39\pm 3$ & 4.3 & 2.18        & 0.041        & Y  & \cite{Stovall2014+GBNCC},\cite{Lynch2018}\\
		\hline
	\end{tabular}
	\label{tab:target_details}
\end{table*}

\subsection{\psra{}}
First discovered as a radio pulsar in a targeted search of unidentified \textit{Fermi} point-like sources \citep{Hessels2011}, \psra{} was later also found to be a $\gamma$-ray pulsar \citep{2013ApJS..208...17A}. In spite of the non-detection of radio eclipses (e.g., \citealt{BakNielsen2020}), the inferred lower limit to the companion star mass ($M_{\rm c}>0.017$ M$_{\odot}$) led to its classification as a potential BW. The optical counterpart to \psra{} was first reported by \cite{Breton2013+Modelling}, and modelling of their sparsely-sampled optical light curve suggested the companion may be significantly under-filling its Roche lobe, with a volume-averaged filling factor of $f_{\rm VA}=0.3\pm0.3$ (defined as the ratio of volumes of the companion star to that of the Roche lobe; see Fig. 10 in \citealt{Stringer2021}). The optical counterpart is very faint, with a peak SDSS $i'/g'$ magnitude of 21.7/23.4. A more recent work \citep{Draghis2019} combining previously published photometry with additional $i$-band observations led them to propose a different solution, favouring a higher filling factor ($f_{\rm VA}=0.72\pm 0.04$), as well as to suggest the presence of a feature in the light curve associated with a potential intrabinary shock (IBS; see, e.g., \citealt{Romani2016+IBS}).

\subsection{\psrb{}}

Like \psra{}, \psrb{} was discovered in a search of unidentified \textit{Fermi} point-like sources \citep{Cromartie2016}, while its ephemeris was further refined by \citet{Deneva2021}, which also reported on radio eclipses. A recently published work \citep{Draghis2019} showed the light curve of the optical counterpart of this system, but their limited phase coverage hampered precise light curve modelling. In particular, they estimated a substantially larger distance ($d=2.3$~kpc) to the system compared to that ($d=1.17$~kpc) estimated from the radio dispersion measure (DM) and the YMW16 Galactic electron density model \citep{Yao2017}.

\subsection{\psrc{}}

\psrc{} was first discovered in the Green Bank North Celestial Cap Pulsar survey \citep{Stovall2014+GBNCC}. The pulsar is notable as it has the fourth shortest known orbital period for a pulsar in a binary. Recently, two studies relating to the optical counterpart of the pulsar were published: \cite{Draghis2018+J0636} and \cite{Kaplan2018+J0636}. In these papers, the authors confirm that the system belongs to the BW class and find that the lower limit to the companion mass is $M_{\rm c}>0.01$ M$_{\odot}$, in spite of the non-detection of radio eclipsing features. Both studies also find that the inclination of the system must be relatively low, with \cite{Draghis2018+J0636} proposing a limit of $i<40$\degree and \cite{Kaplan2018+J0636} favouring $i\sim24\pm4$\degree. The discrepancy here arises from the inclusion of an IBS component in the study performed by \cite{Draghis2018+J0636}.

\cite{Kaplan2018+J0636} noticed that, due to its particularly small $P_{\rm orb}=1.6\,{\rm h}$, \psrc{} companion may be unusually dense for a BW system, with a lower limit on the companion density of 43 g cm$^{-3}$. This density implies that the secondary may be the remnant of a helium white dwarf, and that the system was an ultracompact X-ray binary where the MSP was accreting from the white dwarf companion before becoming a BW. Such an evolutionary scenario has been explored in detail in recent years (\citealt{Deloye2003UCXB}; \citealt{vanHaaften2012+Evolution}; \citealt{Sengar2017UCXB}) and used to explain the existence of the planet around PSR J1719$-$1438 \citep{vanHaaften2012+J1719}.

\subsection{\psrd{}}
The BW \psrd{} is the second fastest spinning MSP \citep{2017ApJ...846L..20B}. While no radio eclipses were detected, this study also identified the faint optical counterpart of the system, with the companion peaking at 22.2 mag in SDSS $r'$. As with \psra{}, model fitting to the sparsely-sampled single-band optical light curve suggested that the companion may be underfilling its Roche lobe, with $f\sim0.5$. A more recent study \citep{Nieder2019} detected and performed a timing model of gamma-ray pulsations from this system, as well as presented new photometric observations of its optical counterpart that led to a higher filling factor ($f\sim0.88$). We include in the present work a re-analysis of the HiPERCAM and ULTRACAM multi-band light curves presented in the aforementioned paper, systemically applying the same analysis performed for all sources in this work and attempting to improve on the data reduction for the redder bands, which are affected by fringing issues.

\subsection{\psre{}}
\psre{} was first identified as an eclipsing BW pulsar using the Giant Metrewave Radio Telescope \citep{Bhattacharyya2013}. It shows deep radio eclipses at 322 MHz, with the eclipse depth decreasing at increasing frequencies (at 607 MHz the pulsar is still visible during the eclipse seen at shorter frequencies, albeit with a reduced pulse amplitude).

The optical companion to the system was discovered by \cite{Tang2014Discovery} and has a \textit{g} band magnitude of $\sim 24.8$ at max and $\sim26.8$ at min. From modelling their optical data, \cite{Tang2014Discovery} concluded that standard pulsar heating models do not match the observed light curve, with better results arising when models which contain asymmetric spots on the secondary star's surface are used. They also find that the secondary star is likely underfilling its Roche lobe and propose that the companion is either a hydrogen brown dwarf or the remnant core of a helium/carbon white dwarf, depending on the exact distance to the object. As with \psrc{}, such a difference in the companion's interior structure has a profound implication for the system evolutionary history, with a hydrogen brown dwarf companion likely meaning that the system has evolved from a low mass X-ray binary state, while a helium/carbon core suggests the system may have evolved from an ultracompact X-ray binary system.

\subsection{\psrf{}}
The discovery of \psrf{} as an eclipsing radio pulsar within a binary system comes from \cite{Stovall2014+GBNCC}. Its characterisation was further refined in \cite{Lynch2018}, who also identified the optical counterpart and reported a peak SDSS $i'$ magnitude of $21.6$. No further optical studies have been performed on this target until now.

\section{Observations and data reduction}\label{sec:obs}

We observed all the above systems using the quintuple-beam camera HiPERCAM \citep{Dhillon2016+HCAM,Dhillon2018+HCAMFL,Dhillon2021} mounted on the Folded Cassegrain focus of the 10.4\,m Gran Telescopio Canarias (La Palma). HiPERCAM uses higher-throughput versions of the SDSS filter set \citep{2010AJ....139.1628D}, which we refer to as {\em Super-SDSS filters}: $u_s$, $g_s$, $r_s$, $i_s$, and $z_s$. The dead time between each frame is 0.008\,s, and each HiPERCAM frame is time-stamped to an absolute accuracy of tens of microseconds using a dedicated GPS system. The unbinned pixel scale of this instrument is $0.08 \arcsec /{\rm pix}$.

The data were reduced using the HiPERCAM data reduction pipeline \citep{Dhillon2018+HCAMFL}. All frames were first de-biased and then flat-fielded, the latter using the median of twilight sky frames taken with the telescope dithering. The $i_s$ and $z_s$-band frames were corrected for fringing by subtracting a scaled fringe frame constructed from deep, dithered images of the night sky. Photometric extraction of the light curves was performed using two extraction algorithms, which were applied to each dataset in parallel and their results compared in order to assess their performance and to avoid potential artefacts associated with each particular method. First, an optimal photometry algorithm \citep{Naylor1998+OptPhot} was used to extract the counts from each target, as well as multiple comparison stars. The object aperture extraction radii were set to 1.4 times the full-width-at-half-maximum (FWHM) of the fitted point spread function (PSF), which maximized the signal-to-noise ratio (SNR) of the extracted light curve. Second, a curve-of-growth algorithm (CoG hereafter, \citealt{1990ASPC....8..312H}) was also employed to extract the counts both from the target and the same comparison stars. We considered aperture radii from 0.6 to 2.0 times the FWHM, in steps of 0.05, and picked for each case the aperture value maximizing the SNR of the extracted light curve. The aperture position of the targets relative to one of the nearby comparison stars was determined from a sum of all the images, and this offset was then held fixed during the reduction to avoid aperture centroiding problems. The effect of atmospheric refraction on the relative aperture positions is negligible due to the similarity in colour between the target and comparison stars and the fact that our observations on each night were approximately centred on meridian transit whenever it was feasible. The sky level was determined from a clipped mean of the counts in an annulus surrounding the target stars and subtracted from the object counts.

The instrumental magnitudes of comparison stars were used to remove the effects of varying transparency through ``ensemble photometry'' \citep{Honeycutt1992+EnsemblePhot}. Absolute calibration of the photomeric light curves was performed using the reported magnitudes for the comparison stars in PanSTARRS catalogue ($g'$,$r'$,$i'$,$z'$, \citealt{2016arXiv161205560C}) and SDSS catalogue ($u'$, \citealt{2019arXiv191202905A}). For those fields where there was no reliable comparison star calibrations in the aforementioned catalogues, we employed instead the flux standard for each night. We note that the colour terms for the conversion between the HiPERCAM Super SDSS filters ($u_s$, $g_s$, $r_s$, $i_s$, $z_s$) and their regular SDSS counterparts ($u'$, $g'$, $r'$, $i'$, $z'$) is still not available (Brown et al., in preparation), but given the common wavelength cut-offs for all pairs of filters, we do not expect large correction terms. As such, and for the only purpose of magnitude calibration, we will consider that both sets of filters are equivalent. Small potential deviations from this assumption should be further attenuated during the light curve modelling through the consideration of small calibration offsets between nights (see Sec. \ref{sec:model}).

Table~\ref{tab:obs_details} summarises the observational conditions for each object, while below we note any peculiarities encountered during data acquisition for each object.

\begin{table*}
	\centering
	\caption{Details of the observations. Because each filter in HIPERCAM and ULTRACAM corresponds to a separate CCD, it is possible to read out different filters with integer multiples of the exposure time. This information is given in the readout column. For example, a readout mode of 2,2,2,1,1 with an exposure time of 10 s means the $u_s$, $g_s$, and $r_s$ filters had effective exposure times of 20s, while $i_s$ and $z_s$ were 10 seconds.}	
	\begin{tabular}{l c c c c c c}
		\hline
		Target		            & Date          & Duration  & Exposure      & Readout       & Binning   & seeing\\
		                        & (UTC)         & (hr)      & Time (s)      & ($u_s$, $g_s$, $r_s$, $i_s$, $z_s$)   & (pixels)    & (\arcsec)\\
		\hline\hline
		\textbf{HIPERCAM}&&&&&\\
		\hline
        \psra{}          & 2019-08-26    & 0.8       & 30\ s         &1,1,1,1,1      &4$\times$4 & 1.0\\
                                & 2019-08-27    & 0.8       & 30\ s         &1,1,1,1,1      &4$\times$4 & 0.9\\
                                & 2019-09-03    & 1.0       & 30\ s         &1,1,1,1,1      &1$\times$1 & 0.9\\
        \psrb{}         & 2019-01-12    & 0.6       & 30\ s         &2,2,2,1,1      &4$\times$4 & 0.8-1.2\\
                                & 2019-01-13    & 1.1       & 30\ s         &2,2,2,1,1      &4$\times$4 & 1.2-3.0\\
                                & 2019-09-06    & 2.35      & 30\ s         &2,2,1,1,1      &4$\times$4 & 0.8-1.2\\
                                & 2019-09-07    & 1.2       & 30\ s         &2,2,1,1,1      &4$\times$4 & 0.8-1.2\\
        \psrc{}          & 2018-11-14    & 2.9       & 30\ s         &2,1,1,1,1      &4$\times$4 & 0.9-1.6\\
                                & 2019-01-12    & 1.6       & 30\ s         &2,1,1,1,1      &4$\times$4 & 0.8\\
        \psrd{}          & 2019-01-12    & 0.94      & 30\ s         &2,2,2,1,1      &4$\times$4 & 0.8-1.4\\
                                & 2019-01-13    & 2.10      & 30\ s         &2,2,2,1,1      &4$\times$4 & 1.2-3.0\\
        \psre{}          & 2018-04-17    & 3.5       & 60\ s         &2,2,2,1,1      &4$\times$4 & 0.8-1.2\\
                                & 2018-04-18    & 2.8       & 60\ s         &2,2,2,1,1      &4$\times$4 & 0.8-2.2\\
        \psrf{}          & 2019-06-05    & 3.08      & 30\ s         &2,2,2,1,1      &3$\times$3 & 1.0-1.6\\
        \hline
		\textbf{ULTRACAM}&&&&&\\
		\hline                                
        \psra{}          & 2016-08-25    & 3.2       & 4.5\ s        &3,1,--,1,--  &2$\times$2 & 0.9-1.5\\
        \psrd{}          & 2018-06-03    & 2.88      & 20\ s         &3,1,--,1,--  &1$\times$1 & 0.9-2.0\\
                                & 2018-06-04    & 2.17      & 20\ s         &3,1,--,1,--  &1$\times$1 & 1.4-2.5\\
                                & 2019-03-02    & 2.75      & 10\ s         &3,1,--,1,--  &1$\times$1 & 0.8-1.4\\
                                & 2019-03-03    & 4.19      & 10\ s         &3,1,--,1,--  &1$\times$1 & 1.5-2.5\\
		\hline
	\end{tabular}
	\label{tab:obs_details}
\end{table*}

\subsection{\psra{}}
There were no useful $r_s$ or $i_s$ data from 2019-09-03 due to internal reflection issues with the $r_s$ CCD and a fault with the $i_s$ CCD at this time. Due to poor weather conditions, no flat field frames were taken on 2019-08-26 or 2019-08-27. As such, all observations were flat fielded using data taken on 2019-09-03. 

The data were reduced in the standard manner previously described for HiPERCAM data. We extracted the counts from \psra{} and 7 additional stars. One of these comparison stars, (PanSTARRS\,119250058165357172) which lies  36.31 \arcsec\ to the south-west of the target, was used as the reference for the PSF fits. The presence of a field star located 1.7\arcsec north-east from the target led us to restrict the aperture size used for extracting the source flux to a radius lower than 12 unbinned pixels (equivalent to 0.96\arcsec), in an attempt to limit the contamination by the interloper. 
Given the limited coverage in orbital phase attained with the HiPERCAM observations, we include in this paper complementary observations performed with the high-speed imaging photometer ULTRACAM \citep{2007Dhillon} installed on the 3.5m New Technology Telescope (NTT) at La Silla observatory (Chile). All data were flat fielded using data taken on the same night. A comparison star (PanSTARRS\,119250058117667419) lying 44.88\arcsec\ to the south-west of the target was used as the reference for the PSF fits. Due to the lower spatial resolution of ULTRACAM and in order to avoid contamination from the interloper star, we restricted the aperture size to be smaller than 3 unbinned pixels (0.9\arcsec), a similar limit to that imposed on the HiPERCAM data.

\subsection{\psrb{}}
Observations of the SDSS standards SA\,97$-$249, SA\,98$-$685 and Wolf 1346 \citep{Smith2002+SDSS} were obtained to flux calibrate the data. The sky flats from the night 2019-01-13 were employed to reduce both this and the previous night of data, while flats from 2019-09-06 were employed to reduce 2019-09-06 and 2019-09-07. We extracted the counts from \psrd{} and six additional stars (four stars in $u_s$ due to the lower number of bright sources). One of these comparison stars (PanSTARRS\,100661480356261625) which lies 27.23\arcsec\ to the north-east of the target was used as the reference for the PSF fits.

\subsection{\psrc{}}
\subsubsection{Optical}
Observations of the SDSS standards SA\,97$-$249 and SA\,98$-$685 \citep{Smith2002+SDSS} were obtained to flux calibrate the data. We extracted the counts from \psrc{} and five additional stars. One of these comparison stars (PanSTARRS\,169780990357687253) which lies 40.2\arcsec\ to the north-east of the target was used as the reference for the PSF fits. 

\subsubsection{Archival Infrared observations}
There are archival $K_s$ and $H$ band observations of \psrc{} taken using NIRC2+AO by the Keck telescope from 2013 March 1, which were used in the modelling performed by \cite{Draghis2018+J0636}. There are a total of 27 $H$ band frames and 28 $K_s$ frames, with an exposure time of 60 s per frame. We reduced the data in a similar manner to \cite{Draghis2018+J0636}, and used the same star (2MASS J06360673+5129070) for flux calibration. The fringing pattern in each image was removed by creating a fringe frame, which is possible due to the dithering performed throughout the observations, with the telescope position shifting every 3 frames.

\subsection{\psrd{}}
Observations of the SDSS standards SA\,97$-$249 and SA\,98$-$685 \citep{Smith2002+SDSS} were obtained to flux calibrate the data. Given the absence of sky flats on 2019-01-12, the sky flats from the night 2019-01-13 were employed to reduce both nights of data. We followed the standard approach previously described to reduce the data, with care given to the $i_s$ and $z_s$ bands which were particularly affected by fringing effects. We extracted the counts from \psrd{} and six additional stars (four stars in $u_s$ due to the lower number of bright sources). One of these comparison stars (PanSTARRS\,100661480356261625) which lies 27.23\arcsec\ to the north-east of the target was used as the reference for the PSF fits. 

We complemented these observations with ULTRACAM data, which were reduced in the standard manner previously described for HiPERCAM data. We extracted the counts from \psrd{} and 7 additional stars. One of these comparison stars (PanSTARRS\,100641480265886119) which lies 34.99\arcsec\ to the south-west of the target was used as the reference for the PSF fits. We used the $u$-band calibrated stars in the field of view of the previous HIPERCAM observations to consistently calibrate the zero point in this band for ULTRACAM data, following \citet{Nieder2019}.

\subsection{\psre{}}
Observations of the SDSS standard Ru 152 \citep{Smith2002+SDSS} were obtained to flux calibrate the data. Given the absence of sky flats on 2018-04-17, the sky flats from the night 2018-04-18 were employed to reduce both nights of data. Following the previously described reduction process, we extracted the photometric light curves from \psrf{} and 8 additional stars. One of these comparison stars (PanSTARRS\,167562360257982793) which lies 20.04\arcsec\ to the north-east of the target was used as the reference for the PSF fits.

\subsection{\psrf{}}

Observations of the SDSS standards Ross 106 and G163$-$50 \citep{Smith2002+SDSS} were obtained on 2019-06-03 to flux calibrate the data. Due to the lack of sky flats obtained during the same night of the observations, we employed the set of twilight sky exposures obtained on 2019-06-06. Following the previously described reduction process, we extract the photometric light curves from \psrf{} and 6 additional stars. One of these comparison stars (PanSTARRS\,205002502413513472) which lies 56.68\arcsec\ to the north-west of the target was used as the reference for the PSF fits.

\section{Modelling}\label{sec:model}

We performed an independent analysis of each BW. The light curves in each of the optical bands for a given source were modelled simultaneously using the \textsc{Icarus} software package \citep{Breton2012+Icarus}. The G\"ottingen Spectral Library\footnote{\url{http://phoenix.astro.physik.uni-goettingen.de}} \citep{Husser2013} produced by the \textsc{phoenix} \citep{Hauschildt1999} stellar atmosphere code was used to construct a photometric grid of synthetic atmosphere models with solar metallicity using built in \textsc{Icarus} routines and the corresponding transmission filters (\citealt{2007Dhillon,Dhillon2021}), which have been used in previous works (see, e.g., \citealt{Clark2021}). We decided to employ these instead of the ATLAS atmosphere models considered in other, recent works (e.g. \citealt{Stringer2021}) due to the former reaching a lower range of temperatures, which is critical for the modelling of BWs. They cover a range in temperatures of $T_{\rm eff}=3000-15000 \, {\rm K}$ and surface gravity of $\log g = 2.5-5.5$. The grids are extrapolated beyond this range if required. While this can introduce significant errors (especially since cooler stars tend to have strong molecular features in their spectra which are not included in \textsc{phoenix}), we find this appropriate as long as the majority of the surface elements on the star's surface are above the minimum temperature of the grid.

We used the \textsc{Multinest} (\citealt{MN1,MN2,MN3}) nested sampling algorithm to explore the parameter space. The binary mass function $f(M_{\rm psr}) = M_{\rm psr} \sin^3 i / (1 + 1/q)^2 = q^3 x^3 4 \pi^2 / G P_{\rm orb}^2$ links the pulsar mass $M_{\rm psr}$ to the binary orbital period $P_{\rm orb}$, inclination angle $i$, mass ratio $q$ (defined as $M_{\rm psr}/M_{\rm c}$, where $M_{\rm c}$ is the companion star mass), and pulsar's projected semi-major axis $x= a_1 \sin i$. The pulsar's timing ephemeris provides extremely precise measurements of $P_{\rm orb}$ and $x$. We therefore fit for $M_{\rm psr}$ and $i$, and derived the mass ratio $q$ at each point accordingly. From these parameters, \textsc{Icarus} generates a star whose surface follows an equipotential surface within its Roche lobe. The stellar radius is parametrised by its Roche lobe filling factor, $f$, defined as the ratio between the distances to the star's ``nose'' ($r_{0}$) and the L1 Lagrange point from the companion star's centre of mass. The star's surface temperature is parametrised by the ``base'' temperature, $T_{\rm base}$, defined as the temperature of the star before gravity darkening is applied (for which we apply an index of $\beta = 0.08$, assuming the star has a convective envelope, \citealt{Lucy1967}). All the light curves presented in this work were fitted using the direct heating model (unless otherwise specified). This model assumes that the temperature of a facet of the companion star is dependent only on its base temperature and whatever irradiating flux is incident on the facet from the pulsar. More complex models which allow for a dependence on the temperature of neighbouring facets via diffusion and convection also exist \citep{Kandel2020,Voisin2020}. Related to this, the base temperature matches the so-called ``night'' temperature of the companion (see, e.g., \citealt{Breton2013+Modelling}) for the direct heating models considered in this work, but it might differ if diffusion and/or convection effects are included \citep{Voisin2020}. The pulsar's heating effect is quantified by the ``irradiating'' temperature, $T_{\rm irr}$. This is defined such that a flux of $F_H = \sigma T_{\rm irr}^4$ at the star's centre of mass (a distance $a$ from the pulsar, i.e. $L_{\rm irr} = 4\pi a^2\sigma T_{\rm irr}^4$), is immediately thermalised and re-radiated by the stellar surface. By this definition, the hottest surface element on the star is that at its nose, which has temperature $T = (T_{\rm irr}^4 (a / (a - r_{0}))^2 + T_{\rm base}^4)^{1/4}$ (before applying gravity darkening). Additionally, we define the heating efficiency as $\epsilon = L_{\rm irr}/ {|\dot{E}_{\rm int}}|$; being $|\dot{E_{\rm int}}|$ the intrinsic spin-down luminosity of the pulsar (see Sec. \ref{sec:spindown}). We also fit for interstellar extinction and reddening, parameterised by the $E(g-r)$ of \citet{Green2018+Av}, which is scaled to each of our filter bands using the coefficients given therein for PanSTARRS filter bands. At each point in the nested sampling, the model light curve is computed, and the resulting chi-squared log-likelihood provided to \textsc{Multinest}. 
 
The flux density light curves with the best fit model to each system, as well as the residuals from the fit are compiled in Appendix \ref{appendix:fluxdensitylc}. The resulting corner plots from modelling each system are shown in Appendix \ref{sec:cornerplots}, and the best-fitting parameters are reported in Table \ref{tab:model_results}.

\subsection{Priors}

The assumed priors for each modelled parameter vary from system to system, reflecting the different prior knowledge for each of them. The following is a general description of the priors that will be applied by default to all the systems, while any particularities will be discussed in the following dedicated subsections.

We assume uniform priors on all parameters, except for the distance, extinction and binary inclination angle, unless otherwise stated. The inclination is drawn from a prior that is uniform in $\cos i$, which implies an isotropic distribution of orbital axes after accounting for projection effects on the sky plane, and implies that edge-on orbits are more likely \textit{a priori}. 

Regarding the distance prior, we consider a few different cases below. For those systems with a measured timing parallax, we adopted a prior using a Gaussian likelihood for the parallax. We also consider constraints on the distance inferred from the population of known binary MSPs in the Galaxy. In particular, we use the \citet{Levin2013+GalMSPs} model for the density of MSPs in the Galactic field, which has a Gaussian profile in Galactic radius with width $\sigma = 4.5$\,kpc, and an exponential profile in height $z$ above the Galactic plane, with scale height $z = 0.5$\,kpc. We multiply the distance prior by the model density along the line-of-sight. The transverse velocities of binary MSPs in the ATNF Pulsar Catalogue can be approximated by an exponential distribution with mean value of $100$\,km\,s$^{-1}$. For those pulsars where proper motion ($\mu$) has been measured, we additionally multiply the distance prior by $\exp(-v_{\rm T} / 100\,\textrm{km s}^{-1})$ (where  $v_{\rm T} = \mu d$ is the transverse velocity) to take this into account. If none of the above are available, we include as a prior the observed dispersion measure (DM) after converting to distance values using available electron density models (\citealt{Yao2017}). Given the higher uncertainty on the distance derived from these models, we used a log-Gaussian prior around the YMW16 DM distance, with the width parameter corresponding to a fractional uncertainty of $45\%$, as estimated by YMW16.

For interstellar reddening and extinction, we used a Gaussian prior for $E(g-r)$ according to the dust maps presented in \citet{Green2018+Av}. We always assume that this value does not change with distance, as the extinction in \citet{Green2018+Av} plateaus above $1$\,kpc in the line-of-sight of all our targets. 

 We also attempt to account for systematic uncertainties in the zero point calibration of the photometric light curves by allowing for a small offset between observations spread over different nights during the modelling. We typically apply a maximum offset of $0.05\,\rm{mag}$ in bands $g_s$,$r_s$,$i_s$ and $z_s$. A slightly higher maximum offset of $0.15\,\rm{mag}$ is considered for the $u_s$ band, due to the absence of $u$-band calibrated stars in the HiPERCAM field of view, which leads us to rely on a single standard star for the zero point calibration.

\begin{table*}
	\centering
	\caption{Best-fit values from the \textsc{Multinest} analysis of the data for the six BWs presented in this paper. The reported preferred values correspond to the median, while 2.5\% and 97.5\% percentiles define the errors bars. If a parameter is pushing against the limits set by the priors, we report instead the $97.5\%$ upper or lower limit, accordingly. We also report the $\chi^2$ and the number of degrees of freedom for each of these best fits as indicative of the fit quality (though these do not include priors). We show separately parameters fitted by the model and those derived from the former. The reported $\epsilon$ value in this table includes the Shklovskii correction on $|\dot{E}|$ for all systems with transversal proper motion, with the exception of \psrf{} (see Sec. \ref{sec:spindown}). }	\begin{tabular}{l l l l l l l}
		\hline
		Parameter		                    & \psra{} & \psrb{}  & \psrc{}  & \psrd{}  & \psre{}  & \psrf{} \\
		\hline\hline
        $q$					                & $59^{+10}_{-11}$   & $30^{+7}_{-3}$   & $89^{+10}_{-16}$ &  $61^{+12}_{-8}$ & $61^{+13}_{-11}$ & $30\pm 5$ \\
        $i$ (deg)			                & $42^{+4}_{-3}$          & $32.2^{+2.0}_{-1.5}$  & $24.0\pm 1.0$ &  $56^{+5}_{-4}$ & $47^{+7}_{-4}$ & $57\pm 2$ \\
        $f$			            & $0.36^{+0.18}_{-0.13}$ &  $0.60^{+0.07}_{-0.06}$  &  $>0.95$ &  $0.87^{+0.02}_{-0.03}$ & $>0.96$& $>0.95$ \\
        $T_{\rm{base}}$ (K)	                & $2780\pm 140$ &  $1090^{+200}_{-50}$ &  $1800^{+300}_{-600}$ &  $2300^{+300}_{-1000}$ & $2870^{+180}_{-160}$ & $3130\pm 160$ \\
        $T_{\rm{irr}}$ (K)	                & $4700\pm200$ &  $3430\pm 70$  & $4600^{+300}_{-200}$ &  $6200\pm 300$ & $4730^{+150}_{-140}$ & $8500\pm 500$ \\
        $d$ (kpc)                           & $1.2^{+0.5}_{-0.3}$   & $1.7^{+0.3}_{-0.2}$  &  $1.1^{+0.3}_{-0.2}$ &  $5.7^{+0.9}_{-0.8}$ & $3.1^{+0.5}_{-0.4}$ & $4.7 \pm 0.6$\\
        \hline
        $\rho_{\rm{c}}$ (g cm$^{-3}$)    & $78^{+192}_{-52}$   &  $10^{+3}_{-2}$ & $41.0^{+0.6}_{-0.3}$ &  $2.8\pm 0.1$ & $12.67^{+0.19}_{-0.12}$ & $23.11^{+0.17}_{-0.12}$ \\
        $\epsilon$   & $0.22^{+0.09}_{-0.07}$   & $0.08^{+0.04}_{-0.02}$    &  $0.21^{+0.09}_{-0.07}$  & $0.28^{+0.13}_{-0.08}$   & $0.17^{+0.06}_{-0.05}$   & $0.41^{+0.19}_{-0.14}$ \\
        $M_{\rm comp} \,(M_{\odot})$  & $0.029^{+0.008}_{-0.008}$   & $0.046^{+0.019}_{-0.008}$    &  $0.021^{+0.004}_{-0.007}$  & $0.025^{+0.009}_{-0.006}$   & $0.026^{+0.009}_{-0.007}$   & $0.055^{+0.016}_{-0.014}$ \\
        \hline
        $\chi^2/{\rm d.o.f.}$               & $4059.66/3585$   & $2832.62/2089$    &  $2571.00/2144$ & $10523.42/9201$  & $933.28/809$  & $1939.31/1827$ \\
		\hline
	\end{tabular}
	\label{tab:model_results}
\end{table*}

\subsection{Potential caveats during light curve modelling of spiders}
\label{sec:caveats}
The light curves presented in this work have been fitted with the so-called direct heating model. This is the fundamental model at the heart of most fits implemented in the literature to reproduce spider light curves. Nevertheless, deviations from this model have been discovered in some spider light curves, and as a consequence modifications have been developed to better describe the observations. Perhaps the most common example is the detection of asymmetries (e.g. \citealt{Kandel2020,Stringer2021}), typically attributed to an emission enhancement from either the leading or trailing hemisphere of the companion. More recently, the discovery of variable spider light curves when comparing observations obtained months or years apart (e.g. \citealt{Stappers2001,Dhillon2022} and \citealt{Cho2018}) have also been a subject of debate. These observational features have been modelled following different approaches, including hotspots in the companion star (e.g. \citealt{Romani2016},\citealt{Clark2021}), asymmetric heating by an IBS \citep{Romani2016+IBS}, convection of heat over the companion surface \citep{Kandel2020,Voisin2020}, or modifications of the gravity darkening law \citep{Romani2021}. Additionally, it is worth noting that the atmosphere models employed in this work correspond to solar metallicities, while some studies have suggested that highly stripped BWs and RBs might require non-solar abundances in order to fully describe their spectra (e.g., \citealt{Kaplan2013,Shahbaz2022}). However, the effect of metallicity at cool temperatures is not well characterised yet, specially for BWs, which would require precise modelling of molecular species (not fully implemented in the PHOENIX models employed in this work). Together with the fact that most spider light-curve modelling has traditionally not accounted for this possibility, we decided to not explore this avenue in the present work, but alert the reader to be cautious of possible systematic biases associated with these effects.

Some of the systems analysed in the current work have been previously observed and analysed by other research groups. In some cases, our analysis of these new sets of observations led to different results, which we recognize could be due to a combination of the aforementioned caveats (e.g., if the light curve of the system is variable over time). For the remainder of this paper, and without the intention of undermining the work performed by earlier authors, we will adopt the fitting results presented in our work as long as they have better statistics than previous attempts (evaluated via their $\chi^2/{\rm d.o.f.}$), which was always the case. We do so to have a set of six different BWs, amounting to over a third of the characterised population, systematically analysed with the same set of instruments, telescopes, and light curve fitting models; helping us to better control the impact of some of the previously described caveats in the reported results.  

\subsection{\psra{}}

\begin{figure*}
    \centering
    \includegraphics[width=\columnwidth]{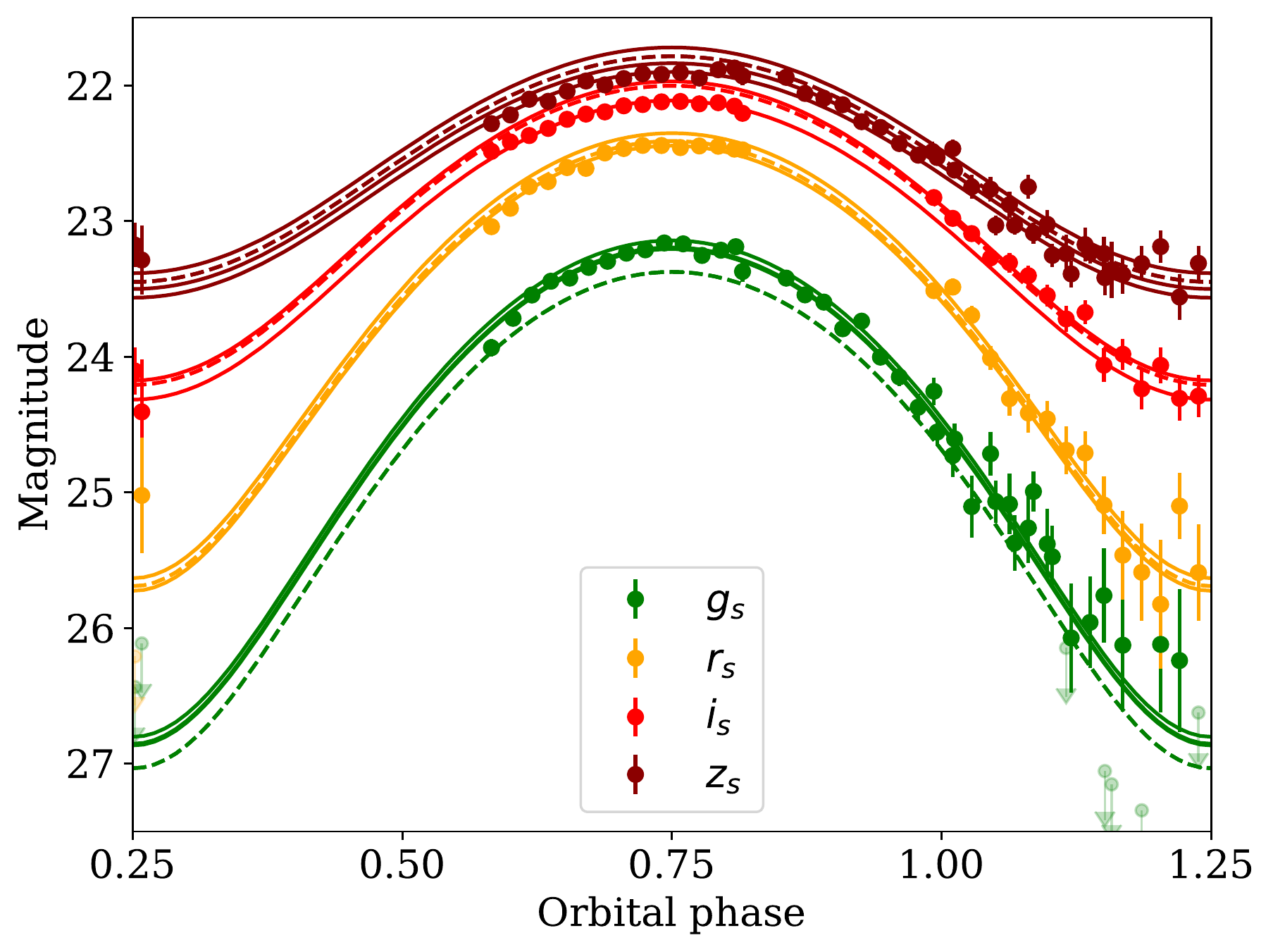}\includegraphics[width=\columnwidth]{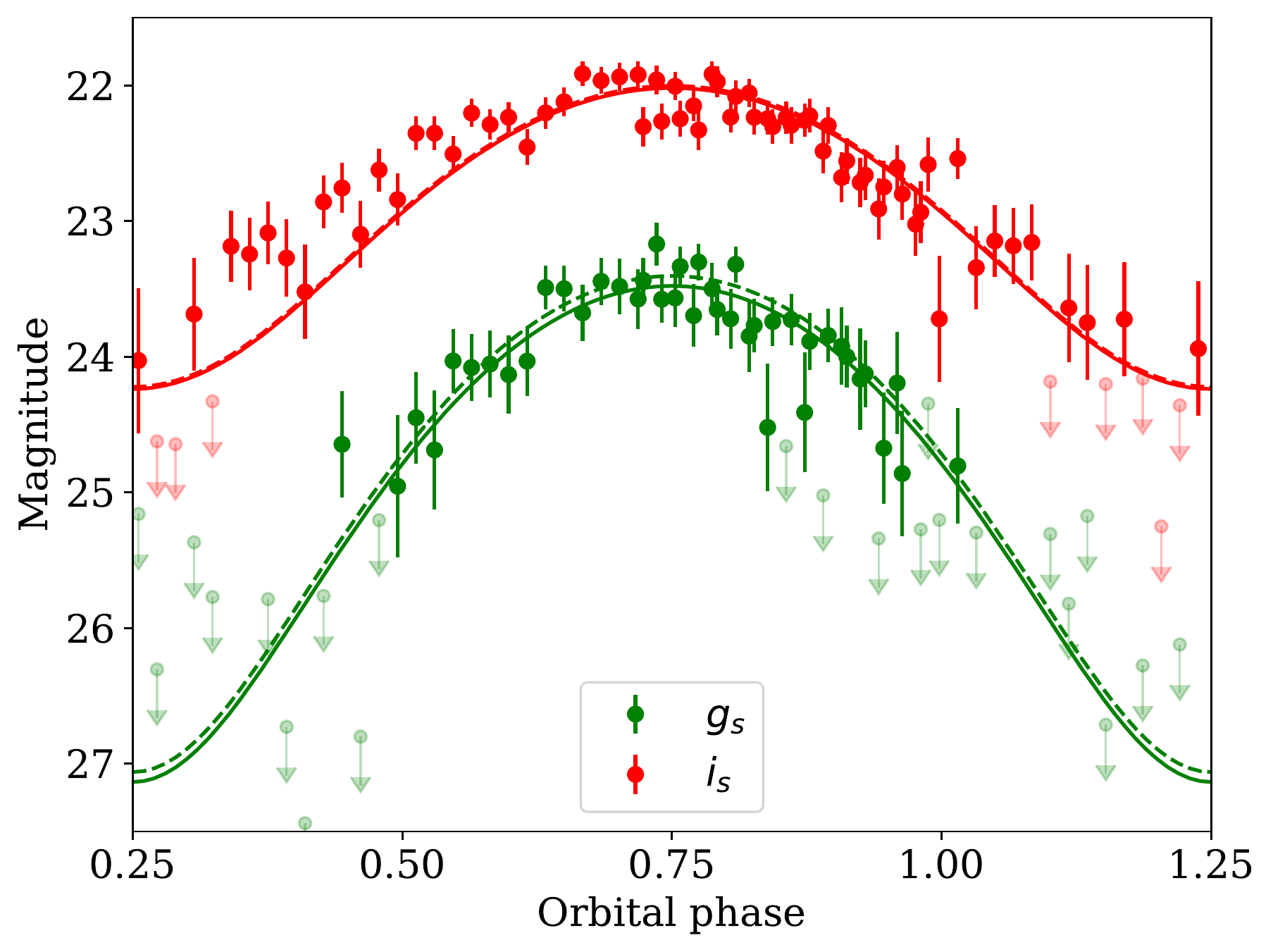}
    \caption{The best-fitting \textsc{Icarus} model for the optical light curve of \psra. Left and right panels correspond to HiPERCAM and ULTRACAM observations, respectively. Dashed lines show the model light curve in each band, while solid curves show the same model but allowing for a small offset in the band calibration so it best fits the data. Due to the simultaneous fit of all datasets, the dashed theoretical model remains the same, while the solid lines differ by simply an offset in magnitude, which varies from night to night. Filled circles correspond to binned magnitudes in 200s stacks, while upper limits (i.e., measurements consistent with null flux at the $2\sigma$ level) are marked as transparent arrows. Note that the fit was performed on the non-binned, flux density data.}
    \label{fig:modelJ0023}
\end{figure*}

We used the ephemeris for the system given by \citet{Arzoumanian2018}, which also allowed us to set priors on the distance from both the proper motion and the timing parallax. We also set an interstellar extinction Gaussian prior of $E(g-r)=0.14\pm 0.03$ from the \citet{Green2018+Av} dust maps. We decided to discard the $u_s$-band light curve during the modelling due to the non-detection of the source at any orbital phase, though we later confirmed that this non-detection was consistent with the best-fit model. The $z_s$-band of the HiPERCAM observations was affected by fringing issues on some nights. Our attempts to correct the data from this effect significantly improved the resulting light curve, particularly on the first two nights of observations, but we nevertheless note that a low amplitude fringing pattern was still detectable in the reduced frames. The small calibration offsets allowed between each night, discussed in the previous subsection, limit the influence of this effect on the modelling results. We used \textsc{Icarus} to model both the optimal and CoG data reductions described in Sec. \ref{sec:obs}, and found consistent best fitting parameters within the calculated uncertainties. However, due to the limitations to aperture size by the presence of an interloper star, the CoG reduction amounts to a fixed aperture extraction. We will hereafter discuss only the optimal extraction results for this system. We also compared the fitting results when allowing for independent calibration offsets for each night of the HiPERCAM data, as well as considering a single offset for the whole dataset, and found consistent results, with the former showing a better fit. The inclusion of ULTRACAM data in the modelling did not significantly change the final results. We present in Figure~\ref{fig:modelJ0023} our best-fit results from the optimal data reduction, allowing for individual night offsets, and fitting simultaneously the HiPERCAM and ULTRACAM datasets.

Due to previous suggestions of an asymmetry being present in the light curve of \psra{} \citep{Draghis2019}, we repeated the fit but including heat diffusion and uniform convection profiles as described in \citet{Voisin2020}. The best-fit parameters obtained were consistent with those derived from the original direct heating model, and it did not show any significant improvement over the former ($\chi^2/{\rm d.o.f.}=4059.64/3585$).

\subsubsection{System parameters}

The best-fit distance for \psra{} is consistent with the timing parallax prior. A previous study of this system \citep{Breton2013+Modelling} reported a similar $T_{\rm base}$, a slightly lower $T_{\rm irr}$ and $i$, and an under-filled Roche lobe fully consistent with that obtained by our modelling (which amounts to $f_{\rm VA}=0.50\pm 0.11$ when using the volume-average definition and $1\sigma$ uncertainty). We note that the small differences between these studies probably arise from a combination of factors, such as different distance priors (as the timing parallax was not known at the time of publication of the aforementioned work), as well as a much better sampled light curve in the data presented in this paper. 

However, further comparison with a more recent paper \citep{Draghis2019} shows remarkable discrepancies in the parameter values. In particular, they propose a much higher inclination ($i=79\pm 13\,{\rm deg}$) and filling factor ($f_{\rm VA}=0.72\pm 0.04$). Before discussing the potential origin of these discrepancies, we would like to acknowledge once more all previous efforts to fit observations of this and any other systems in our sample, and refer the reader to Sec. \ref{sec:caveats} to be aware of known caveats potentially affecting light curve analysis. Assuming that none of these caveats are to blame, we propose the following as the origin of these discrepancies. First and foremost, \citet{Draghis2019} did not set any prior on the distance parameter. This led to their modelling preferring a distance of $d=2.23\pm 0.08\, {\rm kpc}$, doubling the value from the timing parallax reported in \citet{Arzoumanian2018} that we employ as a prior for our models. While the parameter degeneracies prevent a precise assessment of the influence of a larger distance in the remaining fitted parameters, for the system to have the same observed flux, but be twice as far away, would require the companion star to be much larger so as to emit more light. Therefore, it would naturally explain the larger filling factor they report. The effect on the inclination is not as straightforward to assess due to the parameter degeneracy, which we could not properly explore in their fits as they do not provide the corner plot for this particular system. Additionally, it is worth remarking that our light curves have a better sampling over the orbital period than those of \citet{Draghis2019}: we cover all orbital phases with 3 simultaneous bands using ULTRACAM, and complement it by providing photometry in five simultaneous optical bands of HiPERCAM data, which includes a clear detection at minimum for the three redder bands. Our best-fit model produces $\chi^2/{\rm d.o.f.}=4059.66/3585$, a clear improvement when compared with the best previous attempt ($\chi^2/{\rm d.o.f.}=335/62$, which \citealt{Draghis2019} highlight as having the worst reduced $\chi^2$ out of their whole sample). Our results suggest that \psra{} is one of the most under-filled BWs known to date, only comparable with PSR J2256$-$1024 (though the latter suffers from larger uncertainties in the derived parameters, see \citealt{Breton2013+Modelling}). The Roche-lobe under-filling companion star proposed for \psra{} is also consistent with both the non-detection of radio eclipses \citep{BakNielsen2020} and its nearly uniform X-ray light curve \citep{Gentile2014}.

\subsection{\psrb{}}

The ephemeris for this radio eclipsing system was initially determined by \citet{Cromartie2016} and later refined by \citet{Deneva2021}. No timing parallax or proper motion  has been reported for \psrb{}. The DM-derived distance value reported in the former paper ($d\sim 1.17\, {\rm kpc}$, based on \citealt{Yao2017} models) was used to set a broad prior on the distance. We also adopted a Gaussian prior on the Galactic extinction of $E(g-r)=0.14\pm 0.02$ \citep{Green2018+Av}.

\begin{figure}
    \centering
    \includegraphics[width=\columnwidth]{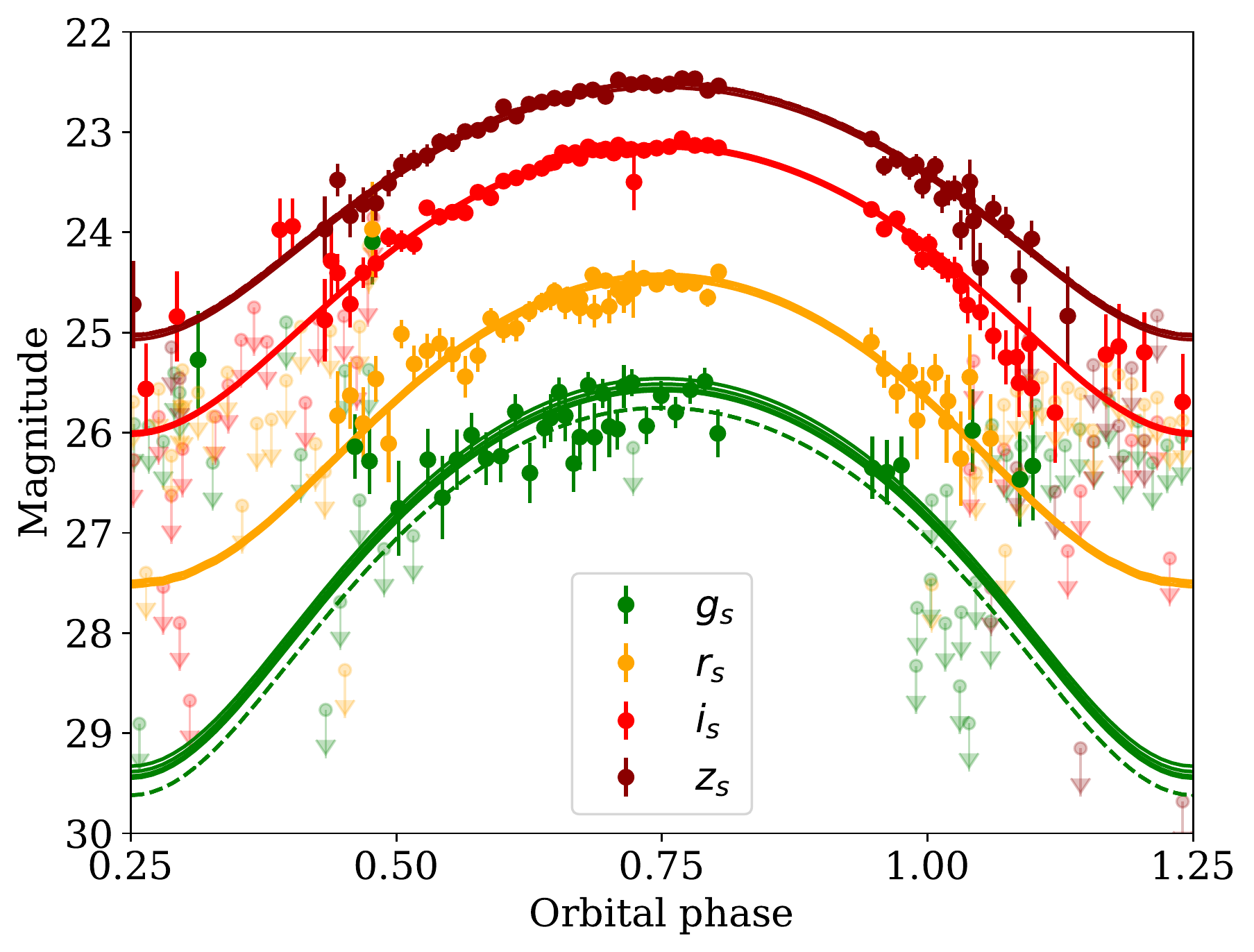}
    \caption{The best-fitting \textsc{Icarus} model for the HiPERCAM optical light curve of \psrb{}. The various lines are as described in Fig. \ref{fig:modelJ0023}.}
    \label{fig:modelJ0251}
\end{figure}

As for \psra{}, we decided to discard the $u_s$-band light curve during the modelling due to the non-detection of the source at any orbital phase. To further justify this decision, we later confirmed that this non-detection was consistent with the preferred model. We also discarded frames with seeing above $2.3\arcsec$ or heavily affected by fringing (even after our best efforts to correct from it), as they introduce artefacts in the observed light curve. In these extreme cases, fringing presents itself as an apparent variable excess over the underlying light curve when the target passes on top of a fringing stripe (due to telescope tracking errors), and affects the redder bands while the bluer filters remain unaffected. The fact that further observations at the same orbital phases but with negligible fringing do not show any hint of such features led us to conclude it is not intrinsic to the system. This affects the $z_s$-band light curve of 2019-01-12 during maximum, as well as most of the 2019-01-13 night in the same band. We used \textsc{Icarus} to model both the optimal and CoG data reductions previously described in Sec. \ref{sec:obs}. The results obtained were consistent with each other within the uncertainties. We will hereafter report the best-fit results (see Figure~\ref{fig:modelJ0251}; also Figure~\ref{fig:corner_plotJ0251}), corresponding to the CoG reduction and allowing for small, individual offsets per night.

\subsubsection{System parameters}

The distance derived from this model prefers a slightly larger value than the DM prior, though still consistent within the reported uncertainties. This could be explained if the local electron density in the line of sight is different to that predicted by \citet{Yao2017}. In this regard, we note that allowing for calibration offsets larger than 0.05 mags during the modelling produces a smaller distance and a slightly better $\chi^2/{\rm d.o.f.}$, while the rest of the parameters remained unchanged. However, we find that this fit systematically underestimates the flux in all bands when compared with the observations, mainly driven by the attempt to accommodate the distance prior. Therefore, we will only consider hereafter the fit requiring smaller offsets.

The temperatures favoured by the best-fit model are particularly low, with $T_{\rm irr}\sim 3400\, {\rm K}$, only just above the atmosphere models' lower limit. Expansion of the models to lower temperatures proved to be challenging due to effects such as the increasing importance of molecular bands when determining the optical emission. Our fit clearly favours low base and irradiation temperatures for this system, and therefore the best-fit parameters presented (in particular, $T_{\rm base}$) should be taken with caution.

We can compare our results with the recently published work from \citet{Draghis2019}. In their paper, they present a light curve only covering $\sim 40\%$ of the orbit (close to the maximum), and show best-fitting results remarkably different to those derived in the present study. In particular, they favour a larger inclination ($52\pm 10\, {\rm deg}$) and a close to Roche-lobe filling solution, at odds with the results presented here. In an attempt to reconcile both results, we binned our light curve and kept only a small range of orbital phases to emulate the conditions of the \citet{Draghis2019} dataset. Modelling of this data still favours a low inclination consistent with the results from our original best-fit (mainly due to larger error bars), but other parameters such as the filling factor appear weakly constrained. Even though we could not reproduce their results (probably due to a combination of the difference in datasets, as well as the different atmosphere models they employ for their analysis), this test shows how incomplete phase coverage can bias the modelling of BWs.

While the results presented here should be taken with caution due to our model atmosphere limitations, our better sampling of the light curve (with over 60 times the number of data points, and covering $\sim 80\%$ of the orbit) led to a better fit with $\chi^2/{\rm d.o.f.}=1.36$ (compared with  $\sim 40\%$ orbital coverage and $\chi^2/{\rm d.o.f.}=2.25$ from \citealt{Draghis2019}). We nevertheless encourage further observations of the system, as well as extension of the model atmospheres to lower temperatures, in order to better understand this BW.

\subsection{\psrc{}}

The ephemeris for this system was obtained from \citet{Arzoumanian2018} which also reported a constraint on the timing parallax of $\varpi = 0.80 \pm 0.33$\,mas. This provides a lower bound on the distance that, together with constraints inferred from the population of known binary MSPs in the Galactic field and their transverse velocities, as well as the total proper motion reported in the aforementioned work, was incorporated in the prior for this parameter. The interstellar extinction prior was implemented as usual from the dust map value \citep{Green2018+Av} of $E(g-r)=0.08 \pm 0.02$. 

\begin{figure}
    \centering
    \includegraphics[width=\columnwidth]{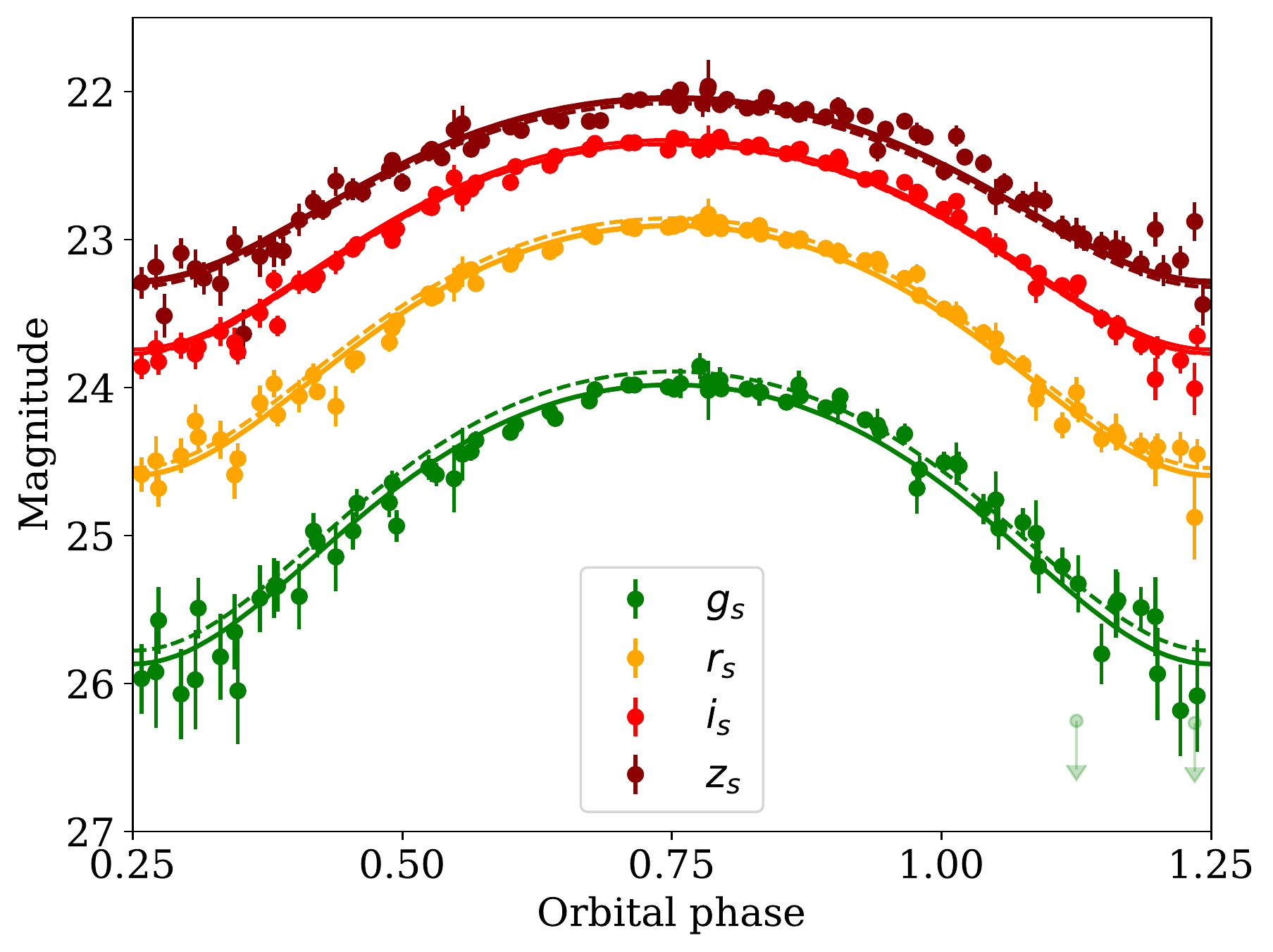}
    \caption{The best-fitting \textsc{Icarus} model for the HiPERCAM optical light curve of \psrc{}. The various lines are as described in Fig. \ref{fig:modelJ0023}.}
    \label{fig:modelJ0636}
\end{figure}

As for \psra{} and \psrb{}, we decided to discard the $u_s$-band light curve during the modelling due to the non-detection of the source, and we later confirmed that it was consistent with the preferred model. The available dataset for this source comprises 2 full orbits with HiPERCAM, as well as a sparsely-sampled archival light curve in two near-infrared bands $K_s$ and $H$ \citep{Draghis2018+J0636}. The inclusion of the IR data made no statistical difference to the fit from solely modelling the HiPERCAM data, and resulted in the same binary parameters. We allowed for independent, small offsets for each of the two nights of data, but also checked that forcing a common offset produced consistent results. We present below the results from the CoG reduction of the data, but note that the optimal reduction also produced values consistent with the best-fit parameters from the CoG case. We present in Figure~\ref{fig:modelJ0636} our best-fit results, as well as the corner plot from the modelling in Figure~\ref{fig:corner_plotJ0636} and the best-fitting parameters in Table \ref{tab:model_results}.

\subsubsection{System parameters}

All the parameters derived in this work are in agreement with those found in \citet{Kaplan2018+J0636}. We are able to reduce the uncertainties on most of the parameters: e.g. the inclination precision improved by a factor of 3, while due to our coverage of the light curve during optical minimum, we provide a tighter measurement of the base temperature of the secondary star. The derived orbital inclination of $i=24\pm 1\, {\deg}$ is the lowest measured in a BW to date, in line with the non-detection of radio eclipses on \psrc{}. Only the filling factor shows a slightly different result (but still remaining consistent within $2\sigma$): while they proposed a loosely constrained $f=0.75\pm0.20$ ($1\sigma$), our model clearly favours a Roche-lobe filling solution, which impacts the derived density for the companion star. \cite{Stovall2014+GBNCC} found that the minimum density for the companion should be $\rho \sim 43$ g cm$^{-3}$. \cite{Kaplan2018+J0636} preferred a much higher density due to their Roche lobe filling factor of 0.75, but were consistent with the lower limit of 43 g cm$^{-3}$ at the $3\sigma$ level. We find that our photometry constrains the filling factor to be $>0.95$ (at the 2$\sigma$ level). This constraint is largely provided by the observations taken at $\phi\sim0.5$ and $\phi\sim1.0$, as data at these orbital phases is taken when the ellipsoidal variations in the light curve caused by the tidal distortion of the secondary star are largest. The density of the companion is close to the minimum density, with a value of $41.0^{+0.6}_{-0.3}\, {\rm g\, cm^{-3}}$. Even then, \psrc{} retains the title of BW with the densest companion of our sample, at least until the filling factor of \psra{} can be better constrained.

Comparison with the direct heating model presented in \citet{Draghis2018+J0636} and revisited in \citet{Draghis2019} also produces consistent parameters with those described above. However, comparison with another proposed model within the latter paper which includes an IBS highlights some tensions, as already described in their work. In particular, the preferred inclination by the latter model is significantly larger ($i=40\, {\rm deg}$). We do note that visual inspection of Fig. \ref{fig:modelJ0636} reveals a slight asymmetry in our light curve, which reaches maximum brightness slightly after orbital phase 0.75. For this reason, we also attempted a fit including convection (uniform profile) and diffusion effects, in order to better reproduce the potential asymmetry \citep{Voisin2020}. It produced system parameters fully compatible with those of the direct heating model, with no significant diffusion component but favouring a convection amplitude of $\nu =-950^{+200}_{-300}\,\rm{W\,K^{-1}\,m^{-2}}$ (following the notation in \citealt{Voisin2020}), which would imply a convection flow rotating
in the opposite direction as the star on its orbit (i.e., making the leading hemisphere of the companion to appear brighter). However, as the fitting statistics did not improve significantly with the latter, more complex fit ($\chi^2/{\rm d.o.f.}=2567.89/2141$), we still favour the results from the direct heating model. At this point, it is important to remark that our direct heating model results, derived here from a better sampled light curve (with over twenty times more data points than that of \citealt{Draghis2018+J0636}), results in a much better $\chi^2/{\rm d.o.f.}=1.2$ than \citet{Draghis2019} IBS model fit to a more sparse dataset ($\chi^2/{\rm d.o.f.}=1.75$).

\subsection{\psrd{}}

We use the binary ephemeris first derived from radio timing \citep{2017ApJ...846L..20B} and later confirmed through the detection of gamma-ray pulses \citep{Nieder2019}. For the interstellar reddening and extinction, we used a Gaussian prior with $E(g-r)= 0.05 \pm 0.03$. No parallax or proper motion measurements are available for this system, so the distance prior is solely based on the known MSP Galactic distribution and the YMW16 DM distance. As usual, we compared the modelling results for different cases: using optimal or CoG reduction, as well as allowing or not for individual offsets for each night. The best results are obtained when analysing optimally-extracted data and allowing for a small offset in each band, varying between the HiPERCAM and ULTRACAM datasets, and also between the two separate epochs obtained with the later instrument, to account for calibration uncertainties.

\begin{figure}
    \centering
    \includegraphics[width=\columnwidth]{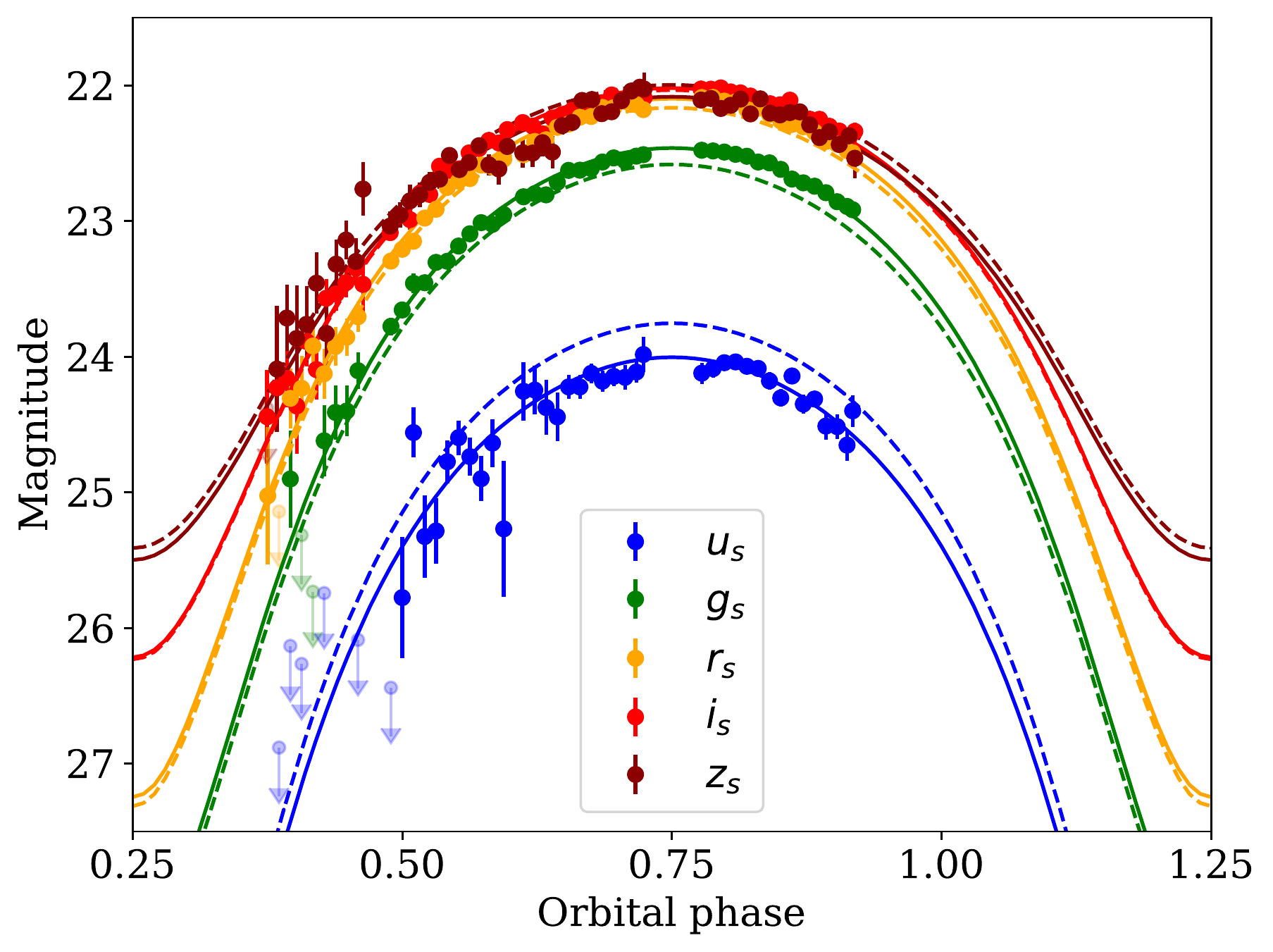}
    \caption{The best-fitting \textsc{Icarus} model for the optical light curve of \psrd. Only HiPERCAM observations are shown here, while ULTRACAM data can be found in  \citet{Nieder2019}. The various lines are as described in Fig. \ref{fig:modelJ0023}.}
    \label{fig:modelJ0952}
\end{figure}

The best-fitting model has been plotted in Fig. \ref{fig:modelJ0952}. The posterior distribution for the model parameters is shown in Fig. \ref{fig:corner_plotJ0952}. The best-fit values for the fitted parameters are collected in Table~\ref{tab:model_results}. The results here obtained are consistent with those previously presented in \citet{Nieder2019}, which given they are based on the same dataset but with a slightly different data reduction (i.e. with an improved fringe correction to the redder bands, see Sec. \ref{sec:obs}), is reassuring.

\subsubsection{System parameters}

The distance preferred by the model is remarkably high when compared with the DM distance estimated using electron density maps \citep{Yao2017}. This result was already noted in \citet{Nieder2019}, who already discussed the potential explanations. They favoured the larger distance obtained from the optical modelling, suggesting the maps are overestimating the electron density in the direction of \psrd{}, based on its gamma-ray efficiency (which would otherwise be unusually low) and the companion star density (otherwise being record-breaking, with over $\sim 100\, {\rm g\, cm^{-3}}$). The absence of radio eclipses for this BW is noteworthy, as it is not favoured by its moderate inclination ($i=56^{+5}_{-4}\, \deg$), high filling factor ($f_{\rm VA}=0.974\pm 0.008$), standard spin-down luminosity ($-\dot{E}= 1.15\cdot 10^{34}\, {\rm erg\, s^{-1}}$) and heating efficiency ($\epsilon =0.28 ^{+0.13}_{-0.08}$).

\subsection{\psre{}}

We decided to fit only the first night of data, where all orbital phases are covered, as the second night is heavily affected by highly variable seeing (1.5-3.0\arcsec). We later checked the validity of our results by applying our best-fit model to this second night and found that it is perfectly consistent, though with a worse $\chi^2$ due to the poorer data quality.

\begin{figure}
    \centering
    \includegraphics[width=\columnwidth]{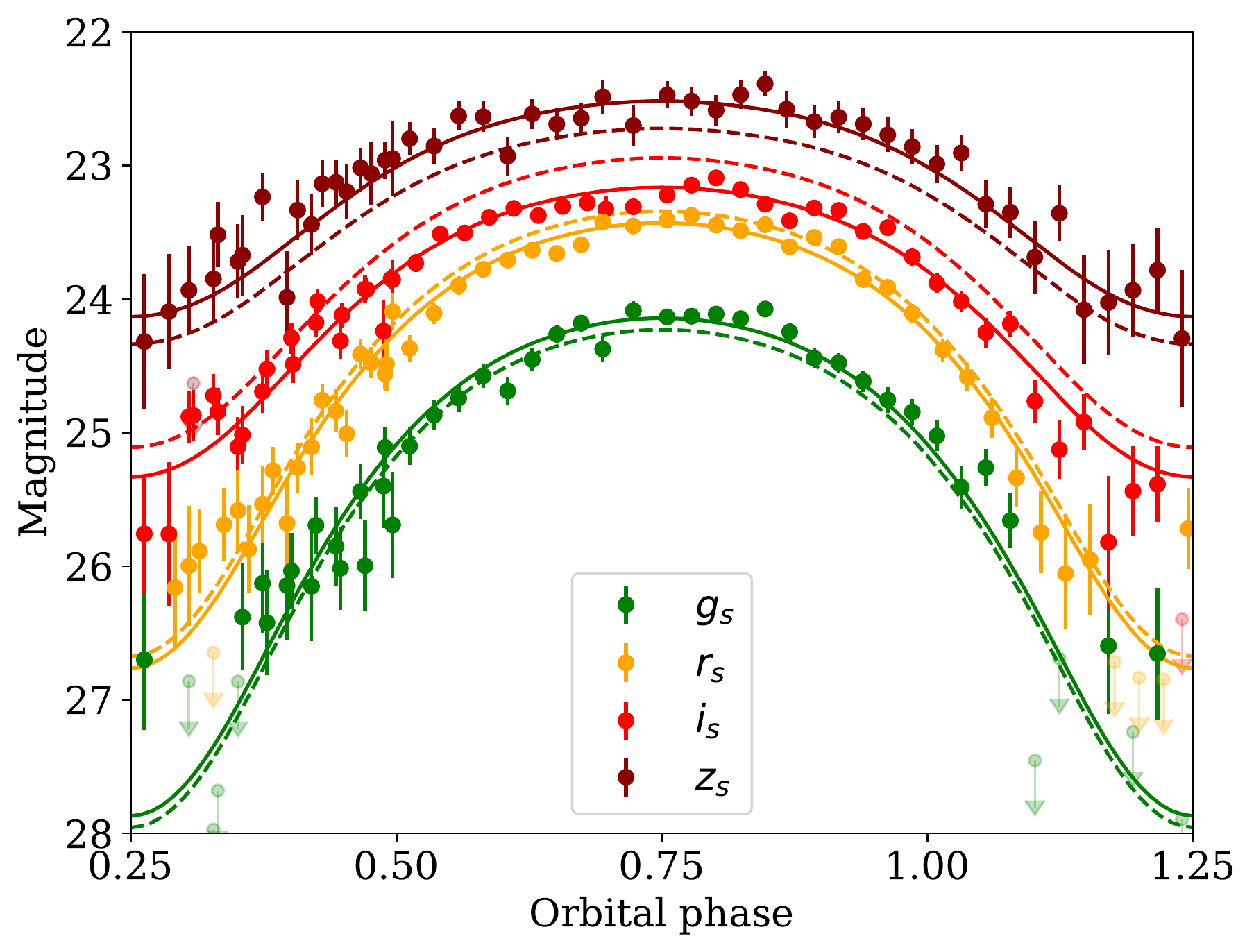}
    \caption{The best-fitting \textsc{Icarus} model for the HiPERCAM optical light curve of \psre{}. Fig. \ref{fig:modelJ0023} description remains valid.}
    \label{fig:modelJ1544}
\end{figure}

The ephemeris used for computing orbital phases is taken from \citet{Bhattacharyya2013}, as well as a prior on the interstellar reddening for this system of $E(g-r)= 0.03\pm 0.02$. Neither the parallax nor the proper motion of \psre{} has been characterised so far, and the proposed distance derived from the DM is $3.0\, \rm{kpc}$ \citep{Yao2017}. We combined this with the information from the known MSP Galactic distribution to construct a broad prior on the distance. The results from the best-fit modelling, corresponding to the CoG reduction of the first night of observations, are described below. The light curve of \psre{} together with the best-fitting model has been plotted in Fig. \ref{fig:modelJ1544}. The posterior distribution for the model parameters is shown in Fig. \ref{fig:corner_plotJ1544}, with the best-fit values compiled in Tab \ref{tab:model_results}.

\subsubsection{System parameters}

As with the other systems in this study, the derived parameters for \psre{} are typical of the BW population. The preferred distance is consistent with the DM prior from the \citet{Yao2017} electron-density models. Our fit adequately describes the observed data with a simple direct heating model, contrary to \citet{Tang2014Discovery}, where they required additional components (i.e. hot-spots) to model the light curve. In order to analyse this situation, we first note that our analysis is based on a different, more complete dataset, evenly covering all orbital phases in five simultaneous bands with a much higher time resolution. On the other hand, the \citet{Tang2014Discovery} light curves are of a longer exposure time, and with 4-bands only simultaneously in pairs ($g$-$I$, $B$-$R$), which produces uneven coverage in orbital phase. A visual inspection shows that the main deviation from the direct heating model appears at the minimum of their light curve, where very low-significance detections are plotted. Our observed light curve appears much better behaved in all bands, and we find that the direct heating model is sufficient to produce a reliable solution for the system. 
The final set of parameters are not consistent with those previously presented, probably due to the aforementioned distinctions. In particular, we find a result for the inclination of $47^{+7}_{-4}\, {\rm deg}$, while their best fit favours either much lower values ($15-30\, \rm{deg}$ for the direct heating model) or slightly higher ($52\, \rm{deg}$ for their spot model). The filling factor derived here is consistent with being Roche-lobe filling, while their best fit favoured a quite under-filled companion ($f=0.39$); which consequently alters the derived distance for the system. Finally, the reported $T_{\rm base}$ and $T_{\rm irr}$ in this work are slightly higher when compared with their models. Given the moderate-to-low orbital inclinations proposed by all the previous models, a Roche-lobe filling solution might be preferred to explain the deep radio eclipses observed in this BW.  

The more intensive phase coverage and higher signal-to-noise of our observations, combined with a better fit using a simpler model, lead us to adopt the results presented here. Nevertheless, we cannot discard the presence of hot-spots in the system. If that were the case, the variability of these structures on typical timescales of a month (e.g. \citealt{Clark2021}) might explain the different light curve observed (as the datasets were obtained 5 years apart). In addition, the dataset of \citet{Tang2014Discovery} is constructed from observations on 4 different nights spread over 4 months, while the data here shown were obtained in two consecutive nights. This might have played a role in the different light curves obtained, whether due to the intrinsic variability of potential hot-spots or to systematic uncertainties not fully accounted for during the data calibration.

\subsection{\psrf{}}

\citet{Lynch2018} provided refined a timing ephemeris that we employ in the light curve modelling of this system. The imposed prior on the interstellar reddening for this system is $E(g-r) = 0.08\pm 0.02$. There is no parallax or proper motion measured for this pulsar. A distance value of $2.1\, \rm{kpc}$ was derived from the DM \citep{Lynch2018}, using the electron density maps of \citet{Yao2017}. This was implemented in our modelling as a broad Gaussian prior, which is then combined with the one derived from the known MSP Galactic distribution.

\begin{figure}
    \centering
    \includegraphics[width=\columnwidth]{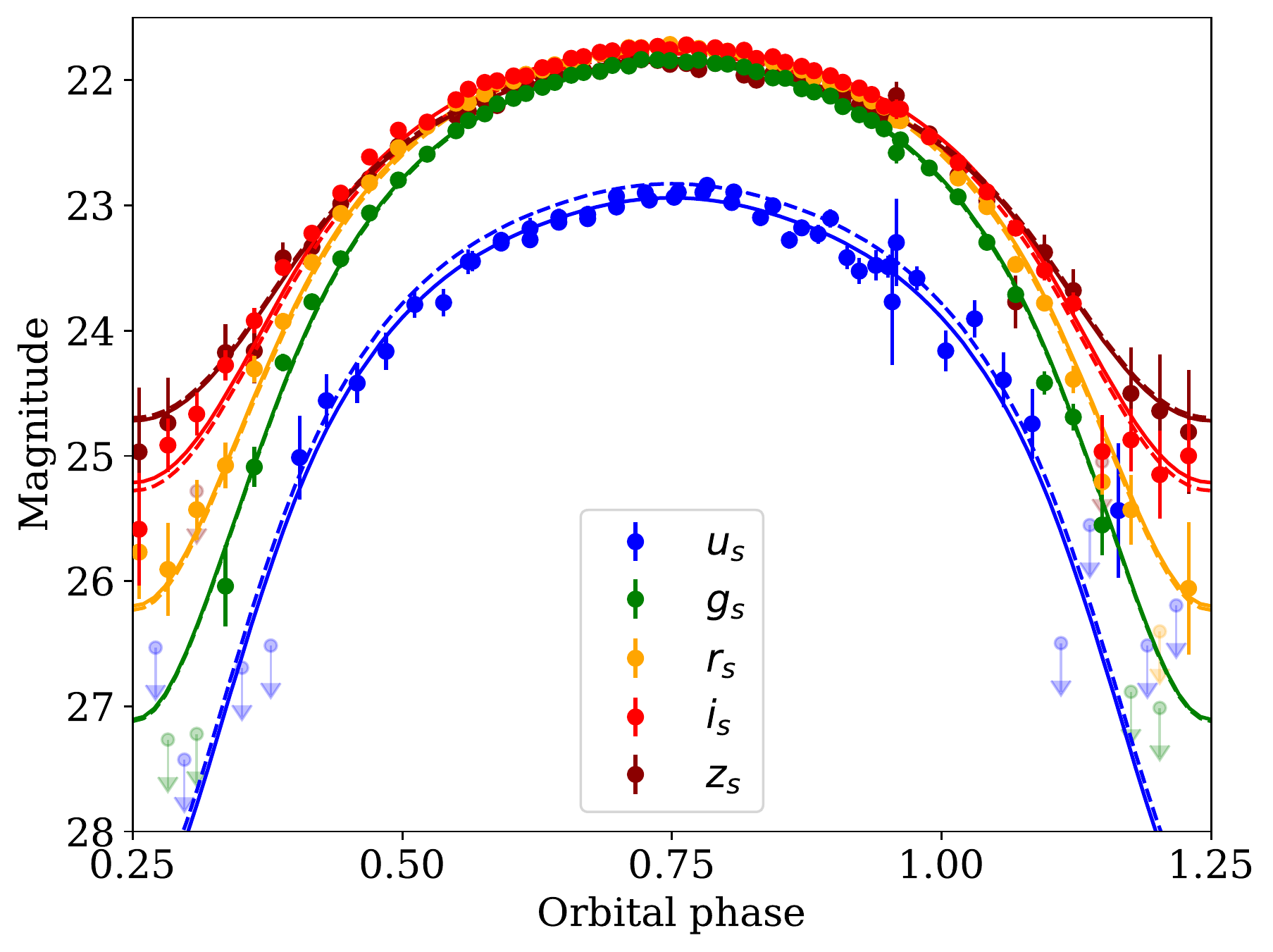}
    \caption{The best-fitting \textsc{Icarus} model for the HiPERCAM optical light curve of \psrf. Fig. \ref{fig:modelJ0023} description remains valid.}
    \label{fig:modelJ1641}
\end{figure}

The best-fitting model to the \psrf{} light curve corresponds to the CoG reduction and it has been plotted in Fig. \ref{fig:modelJ1641}. The posterior distribution for the model parameters is shown in Fig. \ref{fig:corner_plotJ1641}. The best-fit values for the parameters are collected in Tab \ref{tab:model_results}. 

The light curve obtained for \psrf{} is the most complete and has the highest SNR within our sample. For this reason, we also attempted to reproduce it with more complex models, including those considering heat diffusion effects, as well as uniform convection profiles, both following the prescription introduced in \citet{Voisin2020}. The best-fit parameters obtained from these tests were always found to be perfectly consistent with those derived from the direct heating model. None of these tests improved the fit ($\chi^2/{\rm d.o.f.}=1958/1827$), and they favoured convection and diffusion parameters consistent with the direct heating scenario. For all these reasons, we adopt the direct heating model results, at least to the limits imposed by our current SNR.

\subsubsection{System parameters}

This is the first complete light curve provided for the optical counterpart of \psrf{}, and its modelling has allowed us to confirm its BW nature. The distance value previously proposed (based on the analysis of the pulsar DM) is barely consistent within $2\sigma$ with the distance derived here from the optical light curve modelling. A similar mismatch has been previously found in a handful of other systems (including other members of our sample), and it is typically associated with local variations of the electron density which make the DM-derived value less reliable. For this reason, we adopt the derived distance from our modelling of $d=4.7\pm 0.6\, {\rm kpc}$. The remaining derived parameters for this BW are consistent with those of the known BW population (see, e.g., \citealt{Draghis2019}). It is worth noting that the mass ratio ($q=30\pm 5$) is among the lowest of those measured in our sample, and corresponds to a companion mass of $M_{\rm c}=0.054_{+0.016}^{-0.014}\, M_{\odot}$ (limited by the constraints on the pulsar mass). While this is still consistent with the masses typically expected for BWs, if the pulsar were particularly massive, the companion mass might become closer to the RB regime. The irradiation temperature ($T_{\rm irr} = 8500 \pm 500$~K) is also the highest recorded within our sample, while the companion density is roughly half of the densest member of the population (\psrc{}).

These properties are reminiscent of the recently discovered BW pulsar PSR J1555$-$2908 \citep{Frail2018}. A combined photometric and spectroscopic analysis of this system \citep{Kennedy2022} revealed its mass ratio ($q=28.0\pm 0.3$), companion mass ($M_{\rm c}=0.060_{+0.005}^{-0.003}\, M_{\odot}$), distance ($d=5.1^{+0.8}_{-1.1}$~kpc), filling factor (Roche-lobe filling) and a particularly high irradiation temperature ($T_{\rm irr} = 9380 \pm 40$~K). All of these parameters are fully compatible with those derived for \psrf{} in this work (see Tab.\ref{tab:model_results}). On the other hand, the orbital periods of $P_{\rm orb}=5.6$~h (\citealt{Ray2022}) and $P_{\rm orb}=2.18$~h (\citealt{Stovall2014+GBNCC, Lynch2018}), of PSR J1555$-$2908 and \psrf{} respectively, are notably different. This implies that \psrf{} has a smaller and more dense companion, with $\sim$ half the stellar radius and over $\sim 6$ times higher density, than that of PSR J1555$-$2908. The smaller stellar radius is the reason behind the fainter optical counterpart of \psrf{} in spite of a similar irradiation temperature ($g_s\sim 22$ at maximum, i.e. 1.5 magnitudes fainter than PSR J1555$-$2908), which prevents a complete spectroscopic study of the former with the current generation of optical telescopes. The smaller orbital size of \psrf{} might advocate for a stronger irradiation by the pulsar wind when compared with PSR J1555$-$2908. However, their different spin-down luminosity ($-\dot{E}$) compensates for this fact, and finally produces similar heating efficiencies for PSR J1555$-$2908 and \psrf{} ($\epsilon = 0.41^{+0.19}_{-0.14}$ and $0.32\pm 0.01$, respectively). Therefore, under similar irradiation conditions, it appears that the companion star mean density does not have a critical effect on the heating efficiency. This supports the scenario of most BWs companions having similar stellar envelopes, while their cores accommodate most of the mass and density differences.

\section{Discussion}

The BW population discovered so far includes over $\sim 30$ systems for which either radio or gamma-ray pulsations, as well as a low enough minimum companion mass, has been established. Light curve modelling has now been performed for 17 of them and their system parameters derived, while studying the remaining BWs has been typically hampered by a too faint optical counterpart or a crowded field of view (e.g., those found in globular clusters). In this work, we present the results from simultaneous multi-band optical observations of 6 such systems, including the first complete light curve observed for \psrf{} and a significant improvement on the data quality for the remaining targets. 

At this point, it is worth compiling possible caveats of determining BW parameters from light curve modelling. First and foremost, we stress here that poor data quality can bias the modelling results. Given the nature of BW companions, with masses of barely a few per cent of a Solar mass, the detection at minimum for most systems is hampered by their intrinsically low $T_{\rm base}$. Additionally, modelling these datasets requires atmosphere models covering a large range of temperatures in order to produce accurate results. This is particularly challenging when reaching the low temperature regime, which calls for some caution when interpreting the derived $T_{\rm base}$ in both this and other works. Finally, modelling of a poorly sampled light curve could lead to heavily biased results for a number of system parameters (see, e.g., the discussion on \psrb{} and \psre{} light curve modelling, Sec. \ref{sec:model}). 

Another noteworthy feature is the presence of asymmetric components in spider light curves, which can not be explained by the direct heating model alone. The increasing number of systems with asymmetric light curves is partly driven by the improvement in the observing facilities over the last decade, which allows for photometric observations with millimagnitude precision. Different authors have tackled this problem during the last few years by proposing modifications to the direct heating model, such as including hot-spots (e.g.  \citealt{vanStaden2016,Clark2021}), an IBS (e.g. \citealt{Romani2016+IBS}) or accounting for heat diffusion and convection in the companion atmosphere (see \citealt{Voisin2020} and \citealt{Kandel2020}). We tested the latter models \citep{Voisin2020} on three of the targets presented in this work: \psra{}, where a potential asymmetry might be present in the redder bands of the light curve (see Fig. \ref{fig:modelJ0023}); \psrc{}, for which a previous study proposed an IBS component might be required; and \psrf{}, which possesses the most precise and complete light curve of our sample. As previously discussed, employing a more complex model did not change the resulting parameters compared with the direct heating case, nor did it significantly improve the fit. For this reason, it will be not considered for what remains of this paper.

Despite the currently limited size of the BW population, we present below a comparison between the known members, focusing on the system parameters derived in this work. We added to the previously modelled sample the remaining 11 BWs for which optical modelling is available in the literature: PSR J1124$-$3653, PSR J1301+0833, PSR J1959+2048, PSR J2052+1219, PSR J2241$-$5236 (from \citealt{Draghis2019}; using spectroscopic radial velocities from \citealt{vanKerkwijk2011} and \citealt{Romani2016} for PSR J1959+2048 and PSR J1301+0833, respectively); PSR J1311$-$3430 (best-fit model from \citealt{Romani2015}); PSR J1810+1744 \citep{Schroeder2014,Romani2021}; PSR 2256$-$1024 \citep{Breton2013+Modelling}; PSR J1653$-$0158 \citep{Nieder2020} and PSR J1555$-$2908 \citep{Kennedy2022,Ray2022}. We did not consider \citet{Stappers2001} parameters  for PSR J2051$-$0827 due to the conflicting results from the light curve modelling presented in their paper (as noted by the authors themselves), which significantly change key parameters such as the filling factor and the inclination due to an asymmetric component. More recently, observations of this system by \citet{Dhillon2022} revealed a symmetric light curve, whose modelling produced a stable solution. We will include their results in our population analysis for completeness. Table \ref{tab:corr} compiles the combined set of parameters employed for the correlations described below.

\subsection{A note about \psrf{} spin-down luminosity}
\label{sec:spindown}

In this work, we have presented a consistent picture of the newly characterised BW \psrf{}. We also calculated the spin-down luminosity ($-\dot{E}=-4 \pi^2 I f_{\rm spin} \dot{f}_{\rm spin}$) for the target, where $I$ is the moment of inertia of the pulsar (set to a canonical value of $I=10^{45}\rm{g\,cm^2}$),  $f_{\rm spin}$ is the spin frequency and $\dot{f}_{\rm spin}$ its first derivative over time. Combined with the irradiation luminosity ($L_{\rm irr}$), this produces the heating efficiency ($\epsilon$) reported in Table \ref{tab:model_results}.

However, the above formula for $-\dot{E}$ does not include the so-called Shklovskii correction \citep{Shklovskii1970}, required when the transversal proper motion of the pulsar is significant, neither that associated to the pulsar acceleration in the Galactic potential (see e.g. \citealt{Nice1995}). The intrinsic spin-down luminosity of the pulsar ($-\dot{E}_{\rm int}$) is then:

$$ -\dot{E}_{\rm int}=-4 \pi^2 I f_{\rm spin} (\dot{f}_{\rm spin}-\dot{f}_{\rm Shk}-\dot{f}_{\rm Gal});\quad\dot{f}_{\rm Shk}=-\mu^2 d f_{\rm spin}/c $$

where $\mu$ is the transversal proper motion, $d$ is the distance to the pulsar and $\dot{f}_{\rm Gal}$ depends on the Galactic coordinates and the distance to the source (see \citealt{Nice1995,Lynch2018}).
This correction is dominated by the Shklovskii term for all members of the BW population, which implies that the observed spin-down luminosity is actually an upper limit to the intrinsic parameter. The largest correction is found for PSR J2052+1219 \citep{Draghis2019}, with $\dot{E}/\dot{E}_{\rm int}\sim 3$. 

However, \psrf{}'s correction results in a much larger effect, effectively making the spin-down luminosity negative. That would imply that the pulsar is instead being spun up, an unexpected situation as accretion of matter from the companion would be at odds with the detection of radio pulsations. Three spiders of the redback kind have been observed to transition between rotation-powered and accretion-powered states (see, e.g., \citealt{2009Sci...324.1411A}), but none of the known BWs has exhibited that behaviour to date. While the prospect of a transitioning BW is exciting, none of the observed properties of \psrf{} (other than the spin-up resulting from the above correction) supports this scenario. Our smooth optical light curve does not reveal any of the flickering typically associated with the presence of accretion discs. Additionally, the optical counterpart of \psrf{} is not detected by all-sky surveys (e.g., PS1, SDSS). This is consistent with a similar peak magnitude to that derived from our observations ($r_s = 21.7 $), as the catalogues limiting magnitudes are $r\lesssim 21$, therefore arguing against a historical brightening due to the build-up of an accretion disc. 

The puzzling spin-up scenario resulting from the Shklovskii correction in \psrf{} was already noted by \citet{Lynch2018}, where they found two other MSPs also affected by this situation. As none of these targets were suspected to experience accretion events, the authors discussed instead the limitation on the parameters defining the correction. For \psrf{}, a distance $d<1.2\, {\rm kpc}$ is required to obtain a null $\dot{E}_{\rm int}$, while $d=0.8\, {\rm kpc}$ produces a correction factor of $\dot{E}/\dot{E}_{\rm int}\sim 3$ (the largest observed for the remaining BW population). However, neither the distance derived from the DM ($d=2.1\,\rm{kpc}$, using YMW16), neither that derived from our optical modelling ($d=4.7\pm 0.6\,\rm{kpc}$) seem to agree with that scenario. In order to further test this possibility with our optical light curve, we repeated the modelling described in Sec. \ref{sec:model} but setting a hard limit of $d<1.2\, {\rm kpc}$ in the priors. Under these conditions, our best fit favours a distance as close to the upper limit as possible, as well as an extremely under-filled companion star ($f\sim 0.20$), attaining $\chi^2/{\rm d.o.f.}=4319.71/1827$ (significantly worse than any of our previous fits). For this reason, we disfavour the low distance values required to reconcile the spin-down luminosity. Assuming the derived $d$ from our optical modelling is correct, the maximum allowed proper motion to avoid the spin-up scenario would be $\mu < 19\, {\rm mas\, yr^{-1}}$, at odds with the measured $\mu = 39\pm 3\, {\rm mas\, yr^{-1}}$ \citep{Lynch2018}. We note that the measurement in \citet{Lynch2018} comes from just 1.4 years of timing data, and on these time scales proper motion estimates can be biased by covariance with other timing parameters. We therefore consider it likely that the proper motion has been over-estimated for this system, and this spin-down conundrum will likely be clarified by future timing measurements. Further studies on the system are also encouraged to independently determine its distance (e.g., through parallax measurement). For these reasons, we decided not to include the $\dot{E}$ value of \psrf{} when comparing the known BW population in the following section.

\subsection{Parameter correlations for the BW population}

\begin{figure}
    \centering
    \includegraphics[width=\columnwidth]{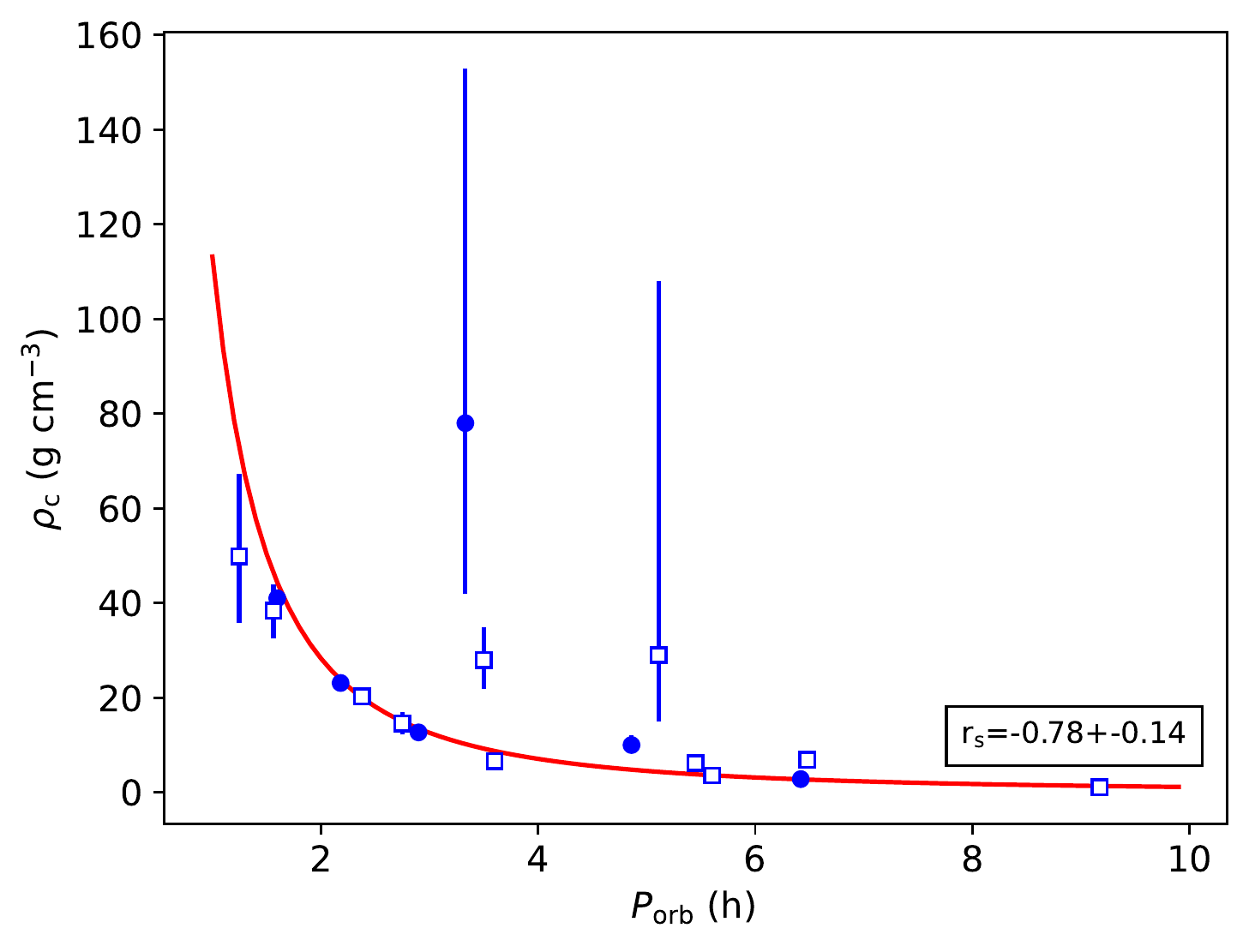}
    \caption{Companion star mean density ($\rho_{\rm c}$) against the orbital period ($P_{\rm orb}$). We include the derived parameters for the BW population with modelled light curves, using the reported value when available in the literature, or else deriving it from the proposed companion mass and radius. Filled circles mark the 6 systems presented in this paper. Empty squares refer to the remaining 11 systems compiled from the literature (references given in the text). Error bars are defined as $1\sigma$ for all the systems for consistency. We also included as a red, solid line the expected correlation for Roche-lobe filling binaries \citep{Faulkner1972}. The Spearman's correlation coefficient is also reported, together with a standard deviation derived from bootstrapping.}
    \label{fig:corr1}
\end{figure}

We searched for correlations between the derived parameters of the BW population defined above. We make use of Spearman's correlation coefficient ($r_s$) to assess the strength of the correlations, and present below the most promising among them, but having always in mind the limited size of the sample. In this regard, we employed the bootstrapping technique to retrieve a standard uncertainty (i.e, confidence level 68$\%$) on the derived $r_s$ coefficient, in order to better assess the influence of individual data points in the derived correlations. We first report on the presence of a clear correlation between the companion star mean density ($\rho_{\rm c}$) and $P_{\rm orb}$ (see Fig. \ref{fig:corr1}, $r_s=-0.78\pm0.14$). This is due to the underlying relation between these parameters for the limiting case of a Roche-lobe filling binary, which can be described with an analytical formula \citep{Faulkner1972}. The fact that many of the BWs have substantially large filling factors (especially when one compares their volume-averaged filling factor) is the origin of the observed correlation. 

\begin{figure*}
    \centering
    \includegraphics[width=\columnwidth]{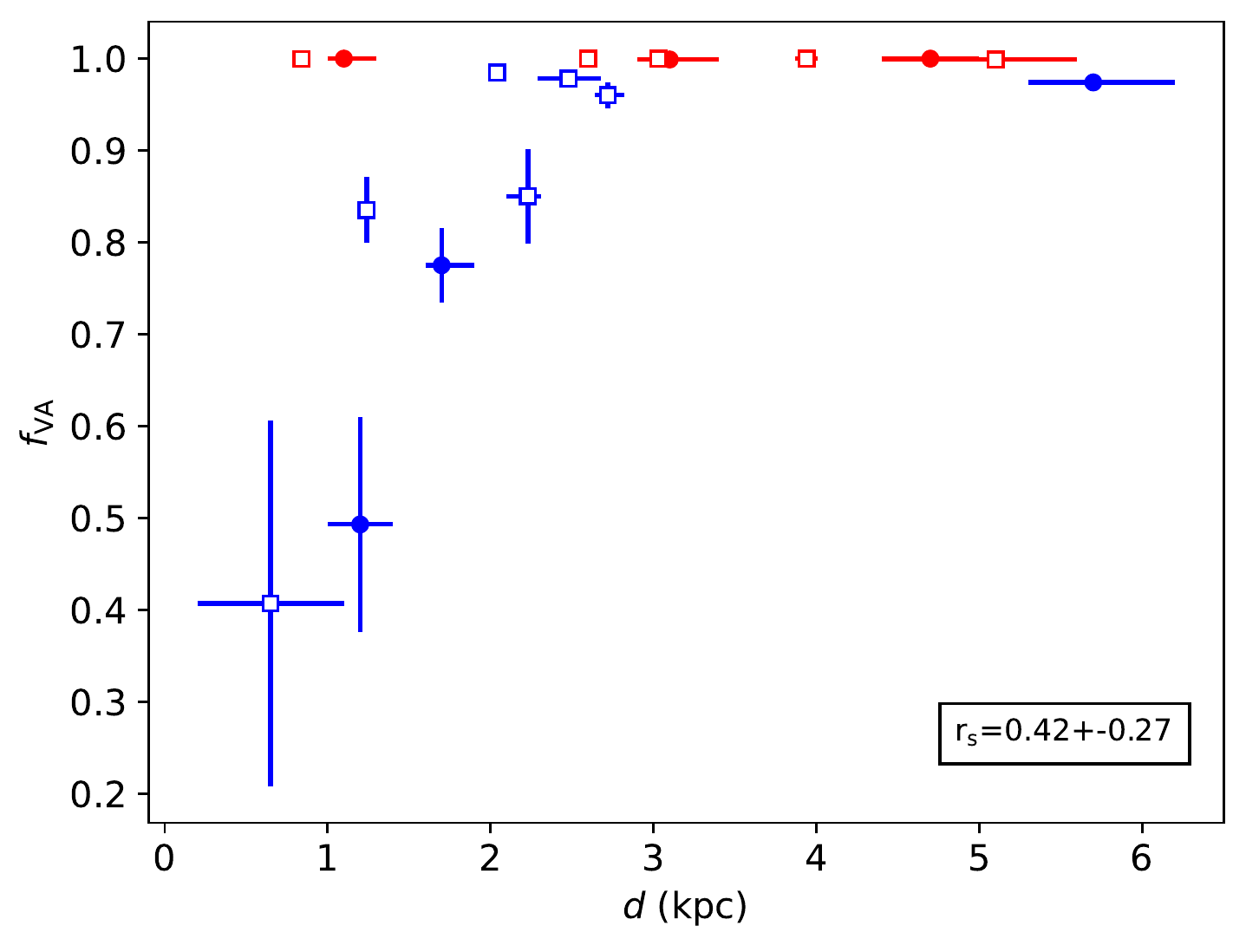}\includegraphics[width=\columnwidth]{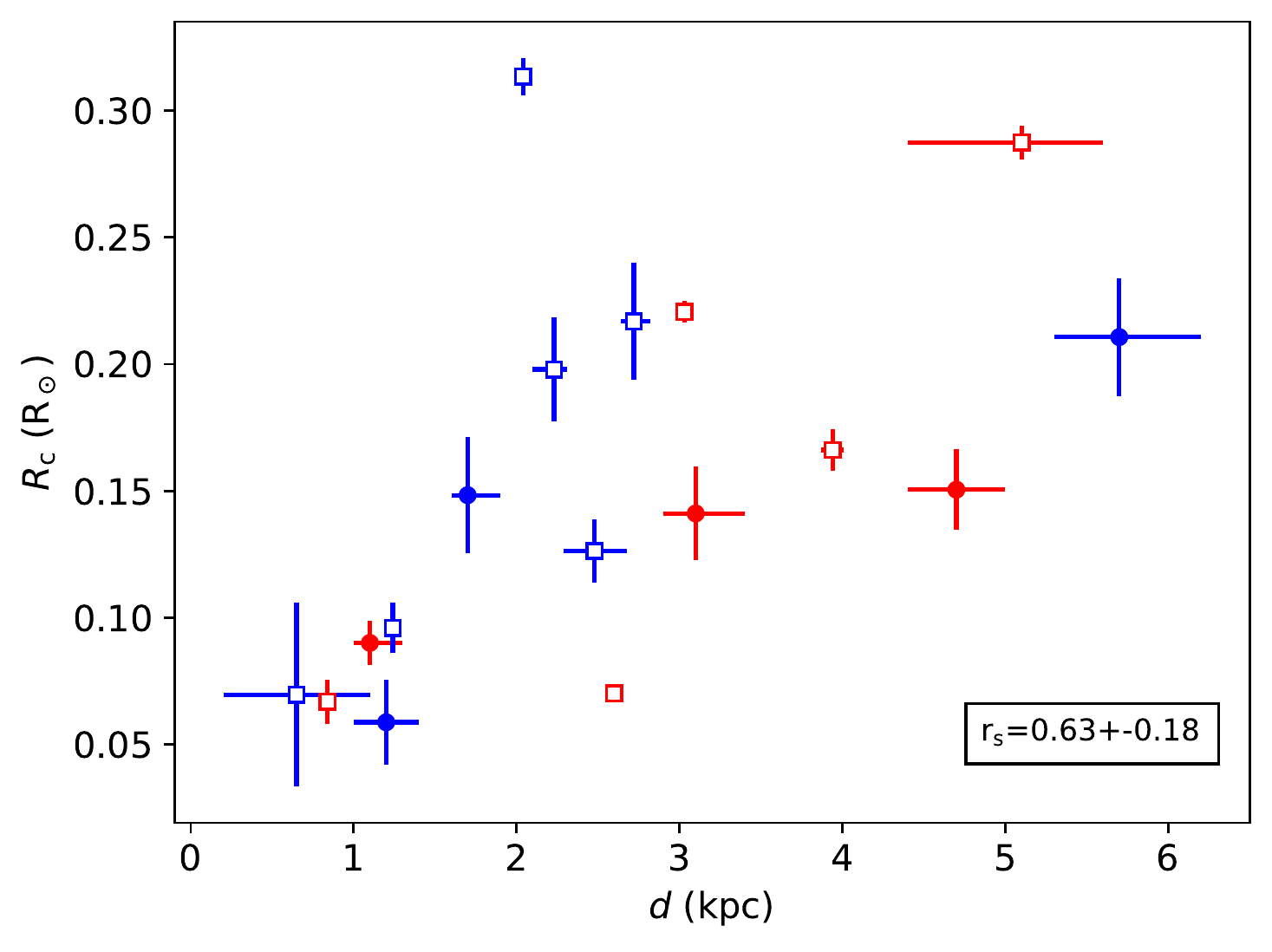}
    \caption{Distance to the system against the volume average filling factor ($f_{\rm VA}$, left panel) and against the companion star radius (right panel). Symbols follow the convention introduced in Fig. \ref{fig:corr1}. Those systems with a $f_{\rm VA}$ consistent with a Roche-lobe filling solution are plotted in red in both plots.}
    \label{fig:corr2}
\end{figure*}

We find weak evidence of a possible trend between the distance to the system ($d$) and the volume-averaged filling factor ($f_{\rm VA}$), with the spread of $f_{\rm VA}$ being narrower and tending asymptotically to unity for larger $d$ (Fig. \ref{fig:corr2}, left panel). To further investigate this, we plotted $R_{\rm c}$ against $d$ (Fig. \ref{fig:corr2}, right panel) and found a similarly positive correlation ($r_s=0.63\pm 0.18$). We marked the Roche-lobe filling systems in red, as they saturate at the maximum allowed $R_{\rm c}$ for the system $P_{\rm orb}$ and therefore may behave differently. A possible explanation for this correlation is that larger companion stars produce brighter optical maxima, and therefore are easier to detect at larger distances. This would naturally bias the observed BW sample to have larger $R_{\rm c}$ at large $d$ values. However, this then raises the question about the absence of large companion stars at lower $d$. The possibility remains that, due to the currently small sample of BWs, we have yet to uncover them, but given that they would produce larger optical variability than their low $R_{\rm c}$ siblings (under similar irradiation conditions), one might expect them to be easier to find. If we assume the known sample is complete for the volume of $d=1.5-6\, \rm{kpc}$ and $R_{\rm c}>0.1\,\rm{R_\odot}$ (i.e., 11 systems), the expected number of BWs of similarly large radius within $d<1.5\, \rm{kpc}$ would be 0.17 (Fig. \ref{fig:corr2}). If we consider instead a closer region ($d=1.5-3.5\, \rm{kpc}$, 7 BWs), the expected number of large and close BWs remains below unity (0.6). Therefore, we cannot confidently claim that the absence of large companion stars at the shortest distances is an intrinsic feature of the population, and might be still due to our limited sample. If future studies prove it true, an observational bias in the original pulsation searches (e.g. related to the BW eclipsing nature, which is a crucial effect hampering the detection of new systems) could be behind it.

\begin{figure*}
    \centering
    \includegraphics[width=\columnwidth]{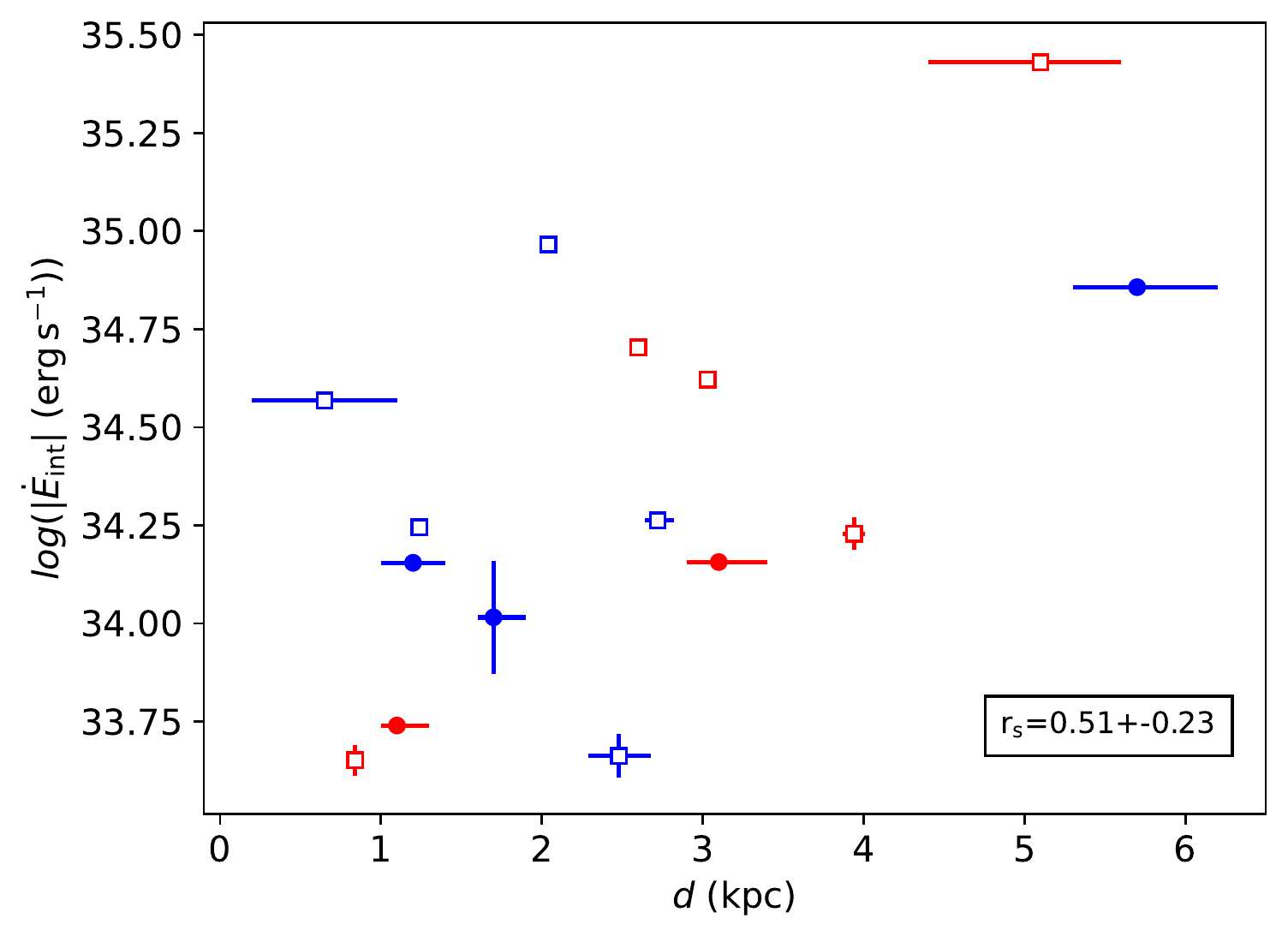}\includegraphics[width=\columnwidth]{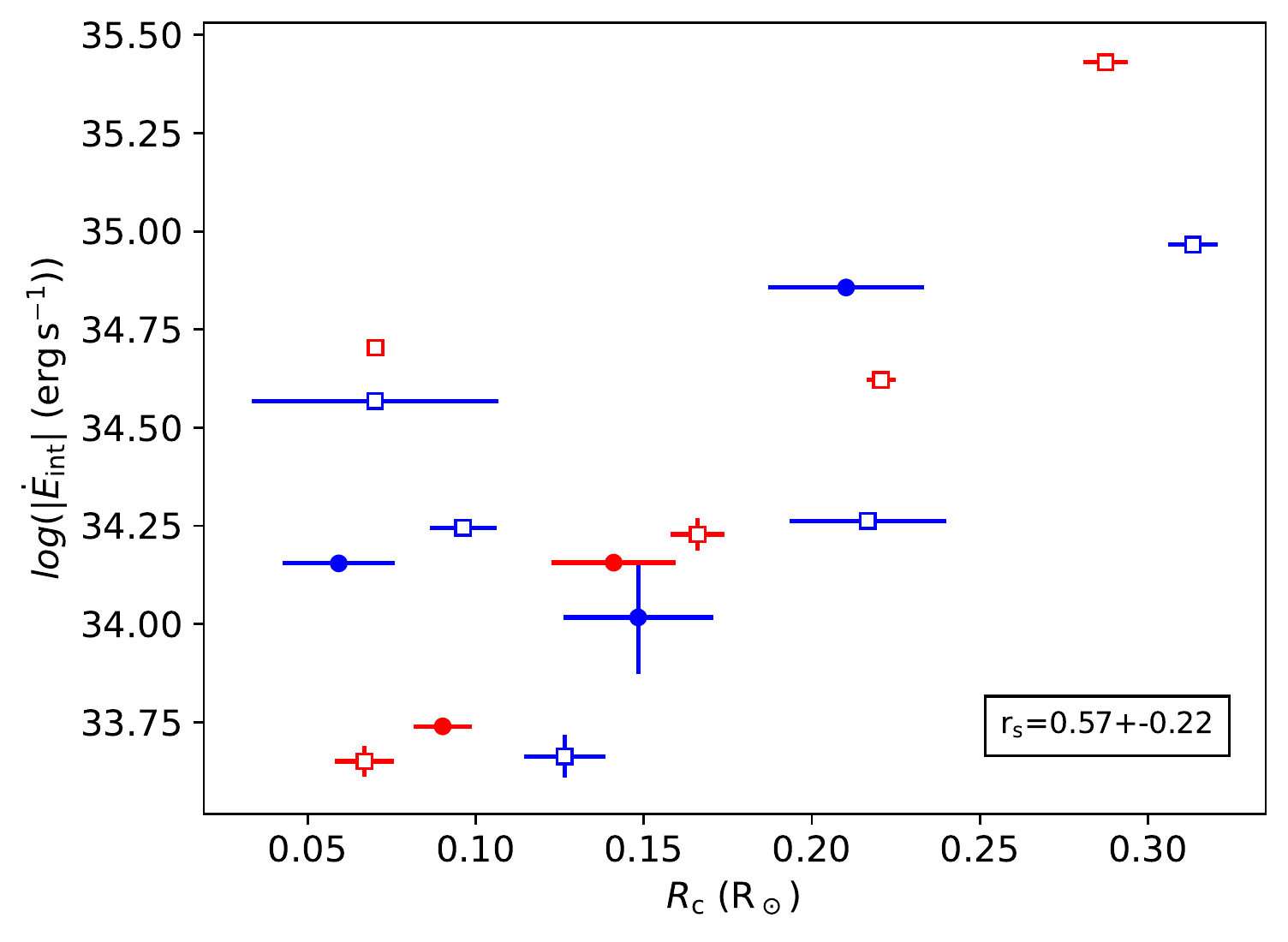}
    \caption{The logarithm of the spin-down luminosity of the pulsar against the distance to the system (left panel) and the companion star radius (right panel). Symbols and colours follow the convention introduced in Fig. \ref{fig:corr2}.}
    \label{fig:corr3}
\end{figure*}

It is worth noting that most BWs have been found through targeted radio searches in fields associated with unidentified gamma-ray sources, with their optical counterpart characterised afterwards. In this regard, an additional observational bias might be at play, as the observed gamma-ray flux is proportional to $\sqrt{\dot{E}}$ and inversely proportional to $d^2$ (see e.g. Fig. 9 in \citealt{2013ApJS..208...17A}). To explore the influence of such a bias in the previously discussed correlations, we plotted the logarithm of the spin down luminosity ($\log{|\dot{E}_{\rm int}}|$, assuming a moment of inertia of $10^{45}\,{\rm g\, cm^2}$) of each system against $d$ and $R_{\rm c}$, respectively, and found marginal positive correlations for both cases (Fig. \ref{fig:corr3}). The high spread in these correlations might be partially explained by the underlying distribution of masses and radii for the NSs (accounting for up to a factor $\sim 4$ in $|\dot{E_{\rm int}}|$). Additionally, higher spin-down luminosities are expected to produce higher irradiation temperatures on the companion (for otherwise similar conditions), which might lead to bloating of the companion star and ultimately larger amplitude for the optical modulations. Together, both of these effects appear as good candidates to explain the observed positive correlations in Figure \ref{fig:corr3} (and potentially, Fig. \ref{fig:corr2} right panel).

\begin{figure*}
    \centering
    \includegraphics[width=\columnwidth]{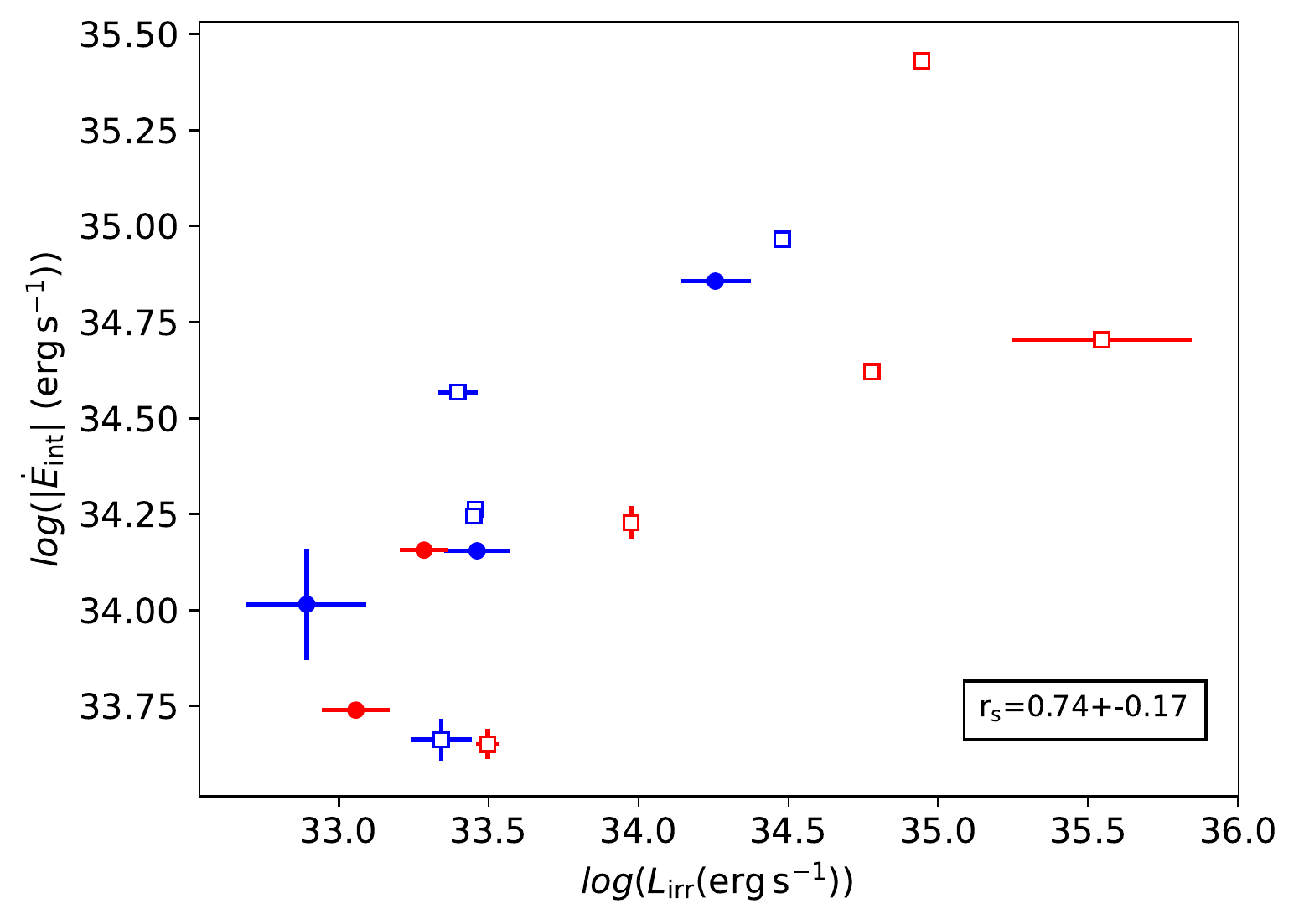}\includegraphics[width=\columnwidth]{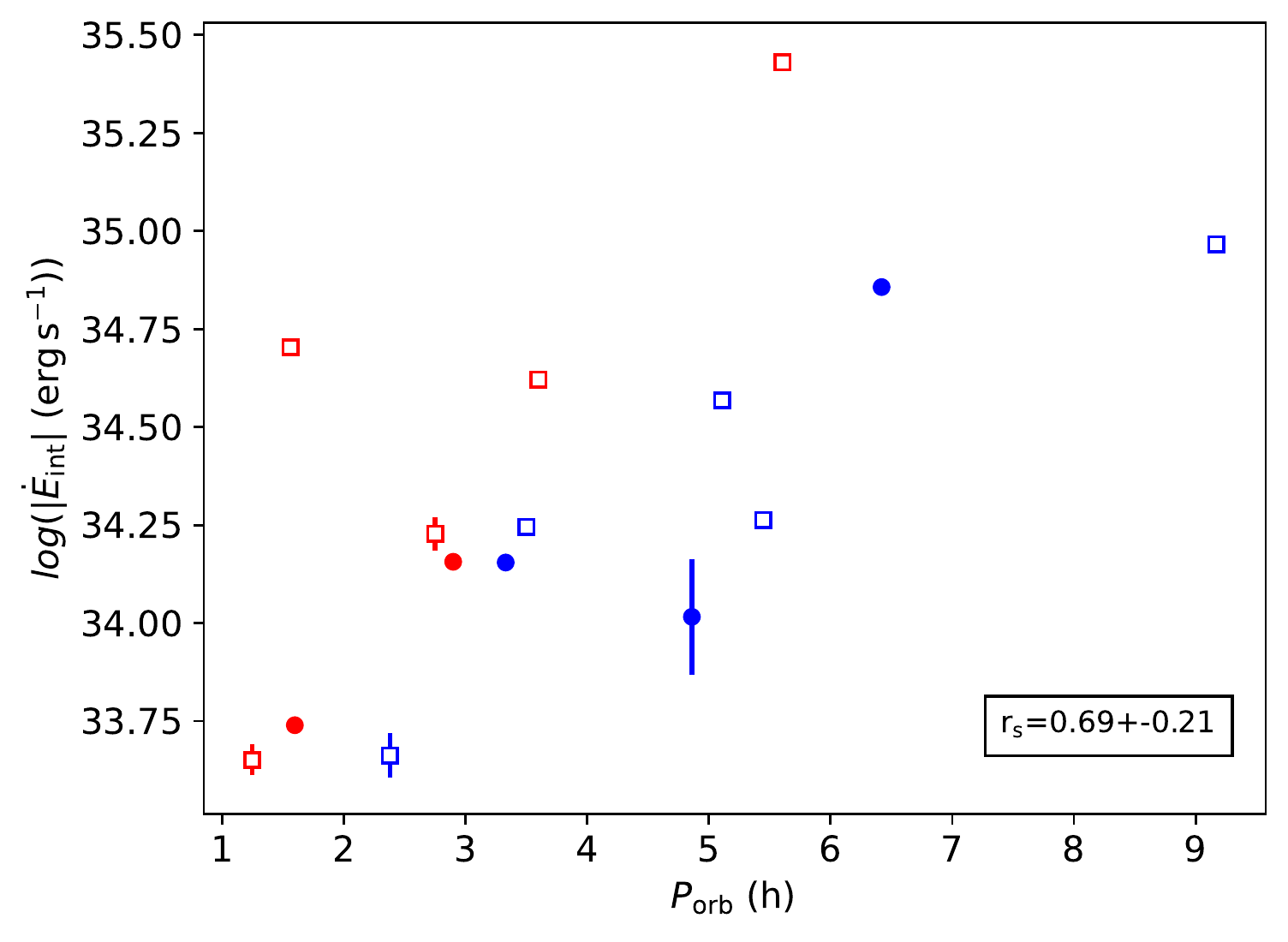}
    \caption{The logarithm of the spin-down luminosity of the pulsar against the irradiation luminosity on the companion star (left panel) and the binary orbital period (right panel). Symbols and colours follow the convention introduced in Fig. \ref{fig:corr2}.}
    \label{fig:corr4}
\end{figure*}

There are two other correlations with $\log{|\dot{E}_{\rm int}}|$ that are worth discussing. First, that with $\log{L_{\rm irr}}$, which is positively correlated with $\log{|\dot{E}_{\rm int}}|$ ($r_{s}=0.74\pm0.17$, see Fig. \ref{fig:corr4}). This is an expected consequence of the companion star's irradiation being powered by the spin-down luminosity of the pulsar. Nevertheless, we would like to remark that, while $L_{\rm irr}$ is obtained from the optical light curve modelling, $\log{|\dot{E}_{\rm int}}|$ is derived purely from radio observations. Therefore, this provides an independent confirmation that the irradiation of the companion star is indeed fuelled by a mechanism connected with the pulsar spin-down luminosity.

The second correlation, is that of $\log{|\dot{E}_{\rm int}}|$ with $P_{\rm orb}$ ($r_s=0.69\pm0.21$, see Fig. \ref{fig:corr4}). We note a positive correlation still holds when comparing with $x$ instead, though with a larger spread ($r_s=0.59\pm0.23$). These might be explained by the binary evolutionary history of MSPs. \citet{Chen2013} proposed that mass loss due to pulsar-driven irradiation is an essential ingredient to widen orbits to the observed periods. Alternatively, other authors (e.g., \citealt{Ginzburg2021}) have suggested that enhanced magnetic braking by the ablated wind also leads to wider orbits, and might be sufficient to explain the range of observed BW periods. In any of these cases, a direct consequence is that pulsars with a higher spin-down energy will be able to induce a higher irradiation and increase the orbital separation.

\begin{figure*}
    \centering
    \includegraphics[width=\columnwidth]{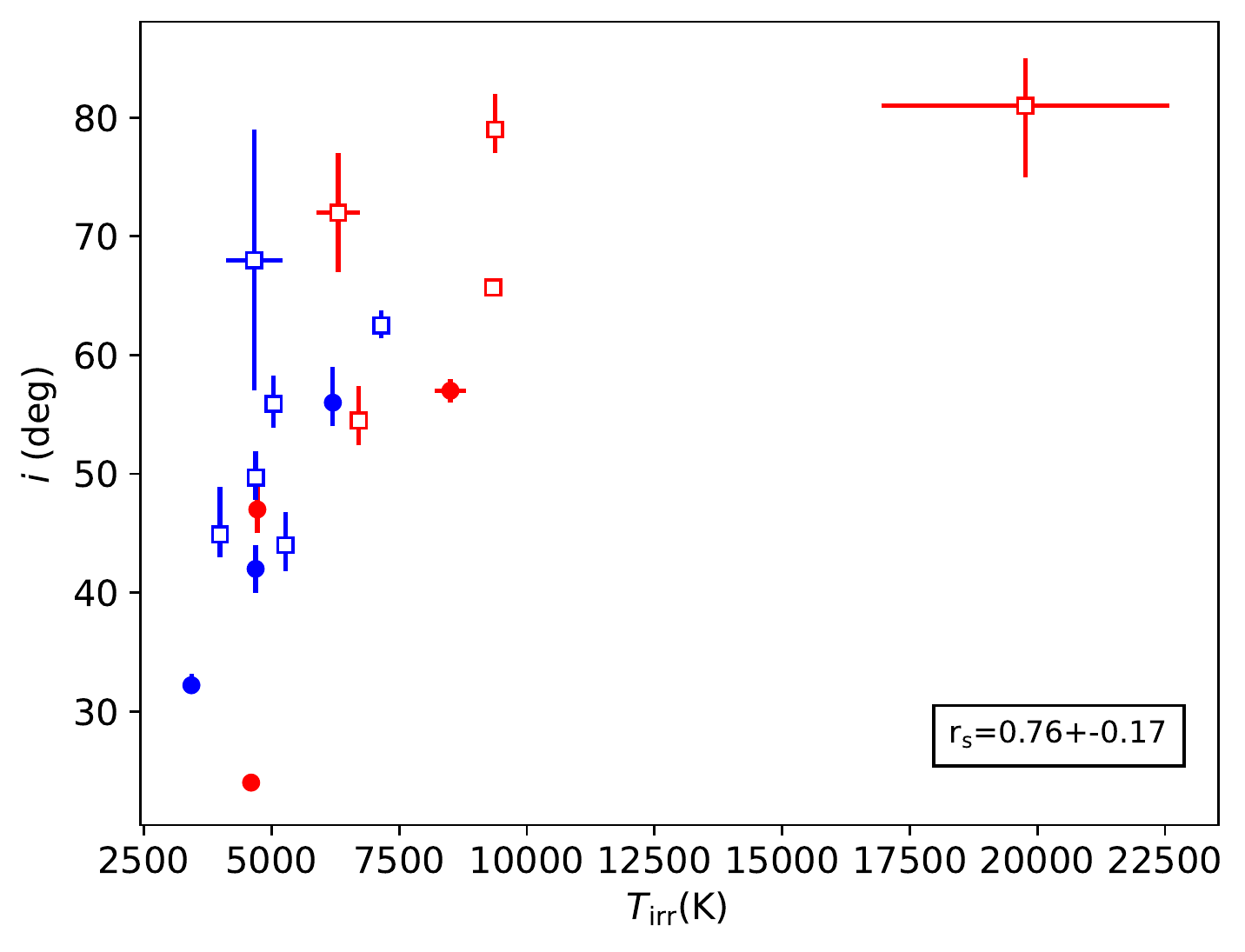}\includegraphics[width=\columnwidth]{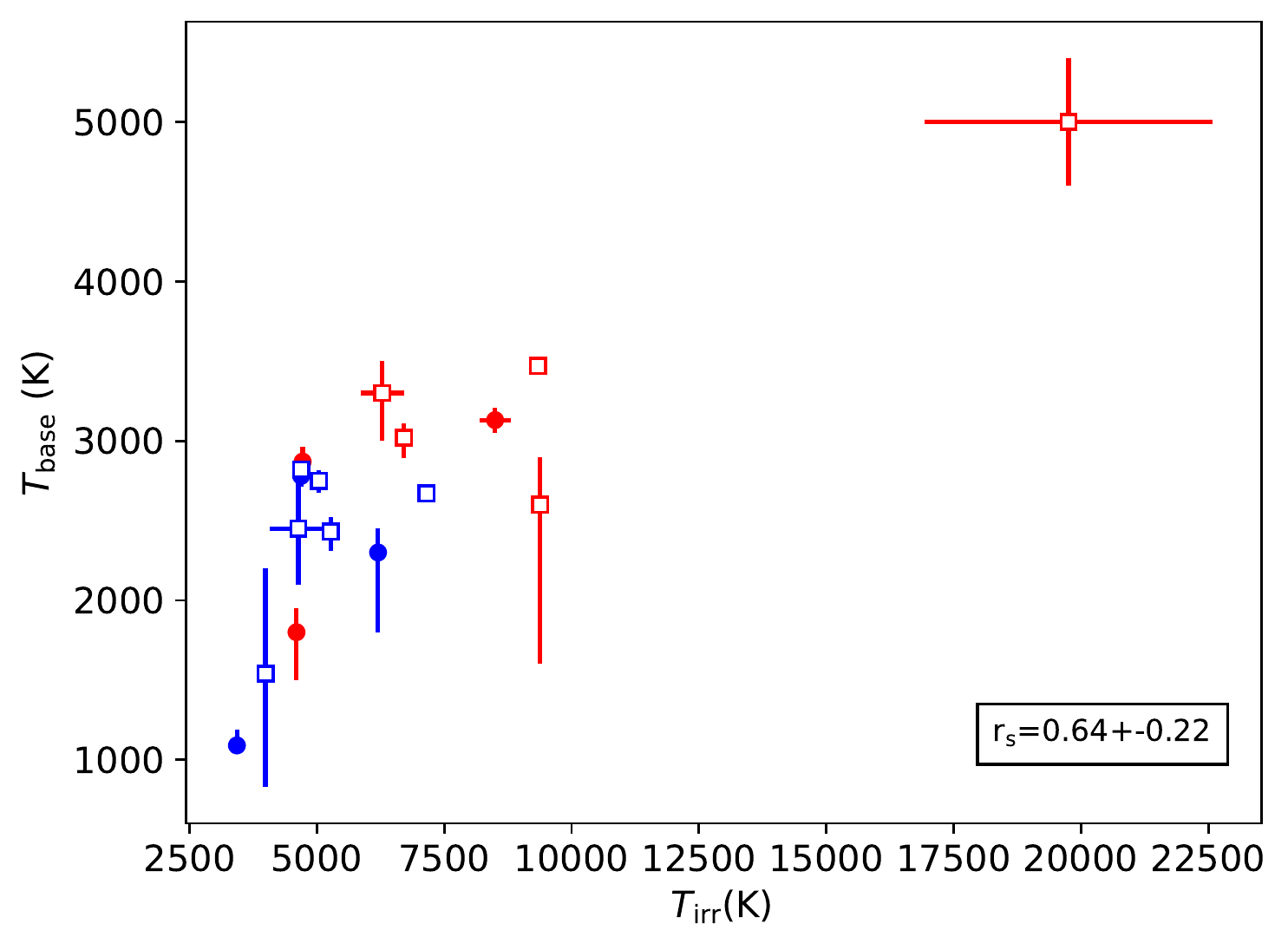}
    \caption{Left panel: orbital inclination against the irradiation temperature. Right panel: base temperature against the irradiation temperature. Symbols and colours follow the convention introduced in Fig. \ref{fig:corr2}.}
    \label{fig:corr5}
\end{figure*}

Last but not least, we also report on the correlations between  $i$ and $T_{\rm irr}$ ($r_s=0.76\pm0.17$), as well as $T_{\rm base}$ with $T_{\rm irr}$ ($r_s=0.64\pm 0.22$), shown in Fig. \ref{fig:corr5}. A correlation between temperatures might arise due to the increase of $T_{\rm irr}$ with $T_{\rm base}$ in order to produce a comparable modulation in the light curve. Alternatively, it could indicate that some of the heating flux is actually redistributed, increasing the overall base temperature of the star. However, the correlation with $i$ poses a challenge. We initially considered a potential bias in the photometric models, as these two parameters are the key drivers of the variability amplitude in the light curves. In that regard, the degeneracy of these two parameters might allow similar light curves to be produced through a combination of either a high $i$ and low $T_{\rm irr}$, or a low $i$ and high $T_{\rm irr}$ (under the assumption of the rest of the parameters being similar). If that were the case, we would expect a negative correlation between the parameters, contrary to that observed in our data. Taking into account that $i$ is not an intrinsic physical parameter of the BW systems, but instead due to the projection of the orbit onto our line of sight, an intrinsic correlation with other physical parameters seems puzzling. Due to the limited sample of BWs, as well as the fact that an isotropic distribution of orbital axes is uniform in $\cos i$, the number of detected low inclination systems is rather low, but critical for the correlation. This forbids us from making absolute claims based on this correlation alone, but given its high Spearman's coefficient (the second highest of those presented), we decided to speculate below about the potential origin of such an intriguing correlation.

BWs are old systems which are thought to have experienced an epoch of accretion in the past (see, e.g., \citealt{2006csxs.book..623T,Chen2013}). This is at the origin of the pulsar spin up, and it is believed to align the spin axis and the orbital axis. In addition, the alignment of the magnetic axis of the pulsar with our line of sight determines our ability to detect pulsations, as the Earth must be swept by the beamed emission. Therefore, one could conclude that the measured orbital inclination serves as a proxy of the angle between the magnetic axis and the orbital axis (also known as the pulsar obliquity, $\chi \sim i$). Following this argument, systems observed at high inclination have their magnetic axis closer to the orbital plane ($\chi\sim ~90\, {^\circ}$). Under these assumptions, the correlation we present would imply a higher irradiation on the companion star for pulsars with the most extreme obliquity. The polar cap opening angle for MSPs is typically of $\sim 10-20\, {^\circ}$, which would contribute to some dispersion in the correlation.

\begin{figure*}
    \centering
    \includegraphics[width=\columnwidth]{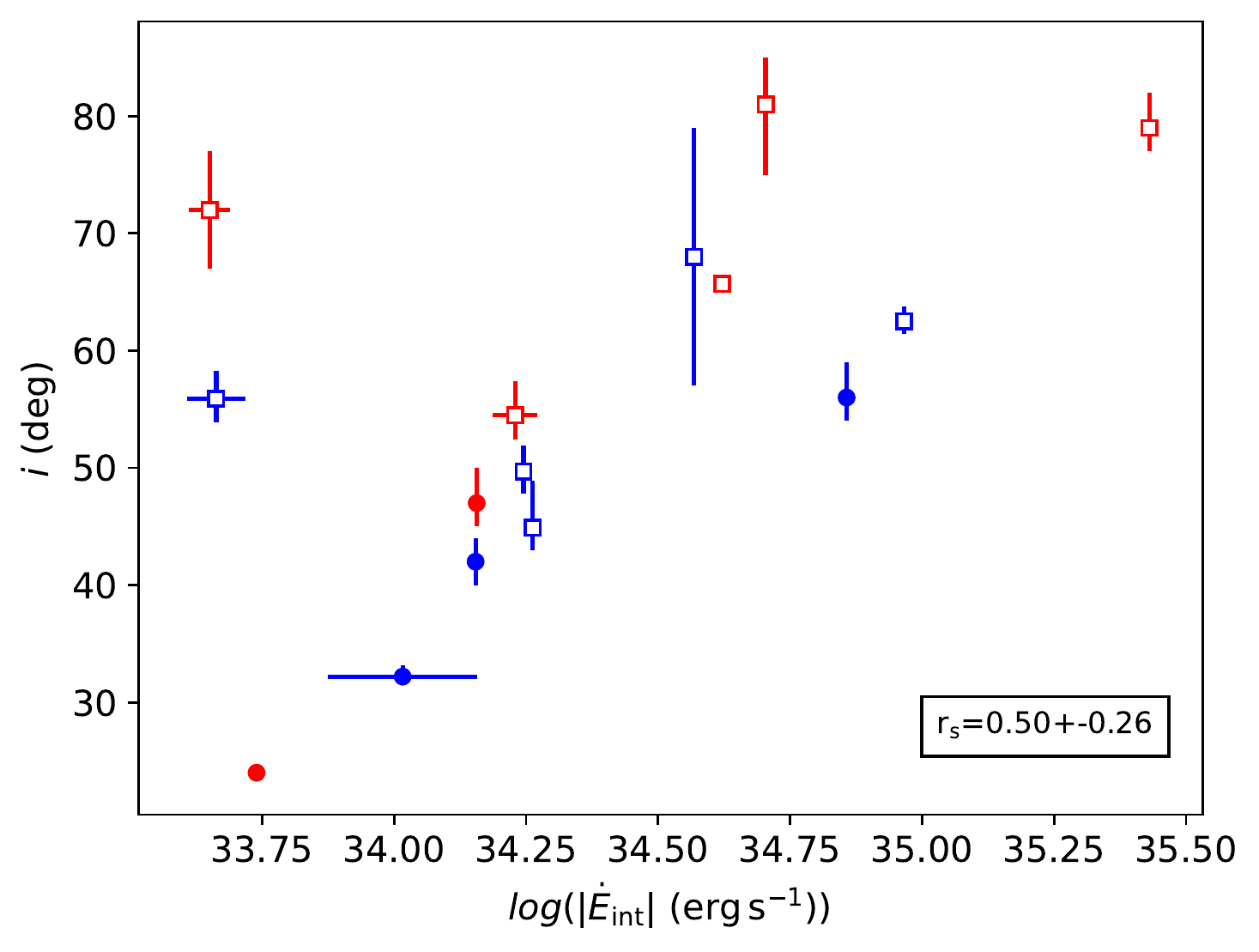}\includegraphics[width=\columnwidth]{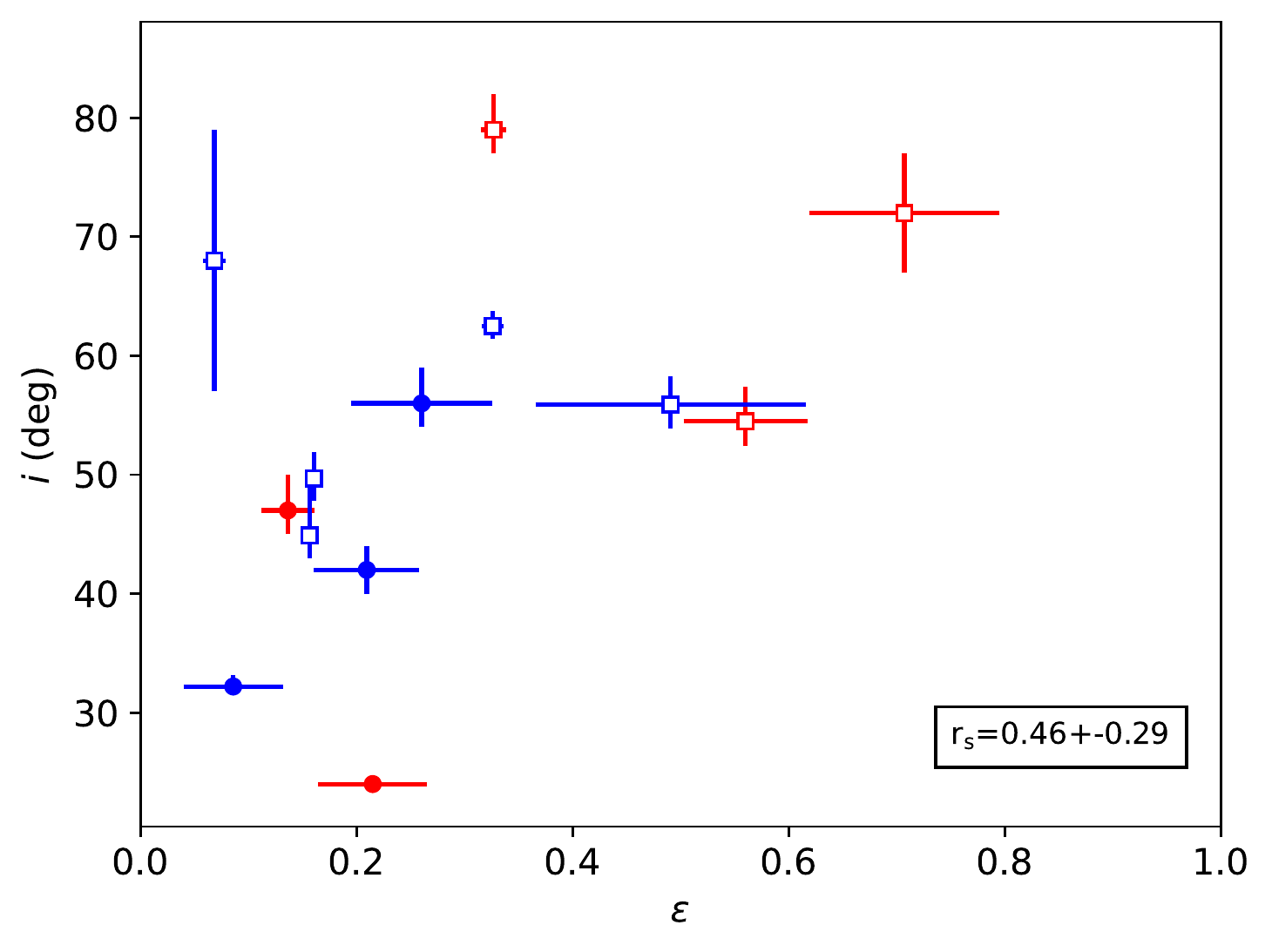}
    \caption{The orbital inclination against the spin-down luminosity of the pulsar (left panel) and its heating efficiency (right panel). Symbols and colours follow the convention introduced in Fig. \ref{fig:corr2}.}
    \label{fig:corr6}
\end{figure*}

It has been shown that for a given spin frequency, a larger magnetic obliquity leads to a larger spin-down power (e.g., \citealt{Spitkovsky2006}, \citealt{Philippov2015}, \citealt{Petri2022}), up to a factor of $\sim 2$ between the aligned ($\chi = 0\, {^\circ}$) and orthogonal ($\chi =90\, {^\circ}$) configurations. On the other hand, the same studies show that the power carried by the pulsar wind is not isotropically distributed but rather concentrated around the spin equator in a way that also depends on magnetic obliquity. In particular, if one assumes that gamma rays are responsible for irradiation then simulated sky maps show a strong dependence on obliquity (e.g. \citealt{Cerutti2016}, \citealt{Petri2022}). It follows that the observed correlation between inclination and irradiation may result either from (i) a correlation between obliquity and spin-down power, or (ii) from a sharper concentration of irradiation power around the orbital plane, which we assume to be identical to the spin equator (see discussion above). In the former case, we expect a correlation between inclination and spin-down power, while in the latter we expect a correlation between inclination and irradiation efficiency. The left-hand panel of Fig. \ref{fig:corr6} shows a tentative correlation between inclination and spin-down luminosity ($r_s=0.50\pm0.26$) over a large range of $\log{|\dot{E}_{\rm int}}|$. This is qualitatively consistent with the theoretical studies proposing that pulsars with larger obliqueness spin down faster, but these propose a much more modest change in the spin-down power ($\sim 2$), far from the $\sim 2$ orders of magnitude observed here. The right-hand panel in Fig. \ref{fig:corr6} shows that a weaker positive trend remains when inclination is plotted against irradiation efficiency ($r_s = 0.46\pm 0.29$, discarding both PSR J1311$-$3430 and PSR J1810+1744 outliers, see Tab. \ref{tab:corr}). These results suggest we cannot clearly attribute the observed correlation between inclination and irradiation power to only one of these effects, and that they might be both at play to some degree.


\subsection{On the detection of radio eclipses}

The presence of radio eclipses in spiders, and particularly in BWs, is one of the earliest features employed to identify new candidates of this elusive population. The origin of the eclipse lies with the ablated material from the companion due to the pulsar irradiation, but the particular details, such as the ablation rate or the geometry of the structure producing the eclipse, are still under debate (e.g., \citealt{Ginzburg2020,2020MNRAS.494.2948P}). It is traditionally assumed that a positive correlation between the orbital inclination and the detection of eclipses should exist, as closer to face-on configurations would require larger covering factors for the ablated material. A search through the literature on the BW population compiled in Table \ref{tab:corr} reveals that eclipses have been found in most of its members. Only four BWs remain without an eclipse detection: \psra{}, \psrc{}, \psrd{} and PSR\,J2241$-$5236; all of them with low-to-intermediate orbital inclinations ($\lesssim 60\,{\rm deg}$). On the other hand, while all high-inclination systems are eclipsing, within the range of $i\lesssim 60\,{\rm deg}$ a comparable number of eclipsing and non-eclipsing BWs exists. Two of the systems analysed in this paper provide good examples illustrating this situation: (i) \psrb{}, where eclipses have been found, but optical modelling suggests a low-to-intermediate inclination (see Sec. \ref{sec:model}); and (ii) \psrd{}, where no eclipses have been reported to date, in spite of its moderate inclination. For these reasons, we conclude that, while the presence of BW eclipses seems favoured by edge-on configurations, these still occur even at low orbital inclinations, favouring an extended geometry for the ablated material. In this regard, it is worth remarking the variable nature of the radio eclipses (changing depth and duration between epochs) and its frequency dependence (which constrains its observation), both critical factors which might lead to a future detection of eclipses in the few remaining uneclipsed BWs (e.g., \citealt{Polzin2019,Wateren2022}).

\section{Conclusions}

We present an optical light curve modelling analysis of six BW systems observed with HiPERCAM at the GTC. This configuration allowed us to better sample the faintest orbital phases, leading in turn to a more precise and less bias-prone determination of parameters. We present the first parameter determination for \psrf{}, confirming its classification as a BW with a particularly high companion mass when compared with the rest of the BW population (close to the RB regime if the pulsar contained in the binary is on the heavy side). Additionally, we revisit the remaining systems and improve on their parameter determinations. Both \psra{} and \psrb{} showed significantly lower orbital inclinations and filling factors when compared to previous studies. \psrc{} was confirmed as harbouring a high density companion, in line with previous works, but favouring the lower end of the available parameter space. While we did not find any clear indication of asymmetries in the light curve, we cannot confidently discard them if higher SNR observations were to be performed. A similar result can be drawn from the analysis of \psre{}, where a previous work included hot-spots in their models in order to obtain a reliable fit, but they are not required for the analysis presented in this paper. Finally, a re-reduction of a previously presented light curve for \psrd{} shows perfectly consistent results with those previously reported, but we still include it for completeness. Comparison of this sample with the full BW population discovered to date shows correlations between some parameters, including an expected relationship between $\rho_{\rm c}$ and $P_{\rm orb}$ for Roche filling binaries. We also highlight the apparent lack of BWs with large companion stars close to Earth, but we cannot confidently conclude if this is an effect of low-number statistics, an intrinsic correlation or an unaccounted observational bias. Comparison of the orbital inclination with the irradiation temperature, the spin-down luminosity and the irradiation efficiency suggests that pulsars with magnetic axis orthogonal to their spin axis might be capable of irradiating their companions to a higher degree. We encourage further studies to increase the size of the BW population in order to confirm if these correlations remain.

\section*{Acknowledgements}

RPB, CJC, MRK, DMS, JS, and GV acknowledge support from the ERC under the European Union's Horizon 2020 research and innovation programme (grant agreement No. 715051; Spiders). DMS also acknowledges the Fondo Europeo de Desarrollo Regional (FEDER) and the Canary Islands government for the financial support received in the form of a grant with reference PROID2021010132 (ACCISI/FEDER, UE); as well as support from the Spanish Ministry of Science and Innovation via an Europa Excelencia grant (EUR2021-122010). MRK also acknowledges funding through a Newton International Fellowship provided by the Royal Society (NIF171019). VSD, ULTRACAM and HiPERCAM operations are funded by the Science and Technology Facilities Council (grant ST/V000853/1). TRM would like to acknowledge support from the STFC, grant ST/T000406/1. Based on observations made with the Gran Telescopio Canarias (GTC), installed at the Spanish Observatorio del Roque de los Muchachos of the Instituto de Astrof\'isica de Canarias, on the island of La Palma, under program IDs GTC102-18A, GTC92-18B,  GTC79-19A and GTC9-19B. Based on observations with the New Technology Telescope collected at the European Southern Observatory, Chile, under programmes 097.D-0933 and 0101.D-0925. The design and construction of HiPERCAM was funded by the European Research Council under the European Union's Seventh Framework Programme (FP/2007-2013) under ERC-2013-ADG Grant Agreement no. 340040 (HiPERCAM). The Pan-STARRS1 Surveys (PS1) and the PS1 public science archive have been made possible through contributions by the Institute for Astronomy, the University of Hawaii, the Pan-STARRS Project Office, the Max-Planck Society and its participating institutes, the Max Planck Institute for Astronomy, Heidelberg and the Max Planck Institute for Extraterrestrial Physics, Garching, The Johns Hopkins University, Durham University, the University of Edinburgh, the Queen's University Belfast, the Harvard-Smithsonian Center for Astrophysics, the Las Cumbres Observatory Global Telescope Network Incorporated, the National Central University of Taiwan, the Space Telescope Science Institute, the National Aeronautics and Space Administration under Grant No. NNX08AR22G issued through the Planetary Science Division of the NASA Science Mission Directorate, the National Science Foundation Grant No. AST-1238877, the University of Maryland, Eotvos Lorand University (ELTE), the Los Alamos National Laboratory, and the Gordon and Betty Moore Foundation.

\section*{Data availability}

The raw ULTRACAM images and associated calibration frames may be obtained by contacting D. Mata S\'anchez or the ULTRACAM team (V. S. Dhillon). The data employed in the correlation plots has been compiled in Table \ref{tab:corr}, available in the online version of this paper.




\bibliographystyle{mnras}
\bibliography{hcam_bws} 




\appendix

\newpage

\section{Compilation of the BW population parameters}

\begin{landscape}
    \begin{table}
	\caption{Compilation of the 17 BWs with parameters derived through optical modelling. All reported error bars correspond to $1\sigma$ for consistency. Parameters derived from radio timing ($P_{\rm orb}$ and $x$) were compiled from the ATNF Pulsar Catalog \citep{Manchester2005} and references therein. As these are typically determined with high precision, we only report here their best value, and refer the reader to the original publications. The remaining parameters were either directly collected from the literature, or else derived from the published parameters by employing the corresponding formulae and a Monte Carlo approach. We note that the parameter $\dot{|E_{\rm int}|}$ (i.e., $\epsilon$) includes the Shklowskii correction for all systems whose proper motion has been measured, and corresponds to an uncorrected value otherwise. We report $f_{\rm VA}=1.0$ when the derived parameter was consistent with a Roche-lobe filling solution within $1\sigma$ uncertainties. }	
    \begin{threeparttable}
		\begin{tabular}{l c c c c c c c c c c c c c l}
		\hline
		Target	 & $P_{\rm orb}$ & $x$  & $T_{\rm base}$ & $T_{\rm irr}$ & $L_{\rm irr}$	& $\dot{|E_{\rm int}|}$  & $\epsilon$ & $\rho_{\rm c}$ & $f_{\rm VA}$ & $R_c$ & $d$  & $i$ & Ref.\\
		& (hr) & (lts)   & (K) & (K) & $(10^{34}\rm{erg} \, s^{-1})$     & $(10^{34}\rm{erg} \, s^{-1})$   & & $(\rm{g \, cm^{-3}})$ & & ($R_\odot$) & ($\rm kpc$) & $(\rm\deg)$ &  \\
		\hline\hline
		J0023+0923 & 3.33 & 0.035 & $2780\pm 70$ & $4690\pm 120$ & $0.30\pm 0.07$ & $1.43\pm0.04$ & $0.21\pm 0.05$& $78^{+75}_{-36}$ & $0.49\pm 0.12$ & $0.059\pm 0.016$ & $1.2 \pm 0.2$ & $42 \pm 2$& 1 \\
		J0251+2606 & 4.86 & 0.066 & $1090^{+100}_{-25}$ &
		$3430\pm 40$ & $0.09\pm 0.03$  & $1.1\pm 0.3$ & $0.09\pm 0.05$ & $10^{+2}_{-1}$  & $0.78\pm0.04$ & $0.15\pm0.02$ & $1.7^{+0.2}_{-0.1}$ & $32.2^{+1.0}_{-0.8}$  & 1 \\
		J0636+5129 & 1.60 & 0.009 & $1800^{+150}_{-300}$ & $4600\pm 150$ & $0.12\pm 0.03$ & $0.549\pm0.007$ & $0.21\pm0.05$  & $41^{+0.3}_{-0.2}$ & $1.0$ & $0.090\pm 0.009$ & $1.1^{+0.2}_{-0.1}$ & $24\pm 0.5$ & 1 \\
		J0952$-$0607 & 6.42 &0.063 & $2300^{+150}_{-500}$ & $6200\pm 150$ & $1.9\pm0.5$ & $7.20\pm 0.02$ & $0.26\pm0.06$& $2.8\pm 0.5$ & $0.974 \pm 0.008$ & $0.21\pm 0.02$ & $5.7^{+0.5}_{-0.4}$& $56^{+3}_{-2}$ & 1 \\
	    J1124$-$3653 & 5.45 & 0.080 & $1500 \pm 700$ & $3990\pm30$ & $0.286\pm 0.010$ & $1.83^{\rm a}$ & $0.156\pm0.005^{\rm a}$ & $6.3^{+1.6}_{-1.4}$ & $0.960\pm0.014$ & $0.22\pm0.02$  & $2.72^{+0.10}_{-0.08}$ & $44.9^{+4.0}_{-1.9}$ & 2 \\
		J1301+0833 & 6.48 & 0.078 & $2430^{+90}_{-120}$ & $5270\pm 110$ & $0.70 \pm 0.06$ & $-$ & $-$ & $6.9^{+1.4}_{-1.2}$ & $0.85\pm0.05$ & $0.20\pm0.02$ & $2.23^{+0.08}_{-0.13}$ & $44^{+2.8}_{-2.2}$ & 2 \\
		J1311$-$3430 & 1.56 & 0.011 & $5000\pm400$ & $20000\pm 3000$ & $40 \pm 20$ & $5.053\pm 0.004$ & $8\pm 4$ & $39\pm 6$ & $1.0$ & $0.0702 \pm 0.0015$ & $2.60^{+0.04}_{-0.05}$ & $81^{+4}_{-6}$ & 2,3 \\
		J1544+4937 & 2.90 & 0.033 & $2870^{+90}_{-80}$ & $4720 \pm 70$ & $0.20\pm 0.03$ & $1.42 \pm 0.02$ & $0.14\pm 0.02$ & $12.67^{+0.08}_{-0.07}$  & $1.0$ & $ 0.141\pm 0.019$ & $3.1^{+0.3}_{-0.2}$ & $47^{+3}_{-2}$ & 1 \\
		J1555$-$2908 & 5.60 & 0.151 & $2600^{+300}_{-1000}$ & $9380 \pm 40$ & $8.8\pm 0.3$ & $26.9\pm 0.6$ & $0.327\pm0.012$ & $3.6\pm0.5$  & $1.0$ & $0.286\pm0.007$ & $5.1^{+0.5}_{-0.7}$ & $79^{+3}_{-2}$ & 4, 5 \\
		J1641+8049 & 2.18 & 0.064 & $3130\pm 80$ & $8500\pm 300$ & $1.9\pm 0.5$ & $4.28\pm 0.04^{\rm b}$ & $0.37^{+0.09}_{-0.08}{}^{\rm b}$ & $23.11^{+0.09}_{-0.06}$ & $1.0$ & $0.151\pm 0.015$ & $4.7\pm 0.3$ & $57\pm 1$ & 1 \\
		J1653$-$0158 & 1.25 & 0.011 & $3300^{+200}_{-300}$ & $6300 \pm 400$ & $0.31\pm 0.03$ & $0.45\pm 0.04$ & $0.71\pm 0.09$ & $49^{+18}_{-15}$ & $1.0$ & $0.067\pm 0.009$ & $0.84\pm 0.04$ & $72 \pm 5$ & 6 \\
		J1810+1744 & 3.60 & 0.095 & $3470 \pm 30$ & $9340 \pm 20$ & $6.00\pm 0.06$ & $4.18^{\rm a}$ & $1.444\pm0.014^{\rm a}$ & $6.6\pm 1.0$ & $1.0$ & $0.221 \pm 0.004$ & $3.03 \pm 0.01$ & $65.7\pm 0.4$ & 7,8 \\
		J1959+2048 & 9.17 & 0.089 & $2670 \pm 30$ & $7150 \pm 20$ & $3.01 \pm 0.04$ & $9.3 \pm 0.3$ & $0.326 \pm 0.010$ & $1.11 \pm 0.17$ & $0.985\pm 0.003$ & $0.313\pm 0.007$ & $2.04 \pm 0.01$ & $62.5^{+1.3}_{-1.1}$ & 2 \\
		J2051$-$0827 & $2.38$ & 0.045 & $2750^{+65}_{-75}$ & $5040\pm 100$ & $0.23 \pm 0.05$ & $0.46 \pm 0.05$ & $0.497 \pm 0.012$ & $20.2^{+0.3}_{-0.2}$ & $0.988\pm0.004$ & $0.127\pm 0.012$ & $2.48 \pm 0.2$ & $56 \pm 2$ & 9 \\
		J2052+1219 & 2.75 & 0.061 & $3020^{+90}_{-130}$ & $6710 \pm 40$ & $0.94\pm 0.03$ & $1.70 \pm 0.16$ & $0.56 \pm 0.06$ & $14\pm 2$ & $1.0$ & $0.166 \pm 0.008$ & $3.94\pm 0.07$ & $55 ^{+3}_{-2}$ & 2 \\
		J2241$-$5236 & 3.50 & 0.026 & $2820\pm60$ & $4690\pm 20$ & $0.28\pm 0.06$ & $1.76\pm 0.04$ & $0.161\pm 0.005$ & $28^{+7}_{-6}$ & $0.83\pm 0.04$ & $0.096 \pm 0.010$ & $1.24^{+0.04}_{-0.05}$ & $49.7^{+2.2}_{-1.9}$ & 2 \\
		J2256$-$1024 & 5.11 & 0.083 & $2400\pm 400$ & $4000\pm 300$ & $0.25\pm 0.04$ & $3.70\pm 0.08$ & $0.068\pm 0.010$ & $29^{+79}_{-14}$ & $0.40\pm 0.20$ & $0.07\pm 0.04$ & $0.65\pm0.45$ & $68\pm11$ & 10 \\
		 \\
		\hline
	\end{tabular}
	\begin{tablenotes}
	\item $^{\rm a}$ Error bars were not provided in the Fermi LAT catalog for some timing parameters, which led to underestimated uncertainties  \citep{2013ApJS..208...17A}.
	\item $^{\rm b}$ Following the discussion in Sec. \ref{sec:spindown}, the Shklowskii correction was not included.
	\item References: [1] This work, [2] \citet{Draghis2019}, [3] \citet{Romani2015}, [4] \citet{Kennedy2022}, [5] \citet{Ray2022}, [6] \citet{Nieder2020}, [7] \citet{Schroeder2014}, [8] \citet{Romani2021}, [9] \citet{Dhillon2022}, [10] \citet{Breton2013+Modelling}
	\end{tablenotes}
	\end{threeparttable}

    \label{tab:corr}
    \end{table}
\end{landscape}

\begin{figure*}
    \centering
    \includegraphics[width=\textwidth]{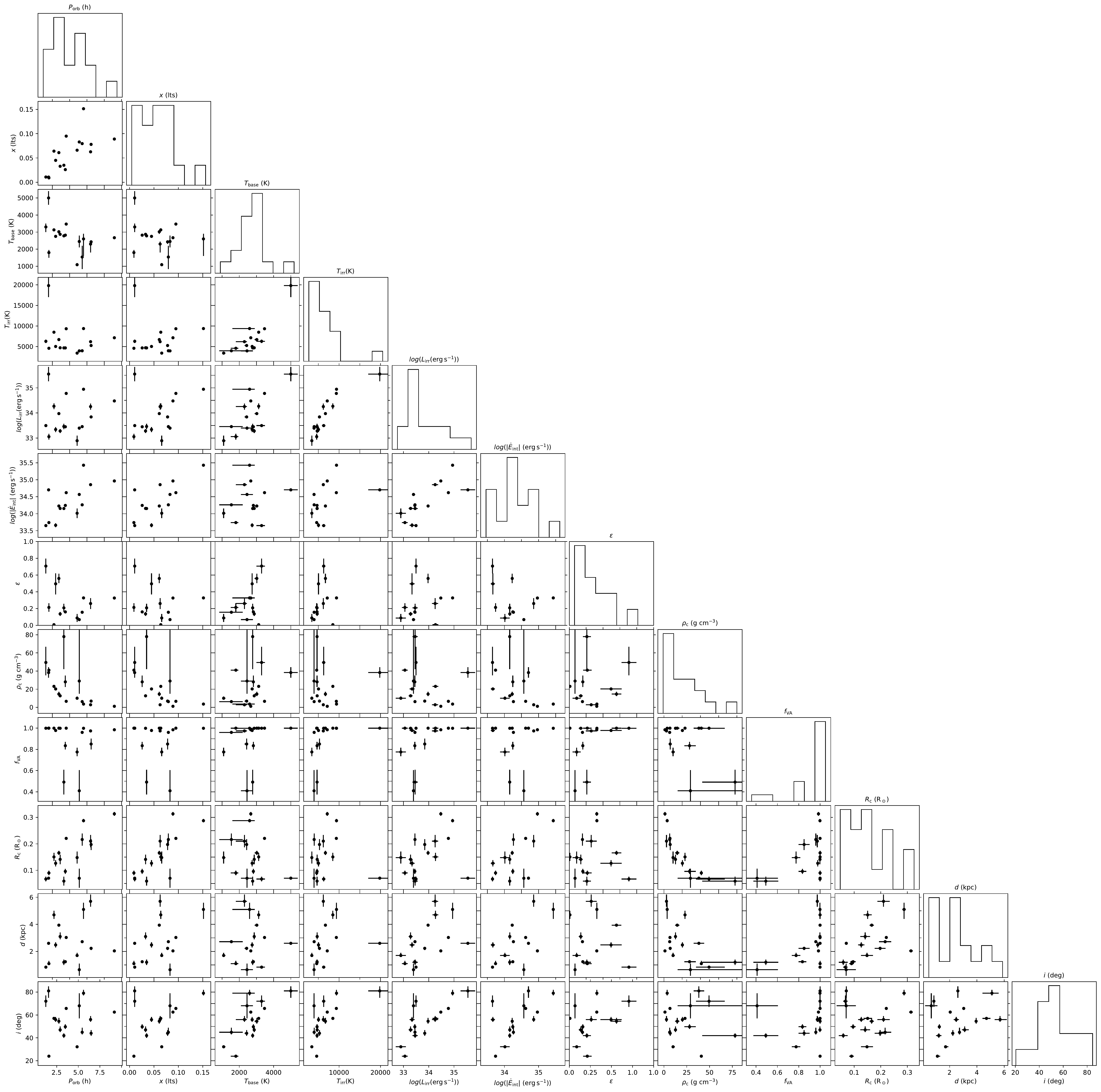}
    \caption{Corner plot of the correlations between pairs of parameters reported in Table \ref{tab:corr}.}
    \label{fig:corr}
\end{figure*}

\newpage

\section{Flux density light curves and fitting residuals}
\label{appendix:fluxdensitylc}
\begin{figure*}
    \centering
    \includegraphics[width=\columnwidth]{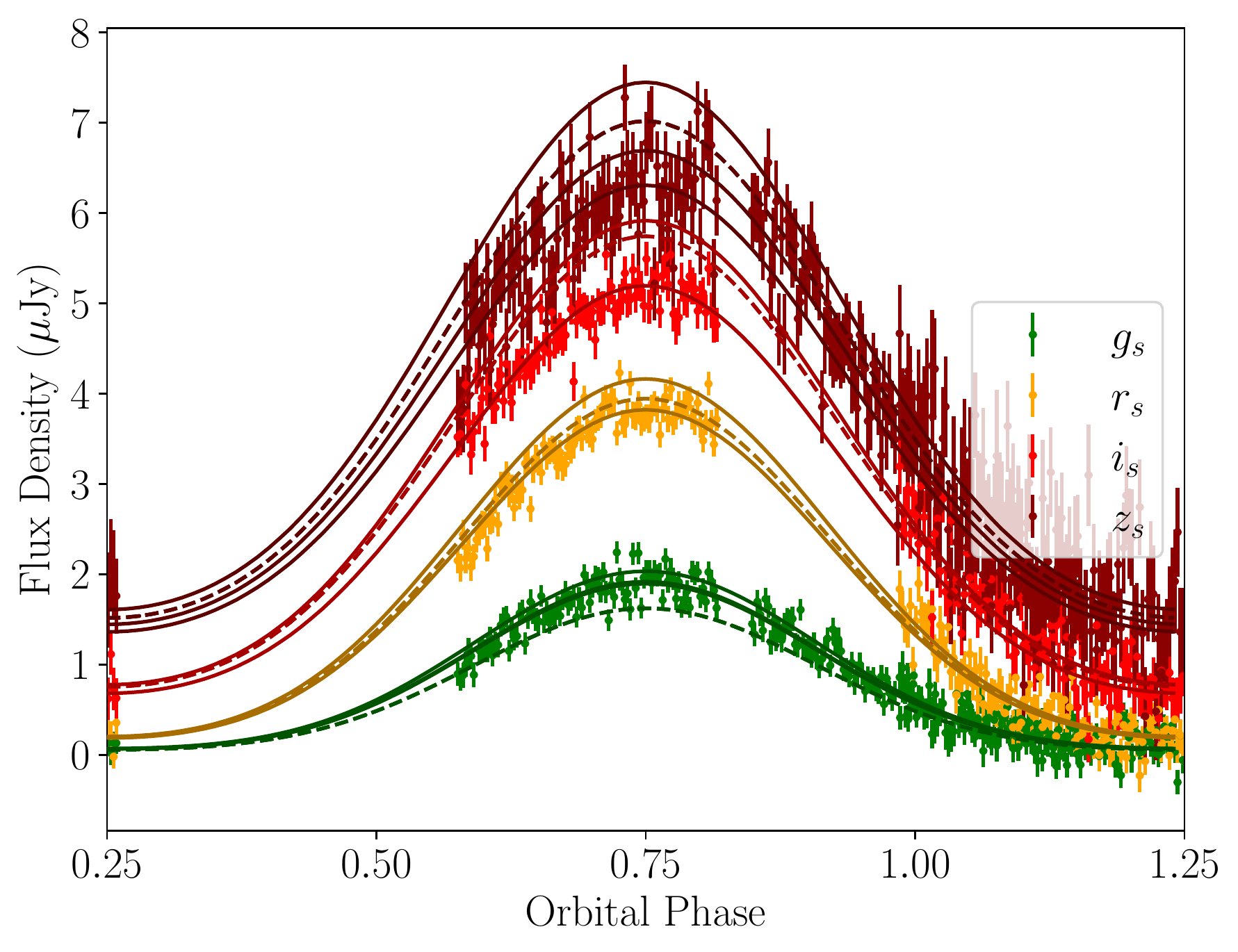}\includegraphics[width=\columnwidth]{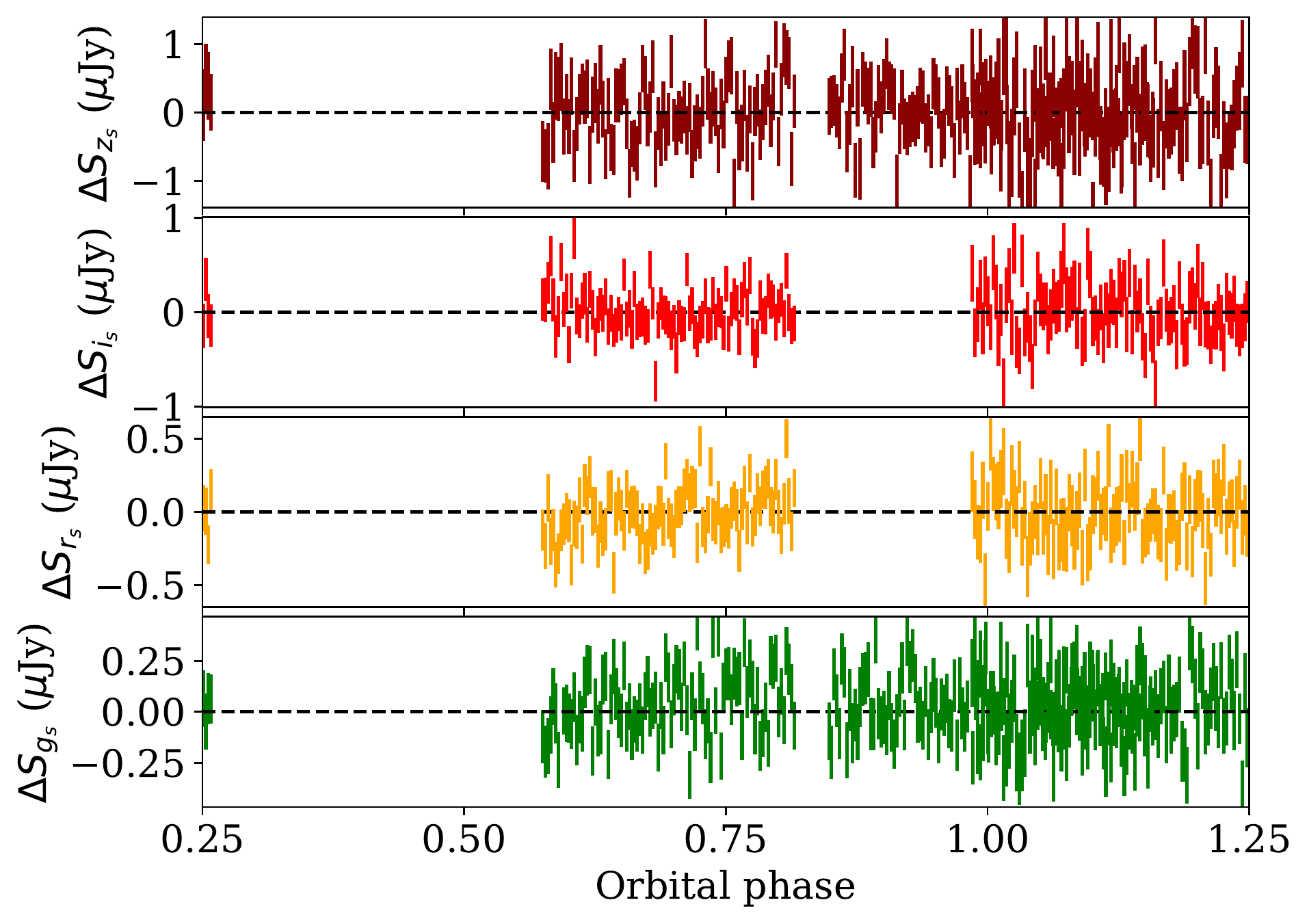}
    \includegraphics[width=\columnwidth]{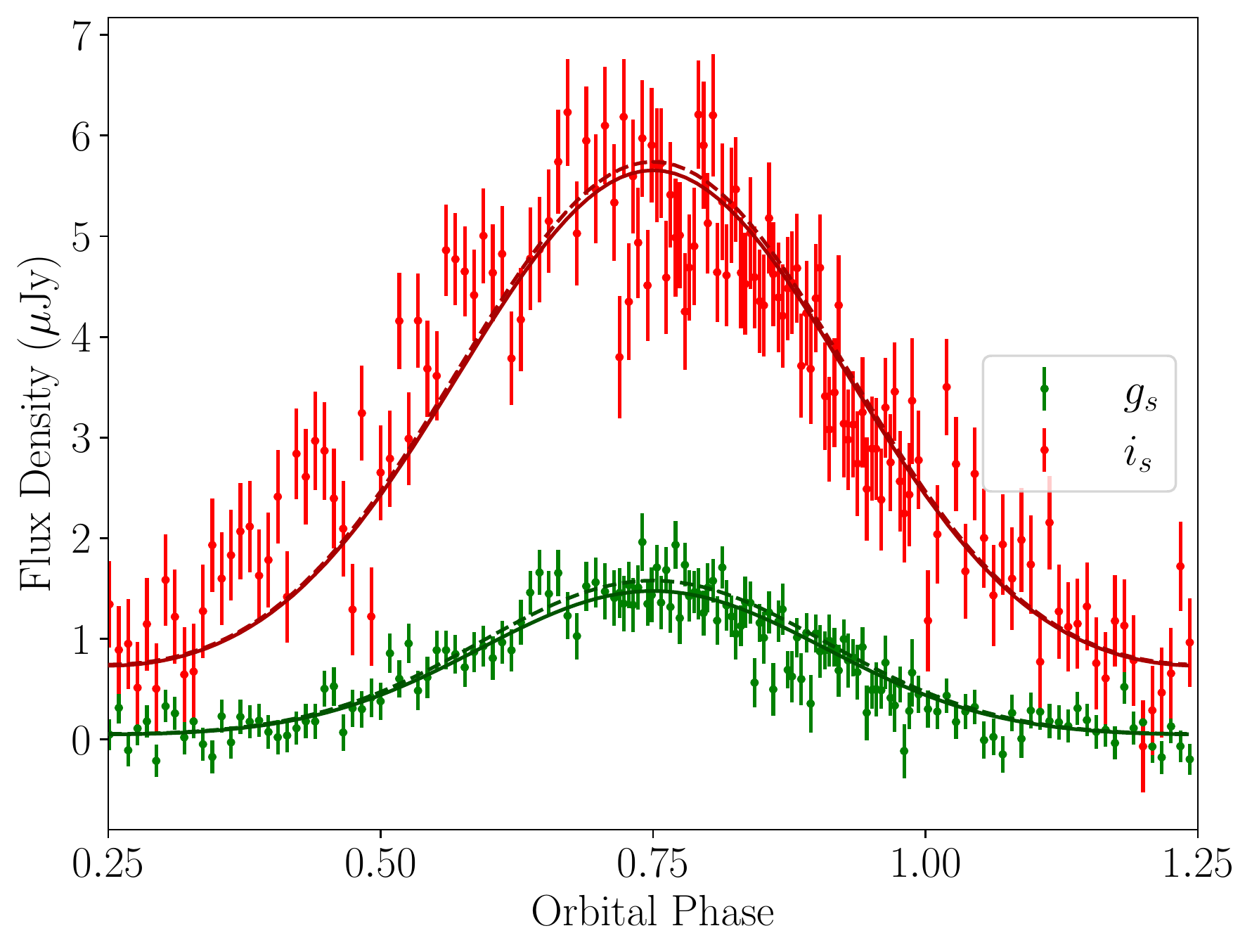}\includegraphics[width=\columnwidth]{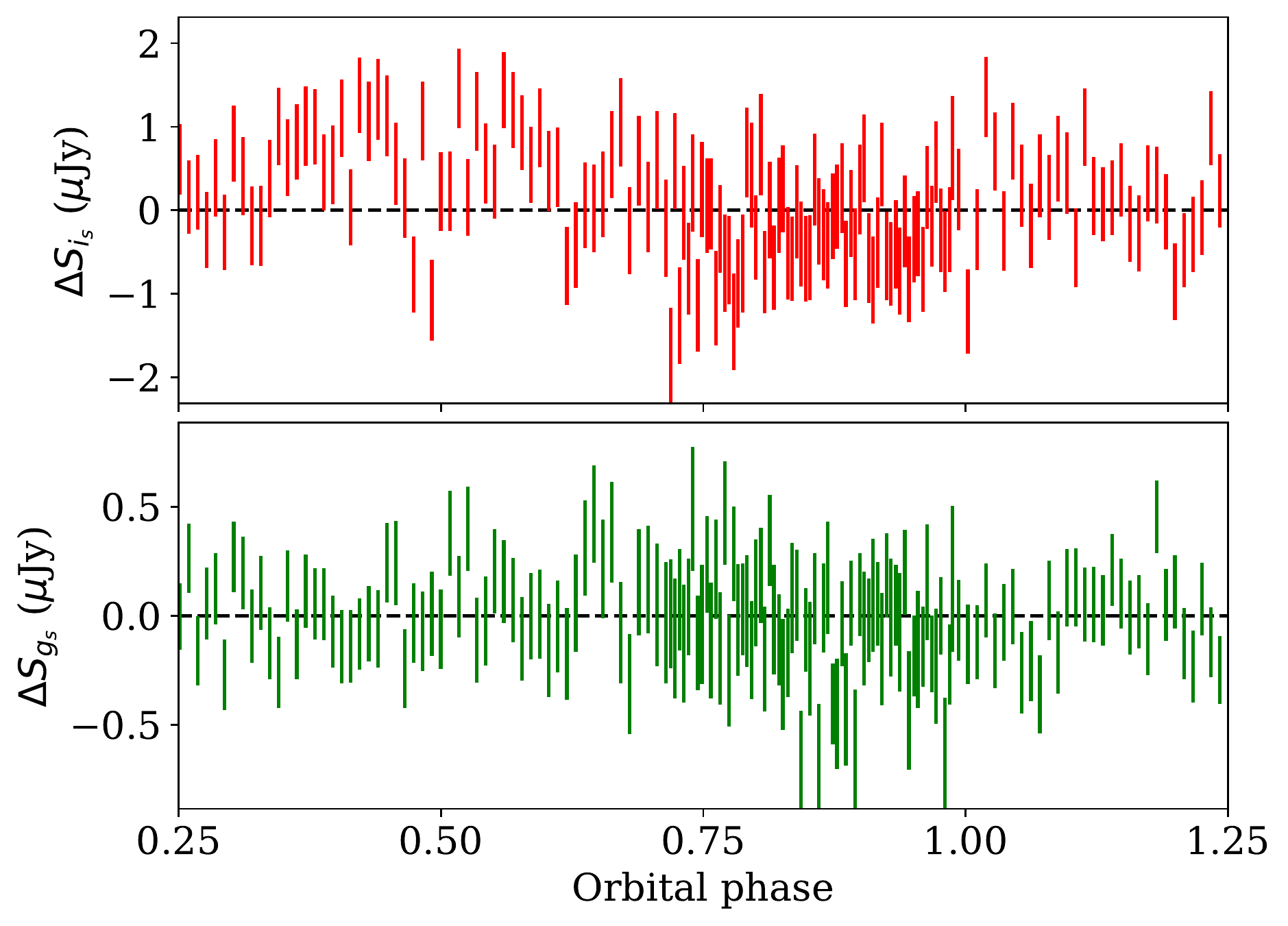}
    \caption{Top-left panel: The best-fitting \textsc{Icarus} model for the HiPERCAM optical light curve of \psra. Dashed lines show the model light curve in each band, while solid curves show the same model but allowing for a small offset in the band calibration so it best fits the data. Due to the simultaneous fit of all datasets, the dashed theoretical model remains the same, while the solid lines differ by simply an offset in magnitude, which varies from night to night. Top-right panel: Residuals resulting from subtraction of the best fit from the observed data. Bottom panels follow the same description but corresponding to ULTRACAM data.}
    \label{fig:modelJ0023flux}
\end{figure*}

\begin{figure*}
    \centering
    \includegraphics[width=\columnwidth]{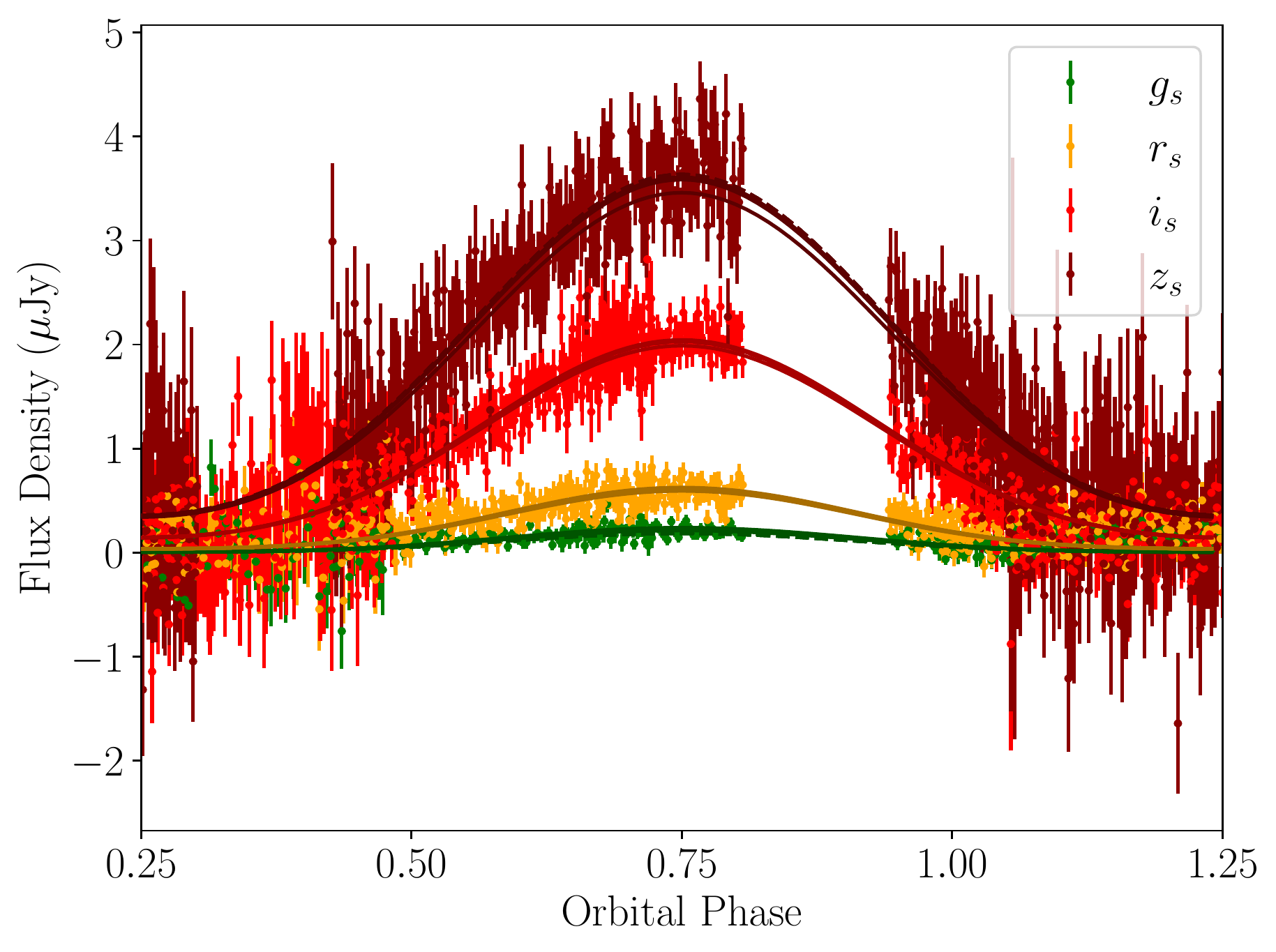}\includegraphics[width=\columnwidth]{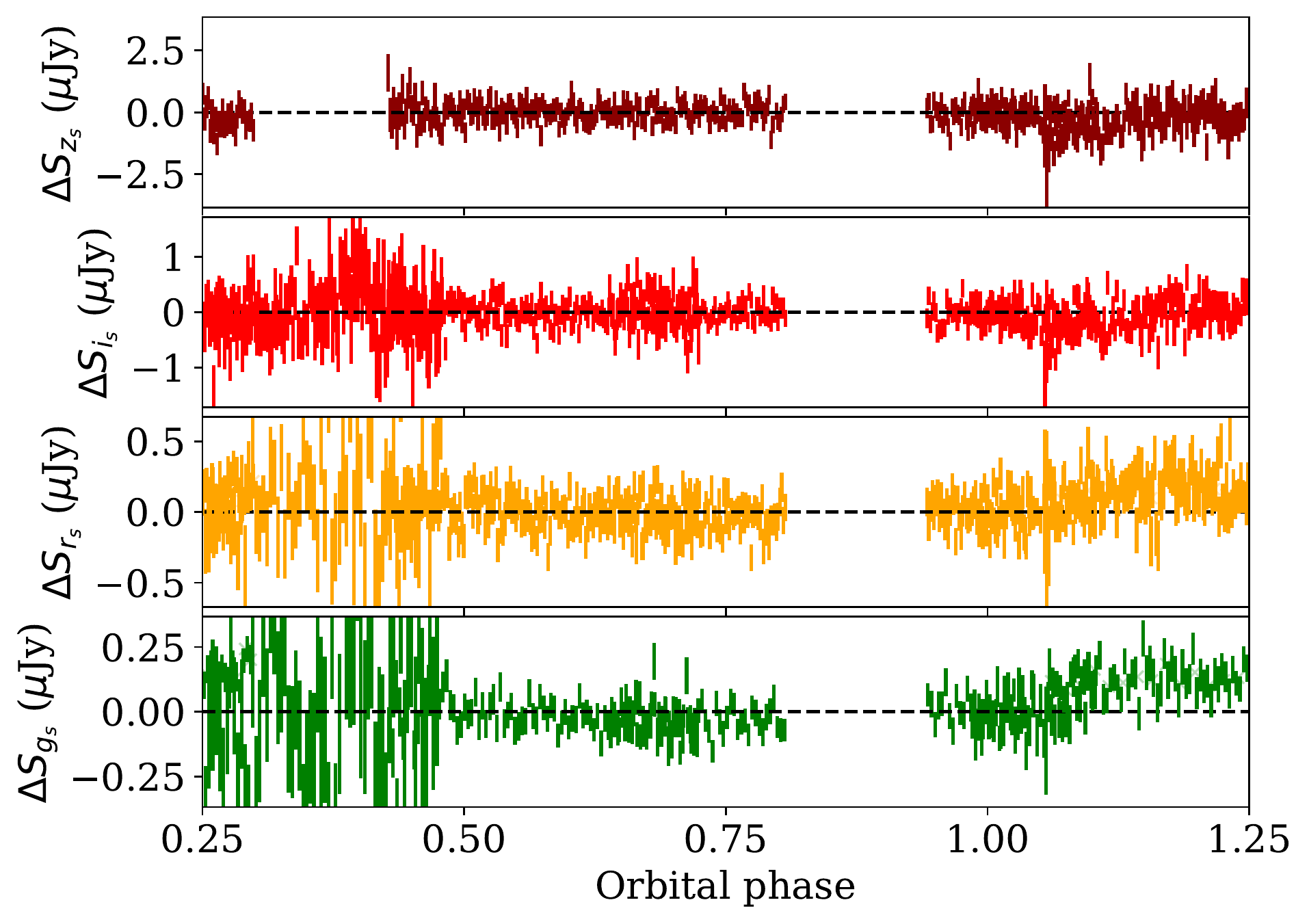}
    \caption{Left panel: The best-fitting \textsc{Icarus} model for the HiPERCAM optical light curve of \psrb. Right panel: Residuals resulting from subtraction of the best fit from the observed data. Fig. \ref{fig:modelJ0023flux} description remains valid.}
    \label{fig:modelJ0251flux}
\end{figure*}

\begin{figure*}
    \centering
    \includegraphics[width=\columnwidth]{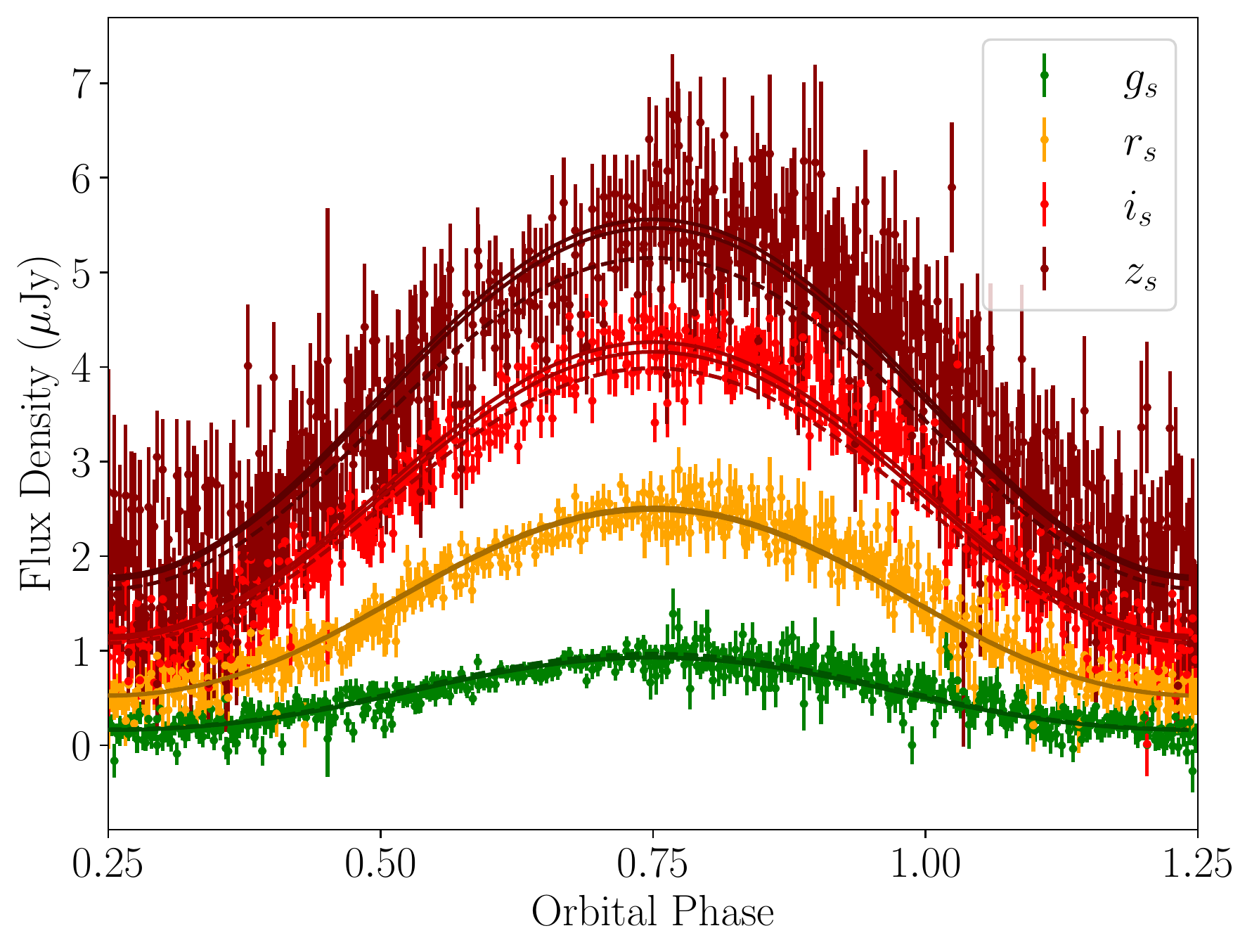}\includegraphics[width=\columnwidth]{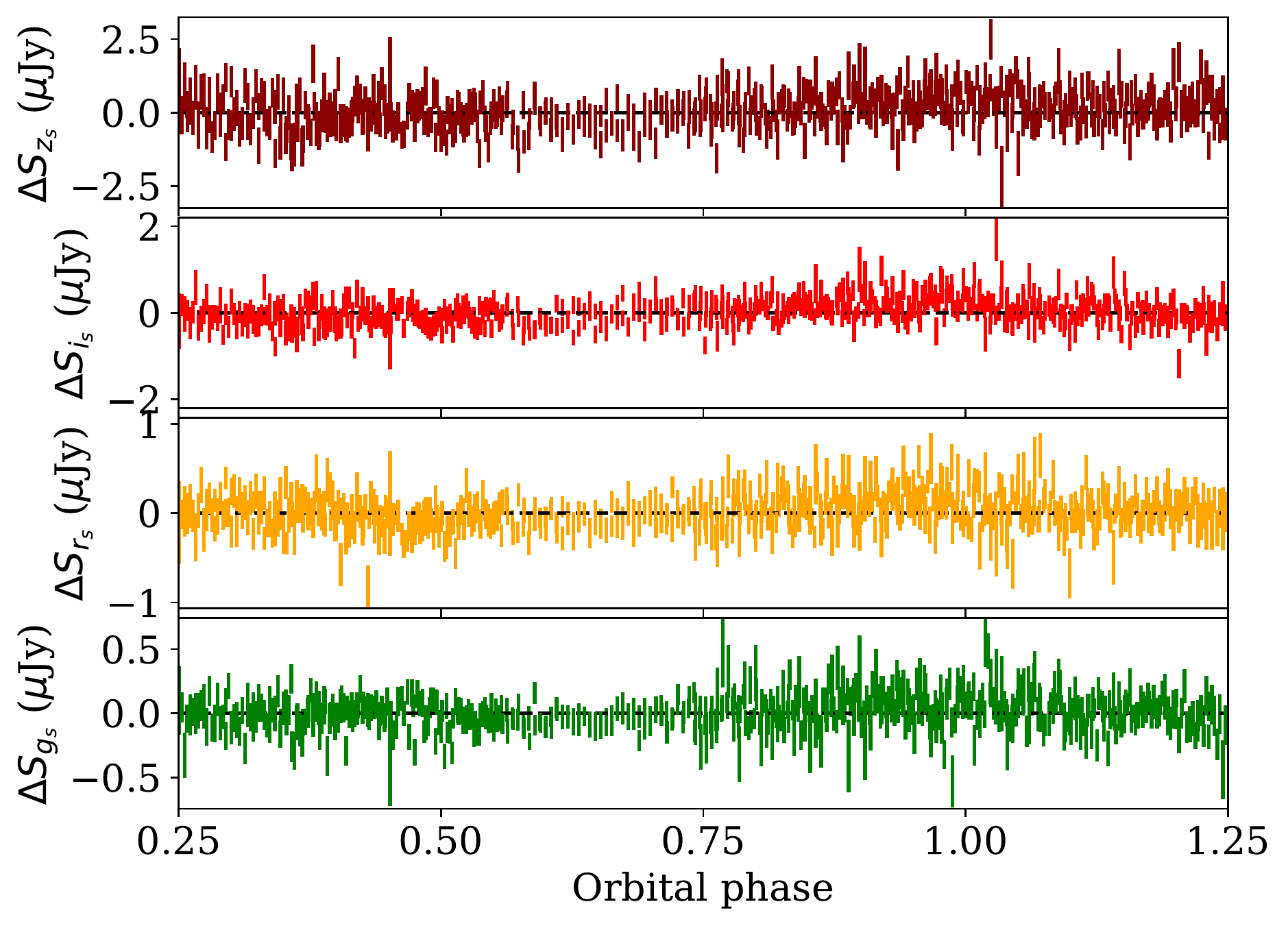}
    \caption{Left panel: The best-fitting \textsc{Icarus} model for the HiPERCAM optical light curve of \psrc. Right panel: Residuals resulting from subtraction of the best fit from the observed data. Fig. \ref{fig:modelJ0023flux} description remains valid.}
    \label{fig:modelJ0636flux}
\end{figure*}

\begin{figure*}
    \centering
    \includegraphics[width=\columnwidth]{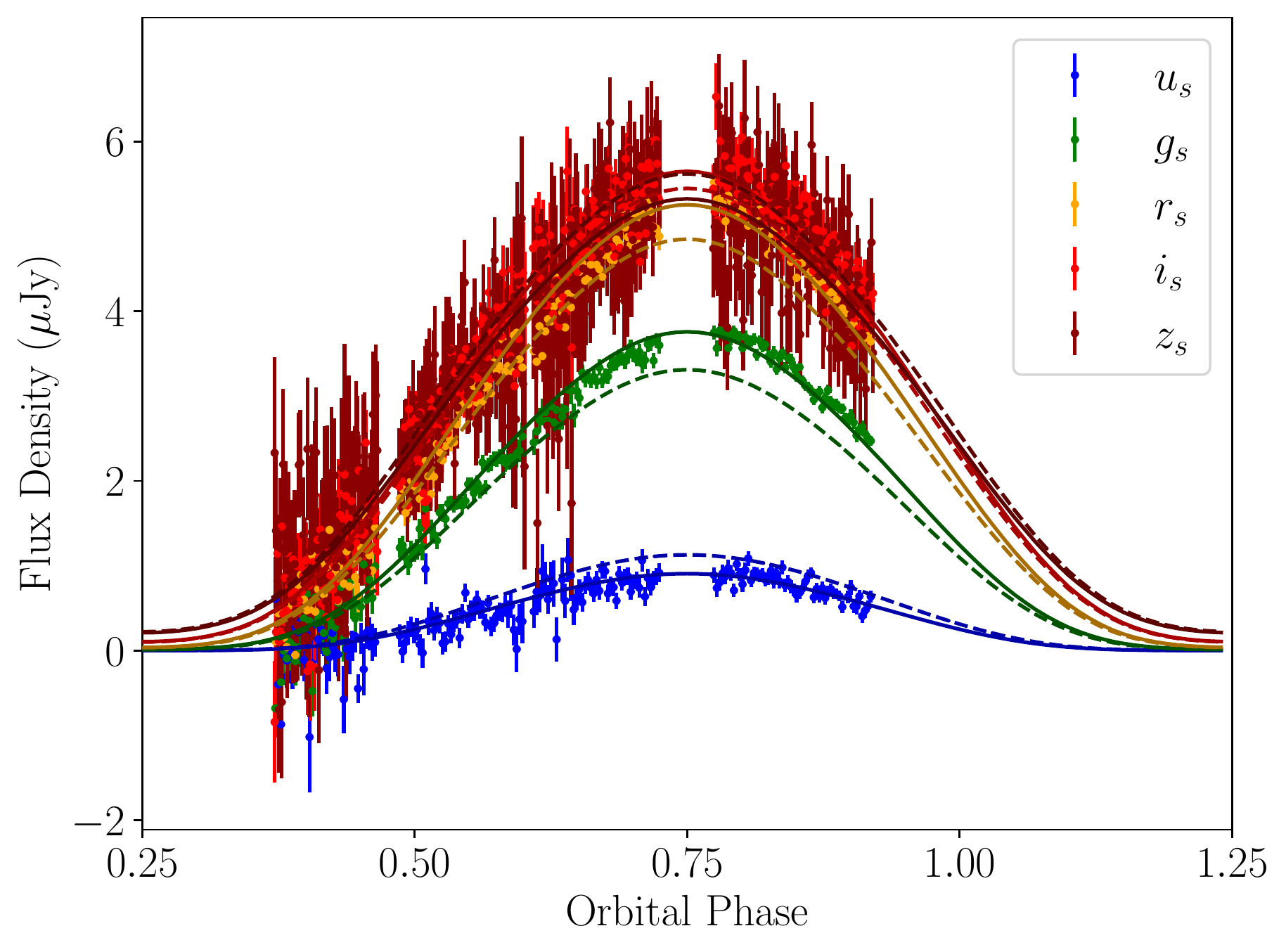}\includegraphics[width=\columnwidth]{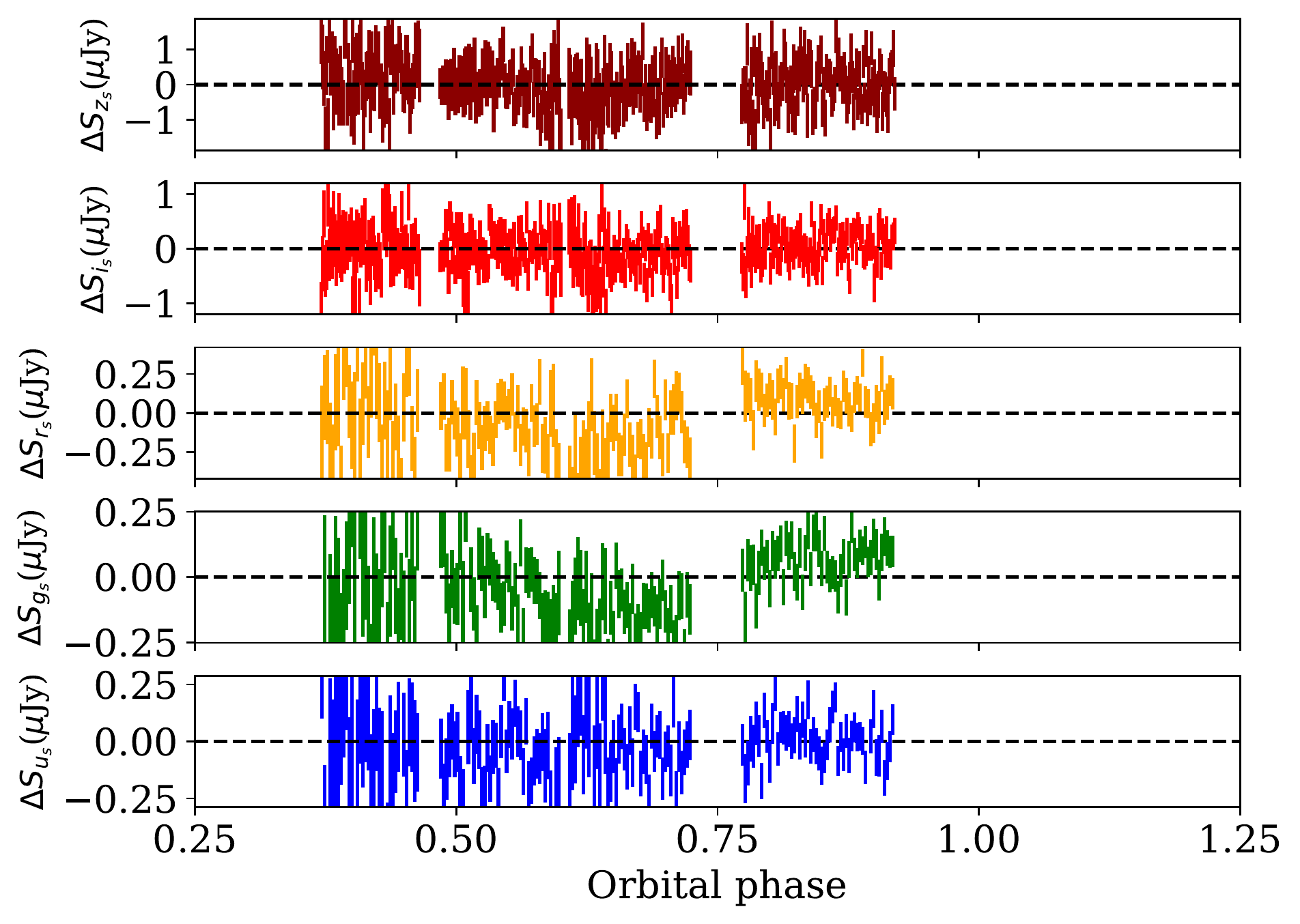}
    \caption{Left panel: The best-fitting \textsc{Icarus} model for the HiPERCAM optical light curve of \psrd. Right panel: Residuals resulting from subtraction of the best fit from the observed data. Fig. \ref{fig:modelJ0023flux} description remains valid.}
    \label{fig:modelJ0952flux}
\end{figure*}

\begin{figure*}
    \centering
    \includegraphics[width=\columnwidth]{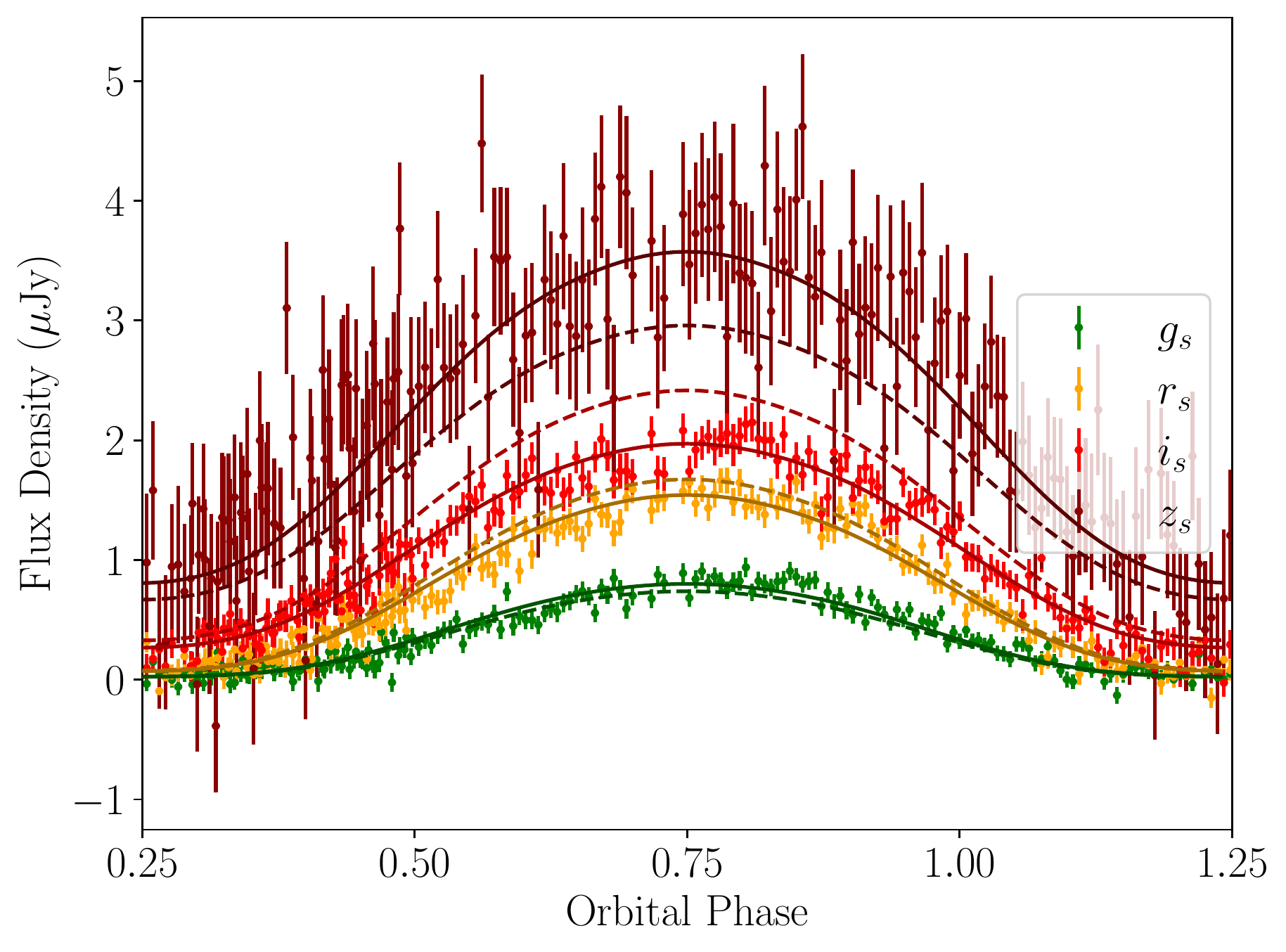}\includegraphics[width=\columnwidth]{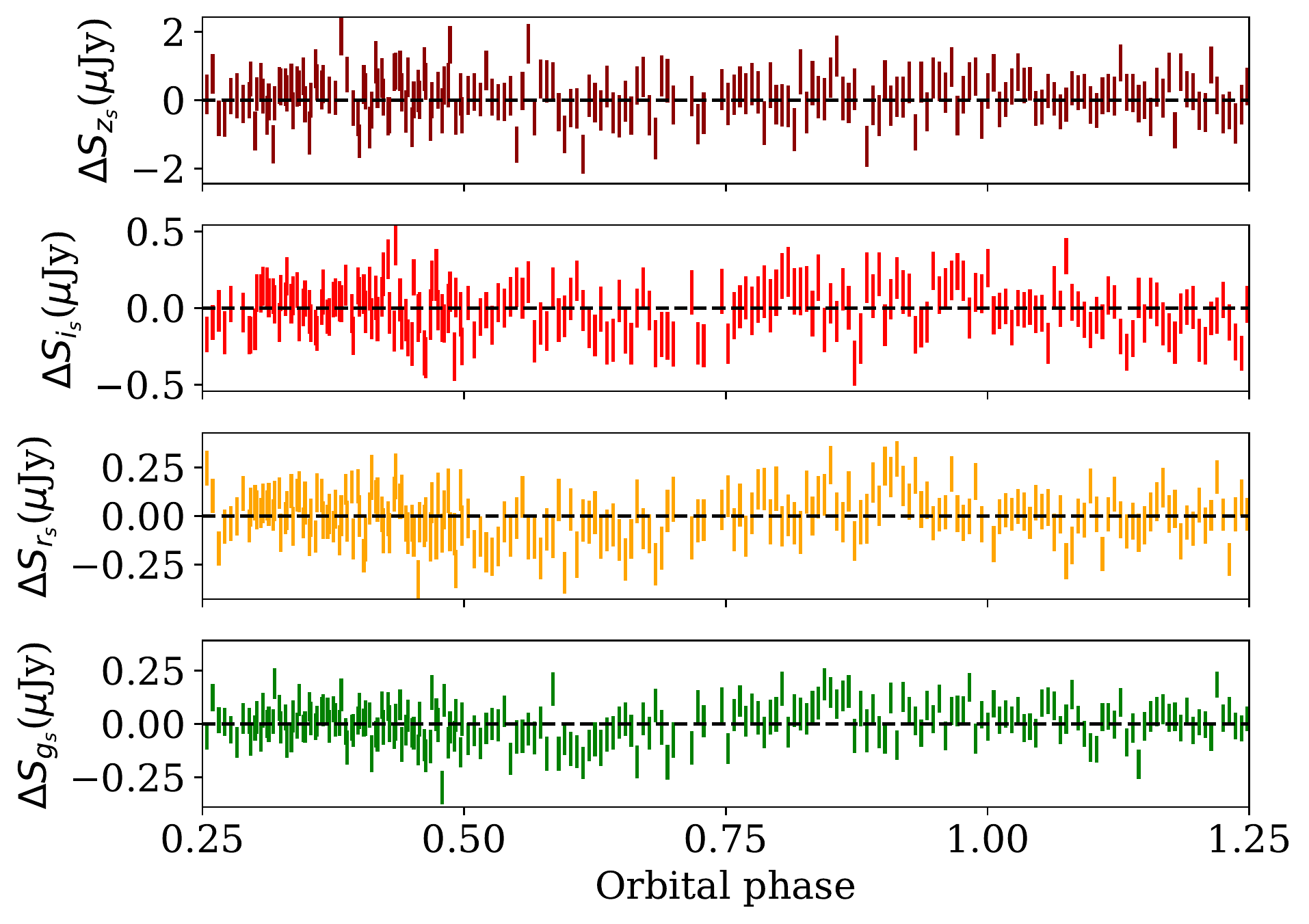}
    \caption{Left panel: The best-fitting \textsc{Icarus} model for the HiPERCAM optical light curve of \psre. Right panel: Residuals resulting from subtraction of the best fit from the observed data. Fig. \ref{fig:modelJ0023flux} description remains valid.}
    \label{fig:modelJ1544flux}
\end{figure*}

\begin{figure*}
    \centering
    \includegraphics[width=\columnwidth]{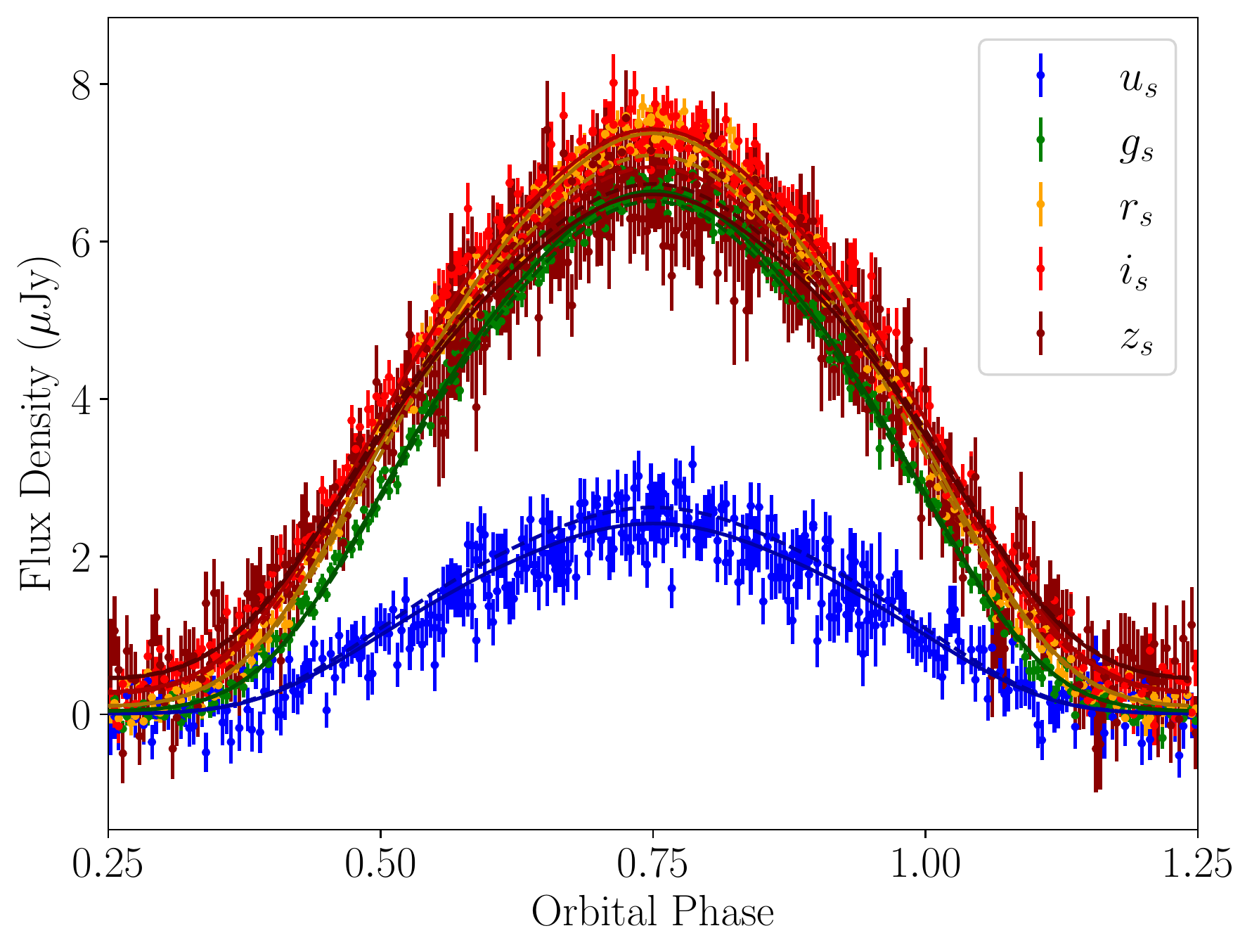}\includegraphics[width=\columnwidth]{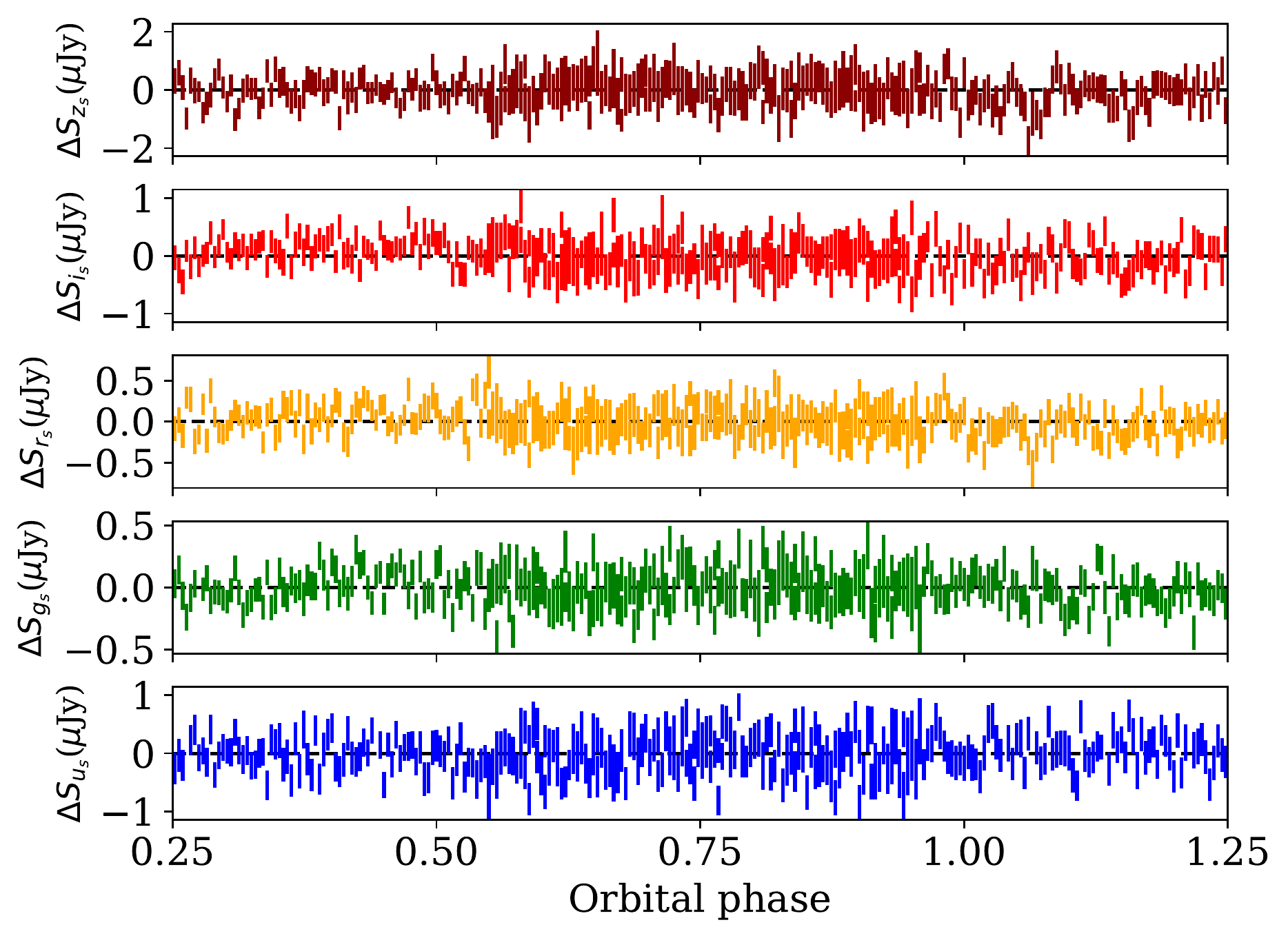}
    \caption{Left panel: The best-fitting \textsc{Icarus} model for the HiPERCAM optical light curve of \psrf. Right panel: Residuals resulting from subtraction of the best fit from the observed data. Fig. \ref{fig:modelJ0023flux} description remains valid.}
    \label{fig:modelJ1641flux}
\end{figure*}

\newpage

\section{Posterior distributions from the \textsc{Multinest} analysis.}
\label{sec:cornerplots}

\begin{figure*}
    \centering
    \includegraphics[width=\textwidth]{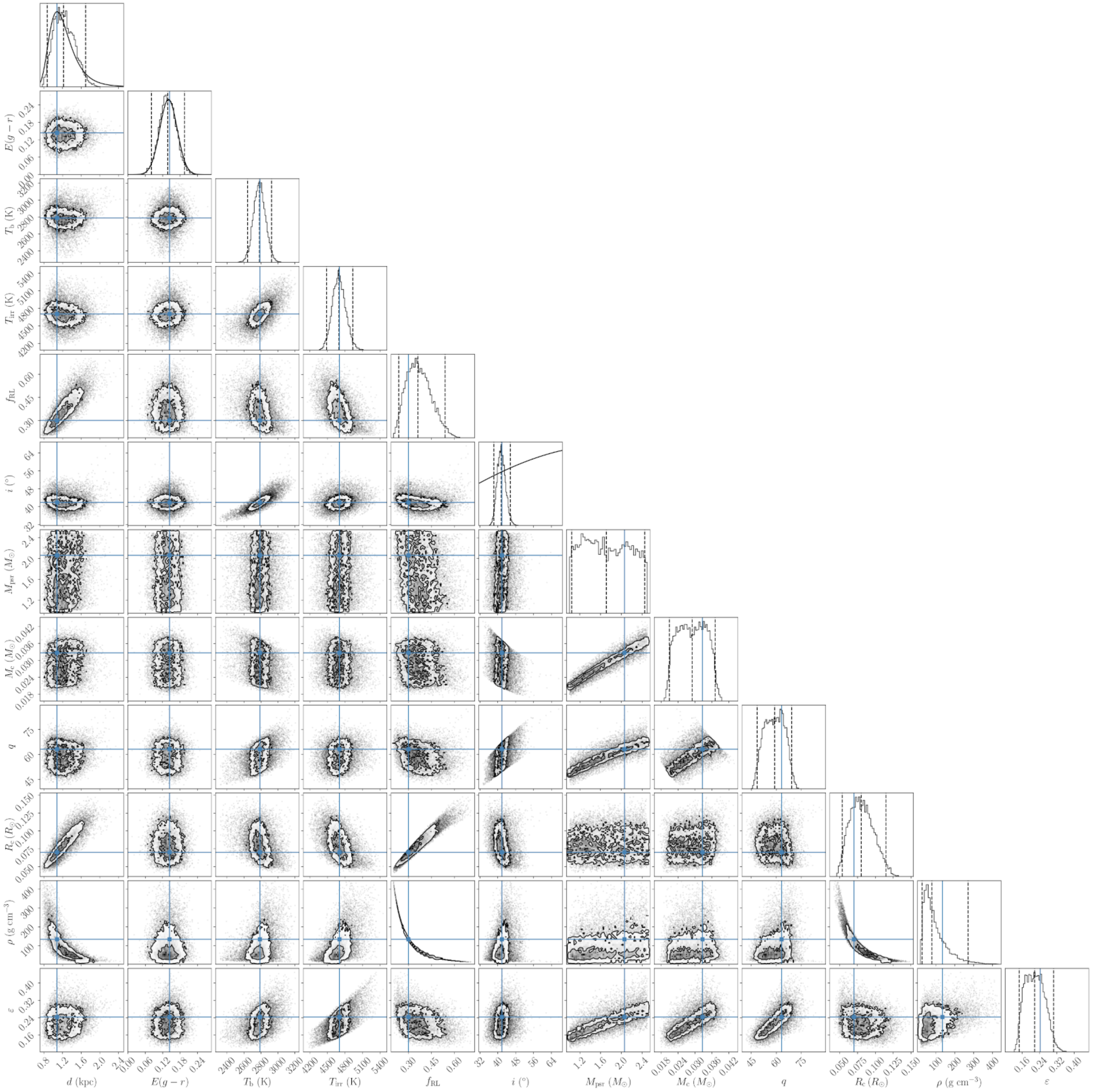}
    \caption{Posterior distributions for the \textsc{Icarus} model parameters of \psra{}. The prior distributions on $d$ and $i$ are shown by black curves over their marginal distributions. The final five parameters $q, M_{\rm c}, R_{\rm c}, \rho$ and $\epsilon$ are derived from the other seven parameters and the pulsar timing ephemeris. On the 1-dimensional marginal distributions, dashed vertical lines indicate the median and 95\% confidence interval. On 2-dimensional conditional distributions, contour lines indicate $1\sigma$ and $2\sigma$ levels. Blue, solid lines mark the maximum likelihood solution.}
    \label{fig:corner_plotJ0023}
\end{figure*}

\newpage

\begin{figure*}
    \centering
    \includegraphics[width=\textwidth]{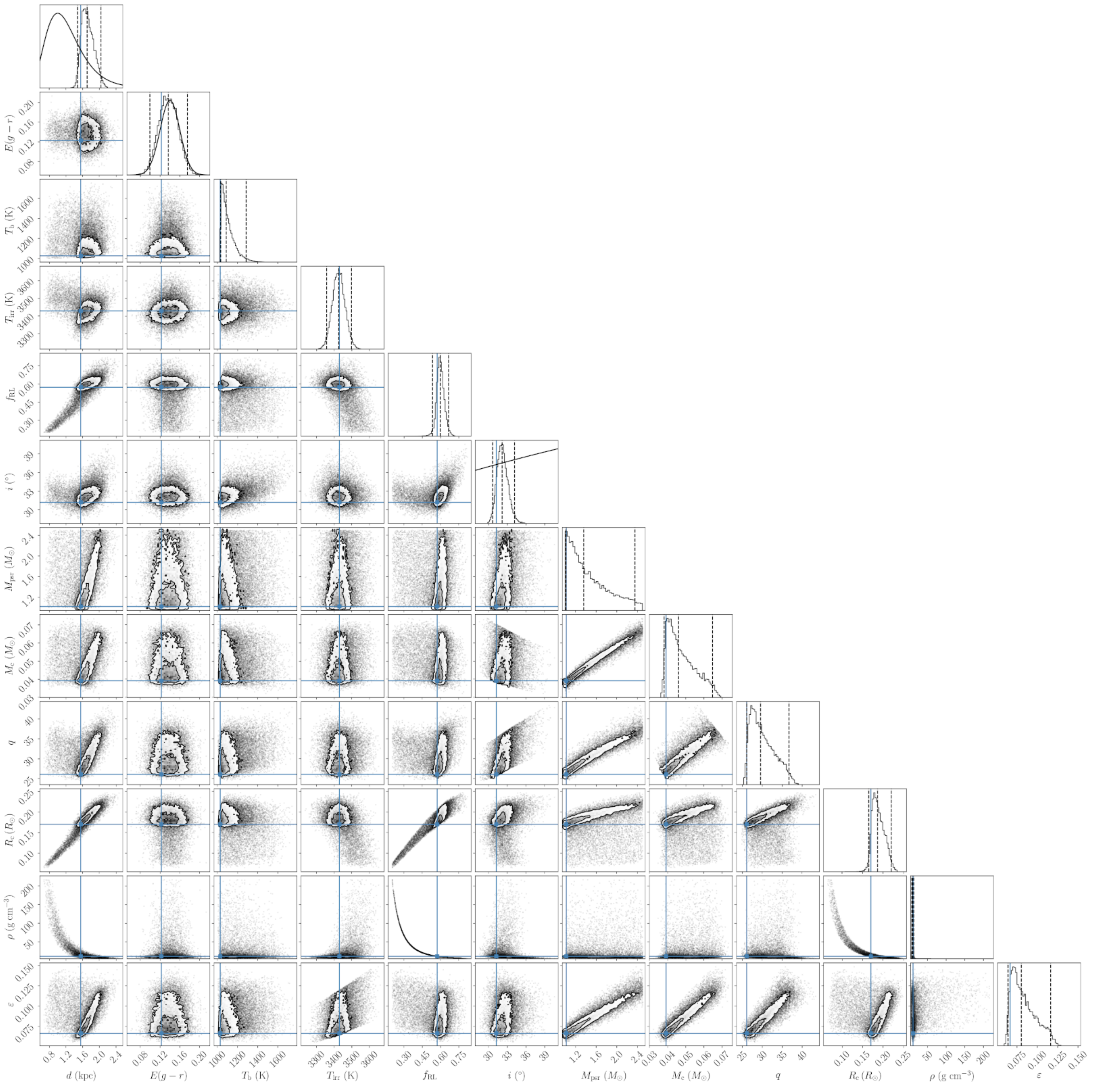}
    \caption{Posterior distributions for the \textsc{Icarus} model parameters of \psrb{}.}
    \label{fig:corner_plotJ0251}
\end{figure*}

\newpage

\begin{figure*}
    \centering
    \includegraphics[width=\textwidth]{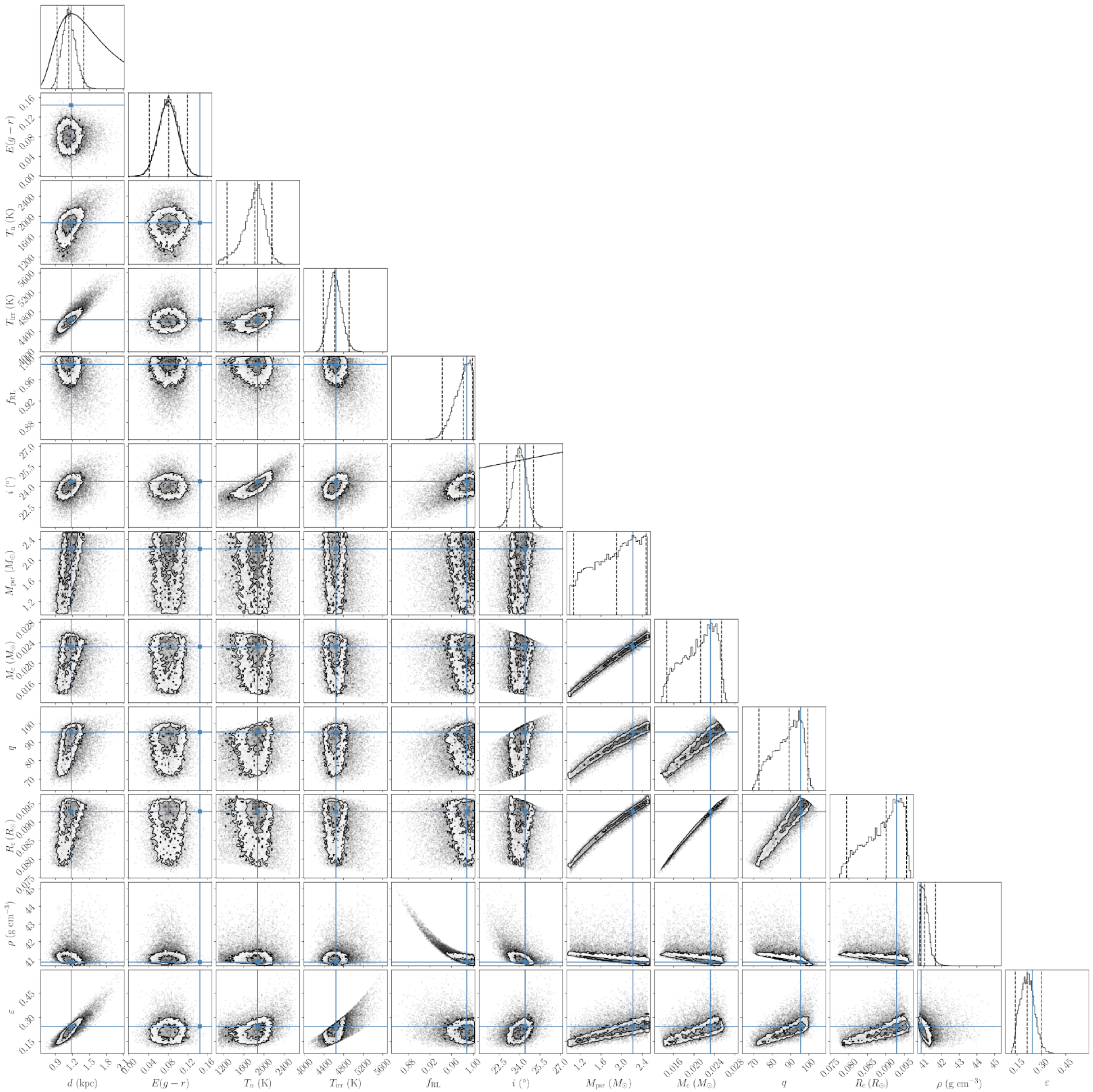}
    \caption{Posterior distributions for the \textsc{Icarus} model parameters of \psrc{}. }
    \label{fig:corner_plotJ0636}
\end{figure*}

\newpage

\begin{figure*}
    \centering
    \includegraphics[width=\textwidth]{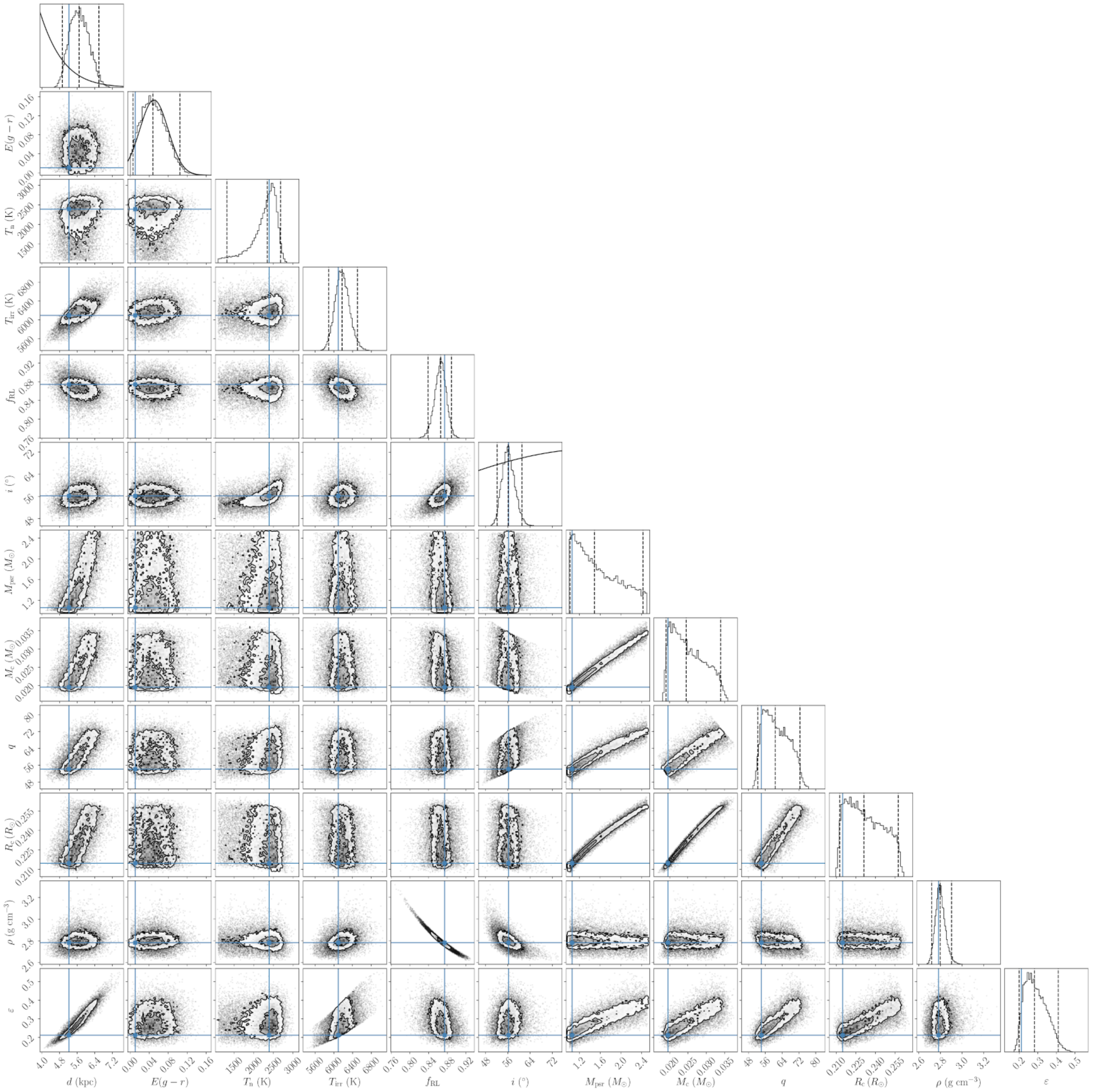}
    \caption{Posterior distributions for the \textsc{Icarus} model parameters of \psrd{}.}
    \label{fig:corner_plotJ0952}
\end{figure*}

\newpage

\begin{figure*}
    \centering
    \includegraphics[width=\textwidth]{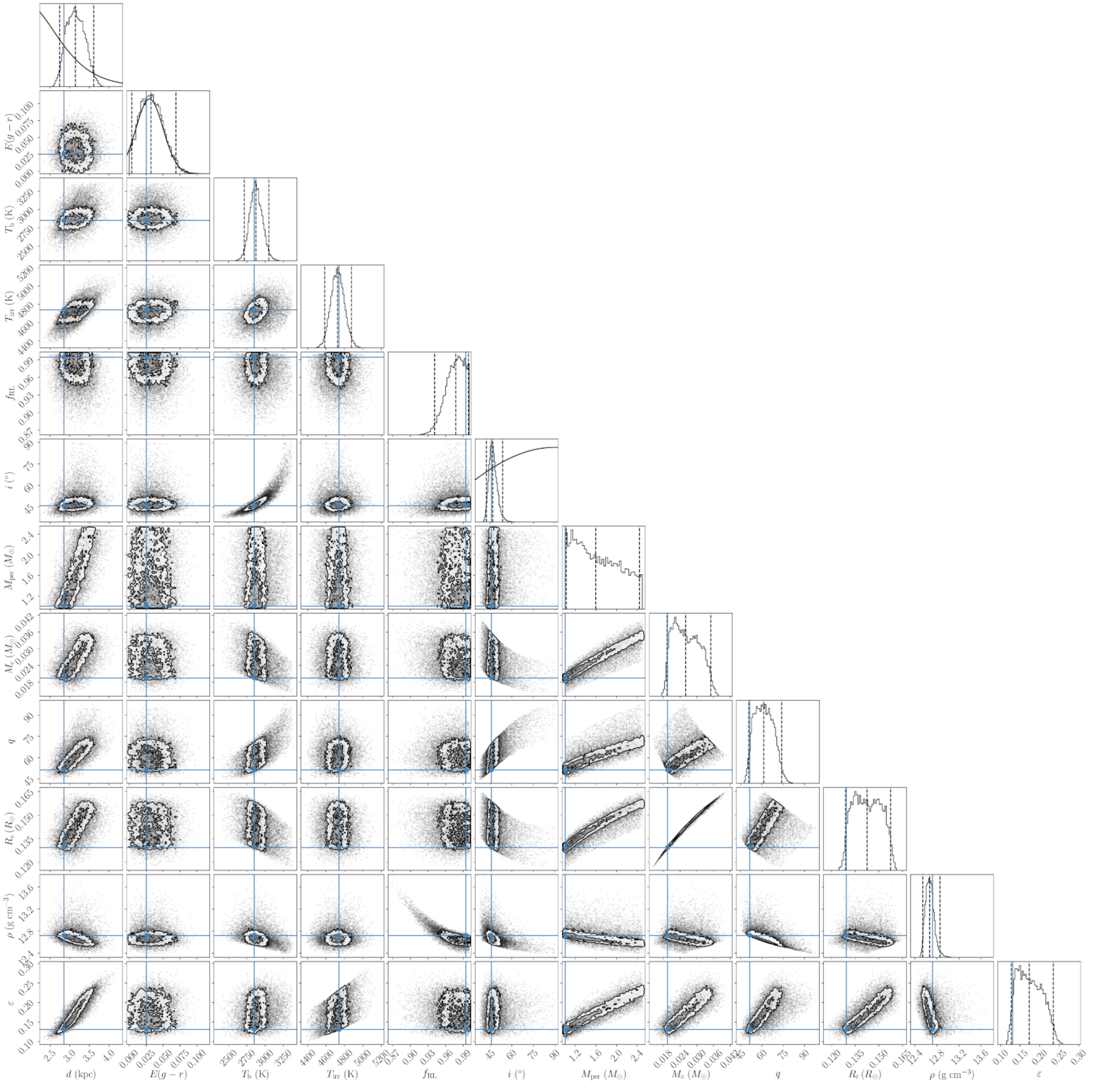}
    \caption{Posterior distributions for the \textsc{Icarus} model parameters of \psre{}. }
    \label{fig:corner_plotJ1544}
\end{figure*}

\newpage

\begin{figure*}
    \centering
    \includegraphics[width=\textwidth]{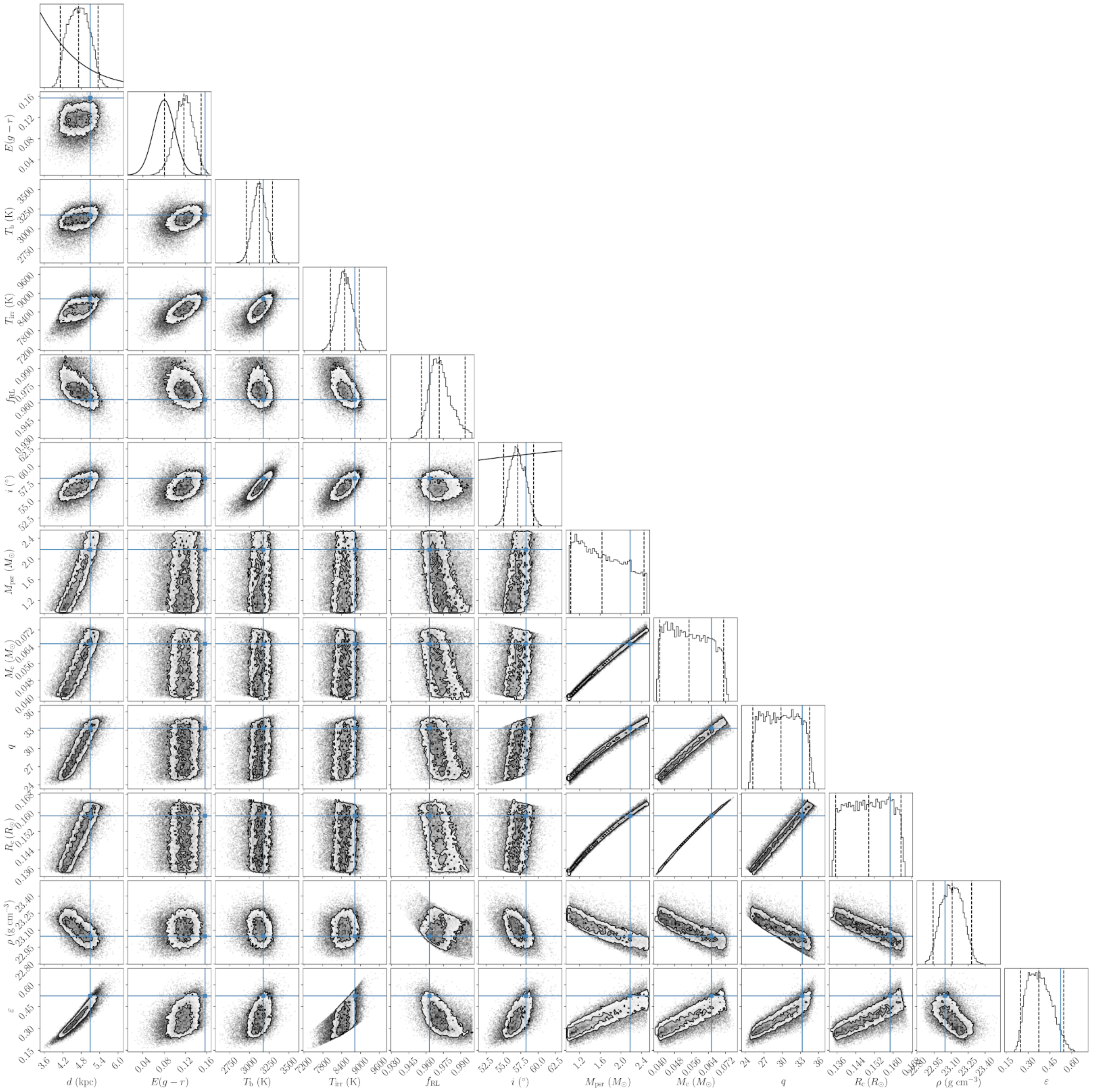}
    \caption{Posterior distributions for the \textsc{Icarus} model parameters of \psrf{}. }
    \label{fig:corner_plotJ1641}
\end{figure*}



\bsp	
\label{lastpage}
\end{document}